# Construction of Multi-Dimensional Functions for Optimization of Additive-Manufacturing Process Parameters

B. Steingrimsson[*,1,2,3], A. Agrawal [4,5], X. Fan[6], A. Kulkarni[7], D. Thoma[8,9], and P. K. Liaw[10]

1. Imagars LLC, 2062 Thorncroft Drive Suite 1214, Hillsboro, OR 97124, USA. Email: baldur@imagars.com.

2. School of Mechanical, Industrial and Manufacturing Engineering, Oregon State University, Corvallis, OR 97331, USA. Email: baldur.steingrimsson@oregonstate.edu.

3. Department of Manufacturing, Mechanical and Engineering Technology, Oregon Institute of Technology, Wilsonville, OR 97070, USA. Email: baldur.steingrimsson@oit.edu.

4. Department of Materials Science and Engineering, University of Wisconsin, Madison, WI, 537006, USA. Email: ankur.agrawal@wisc.edu.

5. Grainger Institute for Engineering, University of Wisconsin, Madison, WI, 537006, USA. Email: ankur.agrawal@wisc.edu.

6. Department of Materials Science and Engineering, The University of Tennessee, Knoxville, TN, 37996, USA. Email: xfan5@vols.utk.edu.

7. Siemens Corporation, Corporate Technology, Charlotte, NC, 28277. USA. Email: anand.kulkarni@siemens.com.

8. Department of Materials Science and Engineering, University of Wisconsin, Madison, WI, 537006, USA. Email: dthoma@wisc.edu.

9. Grainger Institute for Engineering, University of Wisconsin, Madison, WI, 537006, USA. Email: dthoma@wisc.edu.

10. Department of Materials Science and Engineering, The University of Tennessee, Knoxville, TN, 37996, USA. Email: pliaw@utk.edu.

[*] Corresponding authors: baldur@imagars.com, baldur.steingrimsson@oregonstate.edu, baldur.steingrimsson@oit.edu.

**Abstract**

The authors present a generic framework for the parameter optimization of additive manufacturing (AM) processes, one tailored to a high-throughput experimental methodology (HTEM). Given the large number of parameters, which impact the quality of AM-metallic components, the authors advocate for partitioning the AM parameter set into stages (tiers), based on their relative importance, modeling one tier at a time until successful, and then systematically expanding the framework. The authors demonstrate how the construction of multi-dimensional functions, based on neural networks (NNs), can be applied to successfully model relative densities and Rockwell hardness obtained from HTEM testing of the Inconel 718 superalloy fabricated, using a powder-bed approach. The authors analyze the input data set, assess its suitability for predictions, and show how to optimize the framework for the multi-dimensional functional construction, such as to arrive at estimates of AM process parameters consistent with experimentally verified process windows. The authors also compare and contrast the NN-based multi-dimensional functional construction to multi-variate linear regression, to polynomial regression, and to Gaussian process regression (GPR), highlight similarity between the NN-based multi-dimensional functional construction and the GPR, and offer insights into the suitability of each of these methods, for the data set and the application at hand. In terms of the coefficient of determination, $R^2$, for the relative density, a comparatively simple, single-layer NN with 5 or 10 nodes tends to outperform multi-variate linear regression and $2^{nd}$-order polynomial regression, and even slightly outperform GPR, for the primary Inconel 718 HTEM data set studied, both with and without cross-validation. The novelty of the research work entails the versatile and scalable NN framework presented, suitable for use in conjunction with HTEM, for the optimization of AM process parameters of superalloys, and beyond.

# 1. Introduction

Additive manufacturing (AM) entails a process to print parts through the deposition of materials layer-by-layer. The AM techniques for the deposition of materials involving metal powder encompass selective laser melting (SLM), also referred to as laser powder bed fusion (LPBF), electron beam melting (EBM), and directed energy deposition (DED) [1] - [2]. The key principle behind these processes involves the introduction of materials in a wire or powder form and melting these materials, using a high-energy source, e.g., a laser or an electron beam [3].

SLM is the metal AM technique most widely deployed and extensively studied [4] - [5]. SLM owes much of its popularity to the capability to fabricate complex geometrical features, desirable physical or mechanical properties, minimize material waste, as well as to its versatility in fabricating a great number of metals [6] - [9]. In SLM, process parameters can affect physical properties, such as the density, surface roughness, or mechanical strength [10]. Nevertheless, SLM parts can contain processing defects, including anisotropy, residual stress, porosity or inclusions [3], [11] - [13]. This feature prompts interest in defect minimization through process optimization. At least 100 process variables control metal-AM processing with SLM [3], [14]. The most influential parameters, shown in Figure 1, on the physical properties of a component fabricated, are the laser power ($P$), laser scanning speed ($v$), layer thickness ($t$), and the hatch spacing between two adjacent scan tracks ($h$) [15], [16]. A high-throughput experimental methodology (HTEM) entails the production of hundreds of specimens, each treated with a unique set of process parameters [3]. Large datasets produced through high-throughput (HT) experiments may be employed as training sets for a variety of machine learning (ML) techniques [14], including neural network (NN) algorithms, and may be employed to predict ranges for the AM-process parameters. For a certain set of known AM process parameters, applied to a given material, HT can facilitate

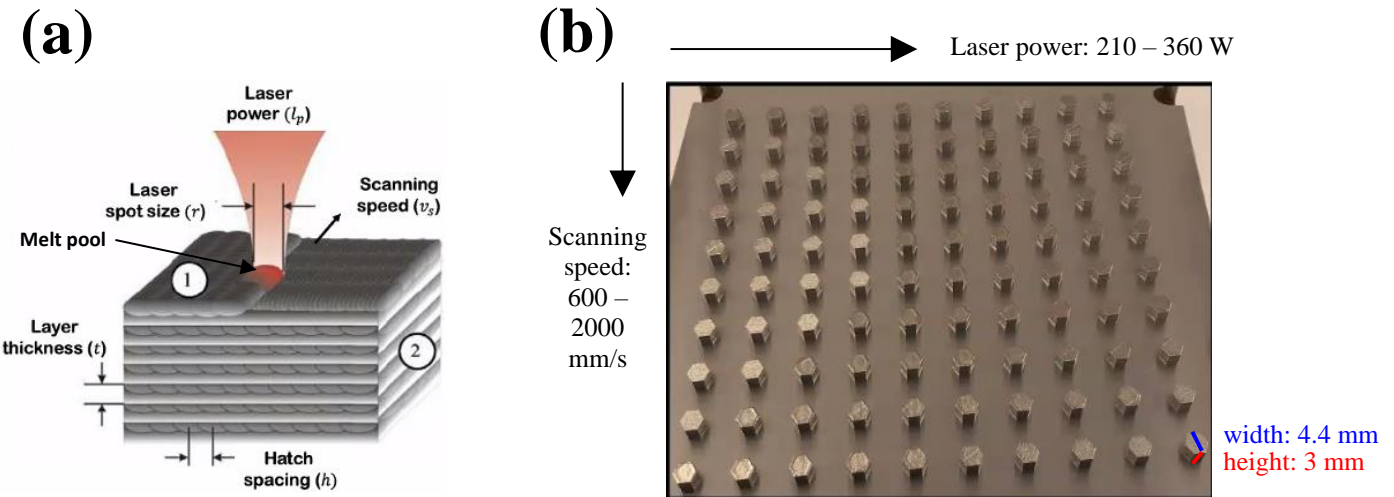

**Figure 1**: The most essential parameters of an SLM process considered for Tier I analysis (a) [16]. 100+ hex nuts at different processing parameters fabricated with an EOS M290 powder bed SLM printer (b) [3].

systematic examination of process-parameter variations.

Processing maps for Power-Velocity (PV) constitute a powerful approach for visualizing the impact of AM-processing parameters on microstructures and performance [3], [11]. Such charts provide useful insights into the defect type and relative population as well as the process window to fabricate specimens of full density. Processing maps for welding, deformation, or alternative manufacturing methods have been engineered to steer users towards the optimal processing conditions. In the case of SLM, processing maps are being utilized to define regions with full density as a function of the laser power and scan speed [17], under the assumption that a higher sample density corresponds to fewer defects (less porosity).

Defining optimal process parameters, which yield the highest density and best properties, is of paramount importance for the successful application of SLM. One of the primary challenges to the successful application of SLM involves determining appropriate process parameters for *achieving maximum density (pore-free) parts* [7]. The selection of adjusted parameter settings, for a specific material, powder feedstock [18], and application, can be difficult and time-consuming. This trend has frequently led to trial-and-error approaches [3].

In regard to the proper and effective definition of process parameters, the authors summarize the challenge as follows: If a vendor sells powder for powder bed AM, the vendor may provide a set of recommended AM-process parameters for the fabrication of a particular, existing alloy, based on past experience of the vendor. A buyer purchasing fabricated alloys may not know if these parameters have been optimized by the vendor, or how they have been optimized. This process involves a major leap of faith (uncertainty, and hence risk). The problem is exacerbated if the buyer is looking for a new alloy to be fabricated, for which no recommended AM-process parameters may exist [19]. So how does the buyer then process the new alloy?

It has been demonstrated that porosity adversely impacts mechanical properties and can result in premature failure of parts [20] - [22]. Many studies exist on characterizations of pores and on how such defects form during AM processing [22] - [26]. Further studies, based on automated HT tensile testing, have been carried out to assess the impact of processing parameters on mechanical behavior of additively manufactured 316L stainless steel (SS) [27] - [29]. Others have concentrated on the control of HT processes by monitoring the melt pool *in-situ* [30]. Moreover, the joint use of HT and ML has been demonstrated for the identification of new metallic glasses [31].

SLM processing maps have been thoroughly explored by Beuth et al. [32]. Here, separate regimes of defects (e.g., lack-of-fusion or keyholing) are analyzed for various laser power and scanning speed duplets. In a few other studies, this approach has been extended and processing maps constructed for different alloys, utilizing computational and experimental methods [33] - [34]. However, according to Ref. [3], the aforementioned studies focused primarily on defects in different processing regimes. Furthermore, there seemingly is need for further comprehension of microstructure evolutions as a result of these varying conditions. Moreover, implementation of processing maps to quickly identify suitable process parameters for novel materials seems slow, even now, and potentially cost prohibitive.

Agrawal et al. have presented a HT methodology to experimentally evaluate viable processing conditions for SLM [3]. Here, both HT and low-throughput (LT) methods are employed for the purpose of rapidly determining processing windows and generating processing maps for SLM of 316L SS [3].

Researchers have also introduced several established, along with relatively-newly derived, dimensionless numbers for AM [35] - [40]. Nonetheless, the numbers tend to rely on data on melt-pool characteristics that may not be easily accessible to AM users and that may generally not be available in advance [7]. In [7], Rankouhi et al. have developed a dimensionless number, which allows a priori prediction of process parameters in selective laser melting for a variety of pure metals and alloys, for the purpose of achieving high density, and for rapidly defining the SLM process parameters for new materials. The predictive and descriptive prowess of the dimensionless numbers has been verified experimentally through manufacturing of bulk samples from various metals or alloys [7]. Furthermore, versatility of the dimensionless numbers has been validated by deploying it on data from the literature generated by other researchers by whom alternative measurement methods and SLM systems have been employed [7].

In the case of [3] and [7], the optimization is two-dimensional (2D), in the sense that it involves optimization of only two parameters: A single property of choice, such as the relative density or hardness, is being optimized as a function of a single parameter, typically the energy density supplied by the laser to melt a unit volume of powder material, defined as the volumetric energy density (VED), or a dimensionless number. In the case of Ref. [3], the relative density of 316L SS, i.e., absolute density divided by a nominal density, is effectively maximized as

$$\max\{\text{relative density}\} = \max_{\text{VED}}\{f(\text{VED})\} \qquad (1)$$

where

$$\text{VED} = \frac{P}{v*h*t} \ . \qquad (2)$$

The authors are here looking to generalize such an approach to AM process optimization, by accounting for a broader set of input parameters, and hence by extending the approach to higher dimensions. Since the process development and material characterization are highly particular to a specific AM system, there is need for generic tools that can establish a procedure for identifying processing-structure-property relationships. While the design and manufacturing process chains are critical to realize the full potential of AM, ML-enabled tools can accelerate materials qualification time with reduced materials testing [41]. Yet, creating ML-based models for AM is challenging, due to the scarcity of sufficient input data and complexity of metal additive manufacturing (MAM) processes [42], [43]. For information on the extent of input data needed for an accurate NN parameter estimation, refer to Section 2.12 below or to the Supplementary Manuscript. Despite challenges related to lack of sufficient input data, researchers have adopted ML-based models for SLM [43], [44], [45], [46], For example, Akbari et al. presented a comprehensive framework for benchmarking ML for melt pool characterization [43]. Akbari et al. introduced physics-aware MAM featurization, versatile ML models, and evaluation metrics to create a comprehensive learning framework for the melt pool defect and geometry prediction, which can serve as a basis for the melt pool control and process optimization [43]. Hong et al. investigated the effects of two key process parameters of the SLM process, namely the laser power and the scanning speed, on the single-track morphologies and the bead characteristics, especially the depth-to-width and the height-to-width ratios, using both experimental and ML approaches [47]. Mycroft et al. focus on the impact of geometry in printability analysis and presented a ML-based framework for determining the geometric limits of printability in AM processes [48]. Maitra et al. proposed a ML-based approach, based on the Gaussian-Process Regression (GPR), to predict the relative density of the as-built Ti-6Al-4V alloy manufactured via SLM, based on the most common input process parameters, such as laser power, scanning speed, hatch spacing, and layer thickness, as well as an integrated input of volumetric energy density [49]. Zou et al. developed an optimized extreme gradient boosting (XGBoost) decision tree model, using hyperparameter optimization with a GridSearchCV method, in an effort to develop understanding of the correspondence of the relative density of SLMed Ti-6Al-4V parts with process parameters [50]. Tapia et al. constructed a surrogate model for LPBF of 316L stainless steel to predict the melt pool depth with an MAE of 10.91 μm of 96 single-track prints from the laser power, velocity, and beam diameter and associated simulations [51]. Scime et al. employed ML-based models for detection of keyholing porosity and balling instabilities for LPBF of Inconel 718 [44]. Lee et al. crafted a data analytics framework for determining melt pool geometry with a coefficient of determination, $R^2$, of 0.94 in a PBF process of single tracks of Inconel 718 and Inconel 625 [46]. Enabling the prediction of LPBF track widths with a coefficient of determination, $R^2$, of 0.93, Yuan et al. employed a ML-based monitoring of LPBF of 316L SS based on in-situ laser single-track videos [52]. Using a pyrometer and high-speed video camera data, Gaikwad et al., furthermore, developed ML-based predictive models of single tracks of 316L SS [53]. In addition, Barrionuevo et al. evaluated melting efficiency in wire-arc AM through ML algorithms, using processing parameters, such as wire diameter, wire feed speed, travel speed, and net power to determine the melting efficiency [54]. Zhang et al. propose an efficient framework to efficiently and effectively determine the processing parameter window (the bounding region in the manufacturing space resulting in near full density, defect-free parts) of a given alloy considered as potential feedstock for AM, a framework integrating design of experiments, physics-based simulation, uncertainty analysis and fabrication and characterization [19]. Upon discovering two types of keyhole oscillation in laser powder bed fusion of Ti-6Al-4V, through simultaneous high-speed synchrotron x-ray imaging and thermal imaging coupled with multi-physics simulations, Ren et al. developed an ML approach for real-time detection of stochastic keyhole porosity generation events with sub-milli-second temporal resolution and near-perfect prediction rate [55]. To fully exploit the potential of ML and release it from dependence on large, labeled, input datasets, Li et al. proposed a physics-informed neural network framework to predict the 3D temperature field, which is generated during the laser metal deposition process [56]. For cases, where an original dataset, which is based on the parameters of defects obtained by micro-computed tomography (μ-CT) prior to fatigue tests, stress level and the fatigue life of additively manufactured Ti-6Al-4V samples is considered too small to train a comprehensive ML model, In addition, J. Hornas et al. proposed a novel approach for dataset augmentation [57].

In this paper, the authors present a generalized framework for the multi-dimensional optimization of AM-process parameters, captured in Figure 2, and one particularly suited for HTEM. The authors present a systematic and rigorous framework for the selection of a NN prediction model and provide related theoretical support. Kolmogorov's Theorem and the Universal Approximation Theorem provide a theoretical basis for the multi-dimensional NN framework presented (see Figure 3). The authors describe how to qualify the input data for prediction, how to tailor the NN structures for optimal fit, and how the multi-dimensional NN framework can be assimilated with established GPR-based Gaussian Approximation (GAP) potentials used in molecular dynamics

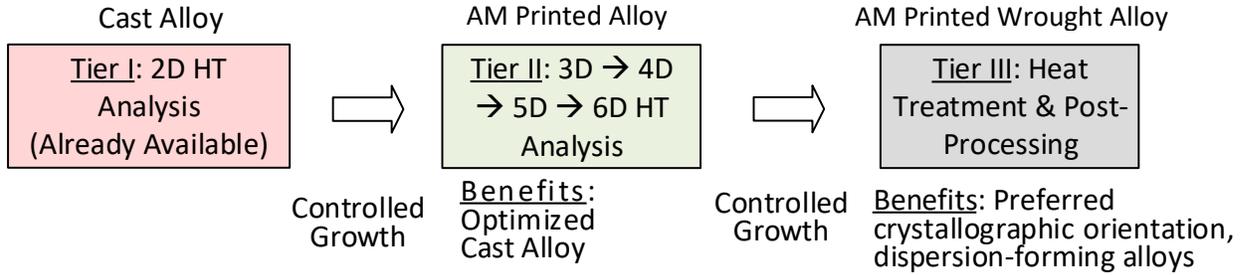

**Figure 2**: Systematic approach to the AM parameter optimization. The approach assumes starting with the core parameters listed in Figure 1 of the main manuscript, modeling them, and then upon success expanding the set of parameters included in the model. In Tier II, the input parameters included in the model are expanded from two (resulting in 3D analysis) to three (resulting in 4D analysis), to four (resulting in 5D analysis), and then to five (resulting in 6D analysis).

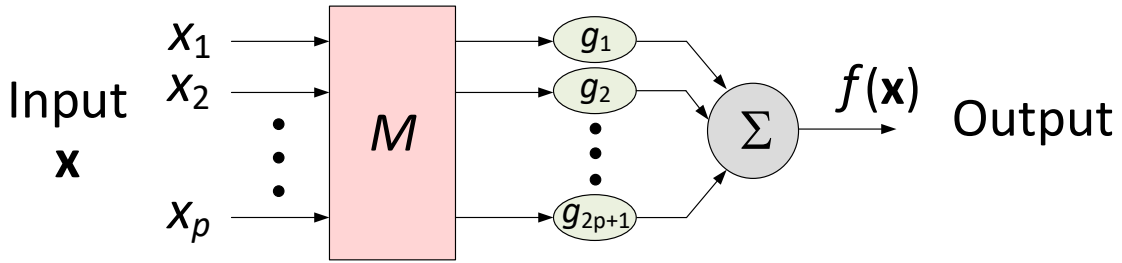

**Figure 3**: Neural network rendering of the Kolmogorov-Arnold Representation Theorem. Adapted from [14], [58].

(MD) or Monte Carlo modeling of material behavior. The authors also present processing windows for Inconel 718 verified through HTEM. The novelty of this research work entails the versatile and scalable NN framework presented, for the AM parameter optimization of superalloys. The impact entails broad applications of the technology within AM and beyond. Compared to today's process-optimization methodologies, the framework presented may enable a paradigm shift in customized manufacturing and accelerated qualification/certification, utilizing the multi-dimensional optimization of process parameters, integrated with the HTEM. Faster optimization to requirements for manufacturing are sought for components to fulfill the varying demands of the operational flexibility for industrial applications. With AM, there is great potential for enabling expedient design-manufacturing iterations; thus, substantially decreasing product cost and lead-time up to 25% from the present baseline [41], [59]. Nevertheless, these benefits cannot be fully realized, due to the potential for unknown AM processing defects, and due to the ensuing effect upon part performance, while in service.

**2. Theory**

*2.1 Modeling the Outputs from an SLM Process Used in Conjunction with HTEM*

Our methodology for combining SLM with HTEM, summarized in Figures 1 and 2, involves an extension of the 2D optimization approach outlined in [3] for the case of 316L SS, where the relative density is optimized as a function of a single parameter (the VED), per Eqs. (1) – (2), to Inconel 718 and to multiple dimensions. Given the significant number of parameters, which impact the quality of AM-metallic parts, the authors advocate for partitioning the AM parameter set into tiers (stages), based on relative importance, as presented in the Figure 2, modeling one stage at a time until successful, and then systematically expanding the framework. The first two tiers in our framework consist, in general, of the parameters exerting the greatest influence on the properties of a printed component, namely the laser power, laser spot size[1], scan speed, layer thickness, and hatch spacing [15], [16]. In addition, Supplementary Table 1 defines tiers for some fifty process parameters characterizing SLM or selective laser sintering (SLS) processes [58], [60]. The emphasis on high-throughput experimental measurements, the outcome of which (large data sets) can be appended to pertinent databases, lends itself well to ML analysis, in particular to sequential learning [14]. Sequential learning consists of iterative ML prediction and experimental verification, where the experimental results from a current verification step are appended to (used to strengthen) a database, that is used for the ML prediction, presumably resulting in improved predictions in the next ML prediction step [14].

---

[1] The laser spot size is included in the first two tiers, in general, since it can exert significant influence on the properties of a printed component [15], [16]. But as indicated below, we assume that the laser spot size is constant for a particular SLM system. Hence, the laser spot size is not accounted for in Eqs. (4) – (5). Similarly, the layer thickness or the hatch spacing can be kept constant for specific build runs.

*2.2 Construction of Multi-Dimensional Surfaces – Motivated by Kolmogorov's Theorem*

The Kolmogorov-Arnold Representation Theorem and the Universal Approximation Theorem provide theoretical basis for function approximation capabilities of neural networks [14], [58] - [61]. According to the Kolmogorov-Arnold Representation Theorem (or superposition theorem) of real analysis and approximation theory, *any* continuous, real-valued mapping can be cast as a two-layer neural network. Kolmogorov and Arnold established, more specifically, that any continuous real-valued function $f(\mathbf{x}) = f(x_1, x_2, \ldots, x_n)$ defined on $[0,1]n$, with $n \geq 2$, can be modeled as shown in Eq. (3) [62] - [63]:

$$f(\mathbf{x}) = f(x_1, x_2, \cdots, x_n) = \sum_{j=1}^{2n+1} g_j \left( \sum_{i=1}^{n} \phi_{ij}(x_i) \right). \tag{3}$$

Here, the $g_j$'s are appropriately chosen functions of one variable, and the $\phi_{ij}$'s are continuously monotonically increasing functions with no dependence on *f*. The schematic in Figure 3 offers a neural network rendering of the Kolmogorov-Arnold Representation Theorem [14], [61], [64]. Similarly, within a given function space of interest, the Universal Approximation Theorem establishes the density of an algorithmically generated class of functions [64].

Motivated by Figure 1 and the Kolmogorov-Arnold Representation Theorem [Eq. (3)], the authors model hardness and relative density, both obtained from HTEM, in the form of multi-dimensional surfaces, as

Relative density $= o_1 = f_1$(laser power, scan speed, layer thickness, hatch spacing) $+ n_1$, (4)

Hardness $= o_2 = f_2$(laser power, scan speed, layer thickness, hatch spacing) $+ n_2$. (5)

Here, $n_1$ and $n_2$ represent noise terms (referred to as the measurement error on the output side, in context with Figures 4 and 5), and are modeled as random variables with zero mean and Gaussian distribution. The relative density is of primary interest, because it yields a direct measure of defect (pore) density relative to its ideal (pore-free) density, facilitating quick comparison across 100+ samples, and because it is subjected to a universal scaling law [7]. The hardness is selected, because it represents a mechanical property and yields an indirect measure of defect levels, since the absence of defects is generally believed to improve the mechanical properties. The hardness is subjected to a scaling law, but not a universal one [3], [7]. However, by employing both HT and LT methods, a predictive mapping between the hardness and the primary dendrite arm spacing can be established, as in the case of 316L stainless steel, by using a Hall-Petch relationship [3]. Further motivated by Figures 4 - 5, and the inherent process variations (time-varying nature of the AM + HTEM system), the authors articulate the primary assumptions and definitions as follows:

1. The mappings, $f_1(\cdot)$ and $f_2(\cdot)$ in Eqs. (4) and (5), represent the noiseless output of the model for the quantities observed, not accounting for measurement error on the output side.

2. Note that the measurement error, defined in the Figure 4(a), only relies on the presence of a physical AM + HTEM system, and can exist in the absence of any type of modeling effort.

3. The authors assume that, through the configuration of the AM printer, the laser power, scan speed, layer thickness, and hatch spacing can be set to a relatively high degree of accuracy (that the measurement error on the input side is small).

4. The measurement errors on the output side, $n_1$ and $n_2$, can be significant, esp. in the case of the hardness, per Figure 6(b).

5. The measurement errors on the output side, $n_1$ and $n_2$, are modeled as Gaussian distributions with zero (0) mean but standard deviation of $\sigma_{meas,1}$ and $\sigma_{meas,2}$, respectively.

6. Modeling error, defined in Figure 4(b), can exist for a single model, i.e., in the absence of model selection.

7. Figure 4(b) captures the definitions of the modeling error and of the testing error. The modeling error is computed from the training samples, but the testing error from the testing set. In cases of no cross-validation or five-fold cross-validation, the authors assume splitting of the input data set into a training and a testing set, according to the classical 80/20 rule of data science[2] [65]. In the case of ten-fold cross-validation, only 10% of each split of the data set is used for testing [66]. We, moreover, assume that the laser spot size is constant for a particular SLM system. Inclusion of the laser spot size as a variable would affect input energy density and might result in deteriorated print quality. Note that, given the four (4) input dimensions, the optimization of the hardness or the relative density would involve the optimization in five-dimensional (5D) space. Note, furthermore, that the model from Eqs. (4) and (5) can, in theory, be extended to other quantities of interest, measured by an HTEM approach, such as to the yield strength (YS). But practical extensions of the model may come down to the specific variables measured by the HTEM approach, and the effort involved in verifying the quantity of interest for the 100+, or even 200+, samples fabricated on a single build plate.

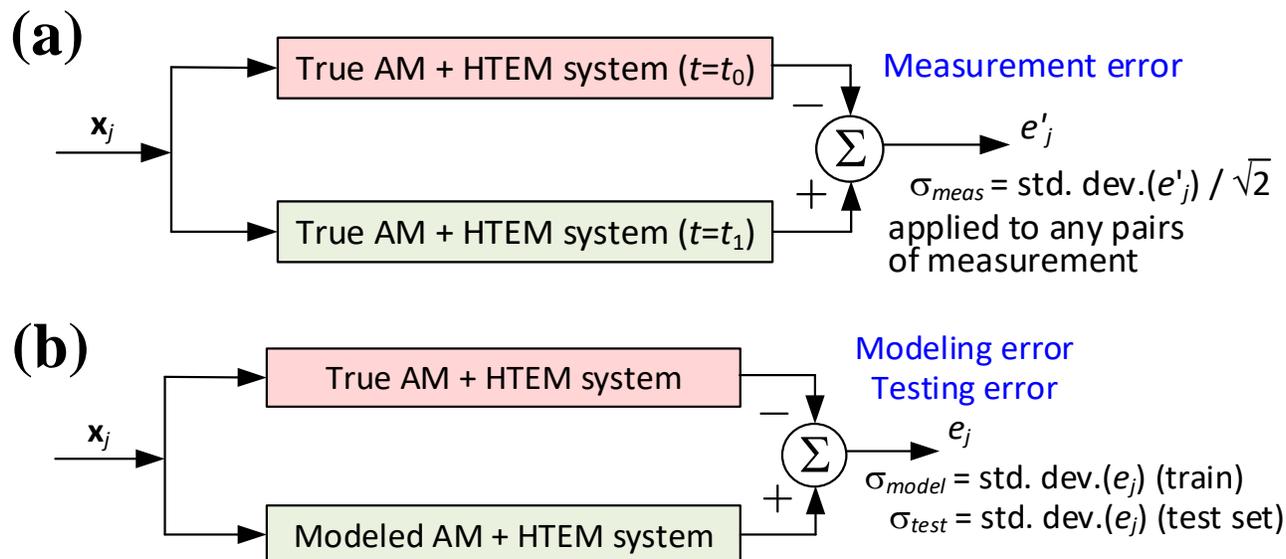

**Figure 4**: (a) Definition of measurement error for a true AM + HTEM system, assuming two, independent fabrication and test runs (build plates). Here, the term "std. dev.($e'_j$)" implies averaging over the testing specimens comprising the build plate. Note that both inputs to the summing junction contain a noise term, hence $\sqrt{2}$ in the definition of $\sigma_{meas}$, according to the algebra of random variables. (b) Definition of modeling error for a modeled AM + HTEM system. The term "std. dev.($e'_j$)" implies averaging over the data points comprising the training set or the testing set, respectively. Note that only one input to the summing junction contains a noise term. This feature explains the absence of a $\sqrt{2}$ factor in the definition of $\sigma_{model}$ and $\sigma_{test}$.

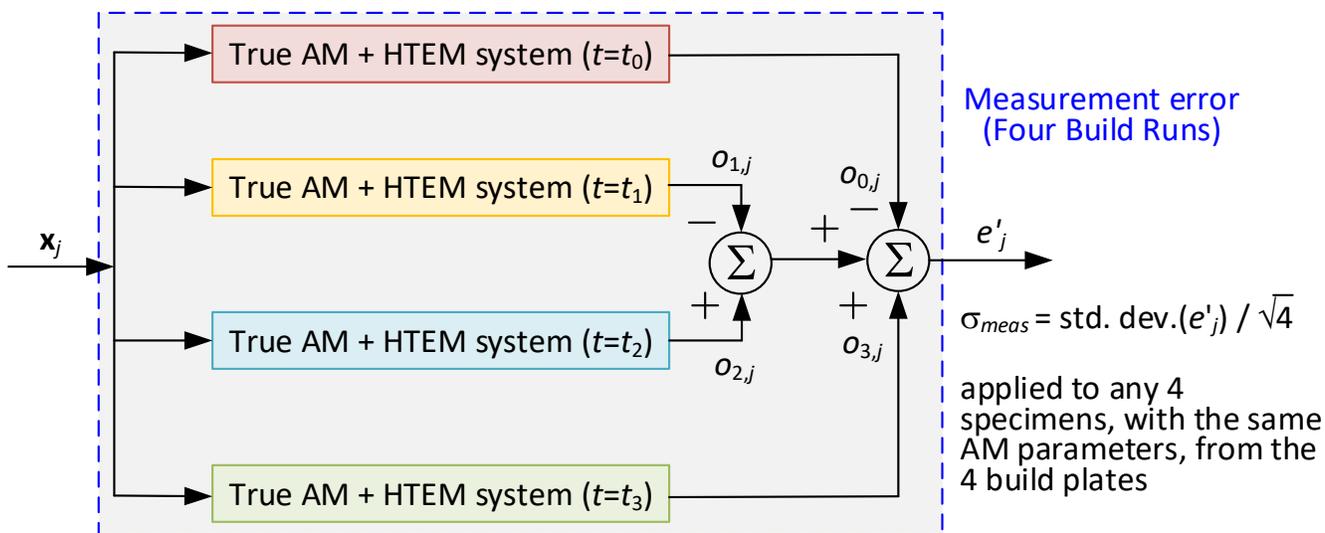

**Figure 5:** Definition of measurement error for a true AM + HTEM system, assuming four, independent fabrication and test runs (build plates).

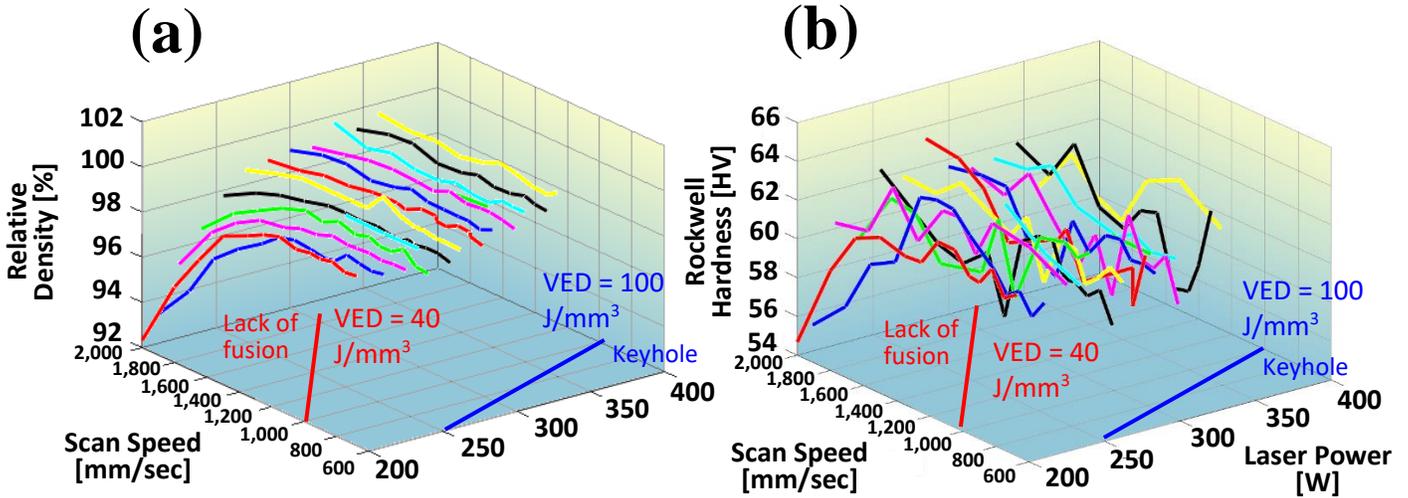

**Figure 6:** Original density and Rockwell hardness data from high-throughput testing of IN 718 samples visualized as dependent on the laser power and scan speed applied. Here, the hatch spacing was set to 0.11 mm and the layer thickness to 0.04 mm. Approximately the same range of AM process parameters (laser power and scan speed), and the same calibration settings for the hardness tester were used to obtain the HTEM results from Build Runs 1 and 2. Different colors for the density and Rockwell hardness curves correspond to various values for the laser power. VED = 40 J/mm³ and VED = 100 J/mm³ correspond to experimentally verified process windows for IN 718 further addressed in Figures 11 - 12 and Figure 16.

---

[2] The 80/20 rule of data science assumes that 80% of the data available is used for training the algorithms under development, but 20% of the data available is retained for testing the algorithms.

Once multi-dimensional functions with the form of Eq. (4) or (5) have been constructed, recommended (optimized) input parameters can be obtained from the multi-dimensional mappings. Assuming that one is looking to identify the AM process parameters that result in the highest hardness or the highest relative density, the recommended input parameters can be determined as

$$(\widehat{\text{laser power}}_1, \widehat{\text{scan speed}}_1, \widehat{\text{layer thickness}}_1, \widehat{\text{hatch spacing}}_1) = \text{argmax } f_1 \quad (6)$$

$$(\widehat{\text{laser power}}_2, \widehat{\text{scan speed}}_2, \widehat{\text{layer thickness}}_2, \widehat{\text{hatch spacing}}_2) = \text{argmax } f_2 \quad (7)$$

Alternatively, a range of recommended input parameters may be determined as corresponding to the relative density, modeled according to a universal scaling law, exceeding, say, 99.5% [7]:

$$(\widehat{\text{laser power}}_3, \widehat{\text{scan speed}}_3, \widehat{\text{layer thickness}}_3, \widehat{\text{hatch spacing}}_3) = \text{argmax } [f_1 > 99.5\%] \quad (8)$$

Then, one can compare the optimal input parameters derived from the hardness mapping ($f_2$) to those obtained from the density mapping ($f_1$), quantify the level of consistency, and verify experimentally.

*2.3 Boundary Conditions Considered for Construction of Multi-Dimensional Surfaces*

For comparison, the authors created a copy of the input data, separate from the original data, which the authors then appended, such as to incorporate the boundary conditions listed below:

1. $\quad$ Relative Density = $f_1$(laser power, scan speed = 0) = 0 $\quad$ (9)

2. $\quad$ Relative Density = $f_1$(laser power → ∞, scan speed) = 0 $\quad$ (10)

3. $\quad$ Hardness = $f_2$(laser power, scan speed = 0) = 0 $\quad$ (11)

4. $\quad$ Hardness = $f_2$(laser power → ∞, scan speed) = 0 $\quad$ (12)

When the laser power → ∞ but the scan speed is kept constant, or when scan speed = 0 but the laser power is kept constant, the relative density and the hardness of the solid phase is assumed 0, since here the alloy is assumed in a molten state. The authors separately analyzed the data sets with or without the boundary conditions included. Figures 8 - 11 all capture results for multi-dimensional density surfaces with the boundary conditions excluded, but Supplementary Table S7 compares modeling accuracy with or without the boundary conditions.

*2.4 Primary Criteria for Selection of Mapping Models*

In the case of two or more candidate mappings for $f_1(\cdot)$, one can select the mapping model that yields the highest posterior probability, $P(H_i|D)$. Here [67],

$$P(H_i|D) \propto P(D|H_i) P(H_i). \tag{13}$$

$H_i$ represents the mapping models under consideration, $D$ the data available, $P(H_i)$ prior probabilities for the mapping model $H_i$, and $P(D|H_i)$ evidence in support of the mapping model $H_i$ [67].

*2.5 Objective Criterion for Qualifying the Fit of a Reconstructed Function – An Ideal Case*

The authors advocate for a quantifiable approach to qualification of mapping models considered:

1. Given a data set, one first needs to estimate the standard deviation of the measurement noise at the output of the physical model ($\sigma_{\text{meas},1}$ and $\sigma_{\text{meas},2}$).

2. When the standard deviation for the testing error, $\sigma_{\text{test},i}$, is roughly the same as the standard deviation for the measurement error, $\sigma_{\text{test},i}$, or within a factor of ~1.1, then one can declare the mapping model good enough:

$$\sigma_{\text{test},i} \leq 1.1\, \sigma_{\text{meas},i} \tag{14}$$

Given a data set, one can first estimate the measurement noise at the output of the physical system ($\sigma_{meas,\text{i}}$) and then the testing error for a trained model ($\sigma_{\text{test},i}$). When the standard deviation for the testing error is roughly the same as standard deviation obtained from the measurements, the model is unlikely to improve much further. This is both how one can measure the quality of the model and estimate the quality of the data. Note that the criterion in Eq. (14) entails fully direct measurable entities. There is no human interpretation (no subjectivity) involved. Note also that ideally at least two build runs are needed for the estimation of $\sigma_{\text{meas},i}$.

*2.6 Criterion for Qualifying the Fit of a Reconstructed Function – A Fall-Back Scenario*

If one is forced to estimate the measurement noise from a single build run, the best option available may involve the following approximation for the standard deviation of the measurement noise, $\sigma_{\text{meas,approx.},i}$:

$$\sigma_{\text{meas,approx},i} \equiv \underset{P_j, v_j}{\text{std. dev.}} \left[ \frac{o_i(P_j + |\delta P|, v_j + |\delta v|)}{4} + \frac{o_i(P_j - |\delta P|, v_j + |\delta v|)}{4} + \frac{o_i(P_j + |\delta P|, v_j - |\delta v|)}{4} + \frac{o_i(P_j - |\delta P|, v_j - |\delta v|)}{4} - o_i(P_j, v_j) \right] \quad (15)$$

Here, $P_j$ and $v_j$ refer to the laser power and scan speed corresponding to a specific point, $j$, in the grid shown in Figure 1(b), $|\delta P|$ denotes the absolute difference in laser power between the grid point, $j$, and the laser power of its nearest neighbor in the grid, $|\delta v|$ represents the absolute difference in the scan speed between the grid point, $j$, and its nearest neighbor in the grid, whereas $o_i$, with $i$ = 1, 2, is defined according to Eqs. (4) and (5), respectively. Note that the authors strongly advocate for generating two build runs, for the purpose of estimating the measurement noise. Two build runs not only help with estimating the standard deviation for the measurement error, but also help safeguard essential consistency in measurements (ensure the absence of oversights during the first build run). Strictly speaking, one can only estimate the mean of a distribution, from a single build run. When the testing error is roughly the same as the standard deviation that one obtains from the measurements (within a factor of ~1.3), in the case of a single build run, one may declare the model good enough, i.e., if

$$\sigma_{\text{test},i} \leq 1.3\, \sigma_{\text{meas,approx},i} \quad (16)$$

(and if there is not a better option available for estimating the standard deviation for the noise measurement). Assuming a first order Taylor-expansion, the metric from Eq. (15) only contains a noise component. But assuming a second order Taylor-expansion, there is also a functional component:

$$\begin{aligned}
\sigma_{\text{meas,approx},i} &\cong \frac{o_i(P_j + \delta P, v_j + \delta v)}{4} + \frac{o_i(P_j - \delta P, v_j + \delta v)}{4} + \frac{o_i(P_j + \delta P, v_j - \delta v)}{4} \\
&+ \frac{o_i(P_j - \delta P, v_j - \delta v)}{4} - o_i(P_j, z_j) \\
&= \frac{\partial^2}{8\, \partial P^2} f_i(P_j, v_j)(\delta P)^2 + \frac{\partial^2}{8\, \partial v^2} f_i(P_j, v_j)(\delta v)^2 + \frac{n_i(P_j + \delta P, v_j + \delta v)}{4} \\
&+ \frac{\partial^2}{8\, \partial P^2} f_i(P_j, v_j)(\delta P)^2 + \frac{\partial^2}{8\, \partial v^2} f_i(P_j, v_j)(\delta v)^2 + \frac{n_i(P_j - \delta P, v_j + \delta v)}{4} \\
&+ \frac{\partial^2}{8\, \partial P^2} f_i(P_j, v_j)(\delta P)^2 + \frac{\partial^2}{8\, \partial v^2} f_i(P_j, v_j)(\delta v)^2 + \frac{n_i(P_j + \delta P, v_j - \delta v)}{4} \\
&+ \frac{\partial^2}{8\, \partial P^2} f_i(P, v_j)(\delta P)^2 + \frac{\partial^2}{8\, \partial v^2} f_i(P_j, v_j)(\delta v)^2 + \frac{n_i(P_j - \delta P, v_j - \delta v)}{4} \\
&- n_i(P_j, v_j)
\end{aligned} \quad (17)$$

Once again, the authors strongly advocate for generating two build runs and not resorting to Eqs. (15) - (17).

*2.7 More on a Systematic Framework for Model Selection – Bayesian Model Comparison and Selection*

The systematic approach presented here is a generalization of the exposition on the theory and application of Bayesian methods for neural networks from [67]. The Bayesian approach to model selection assumes selection of the model which is the most plausible given the data (the model with the highest posterior probability):

$$\arg\max_i P(H_i|D) = \underset{i}{\operatorname{argmax}} \left[ P(D|H_i)\, P(H_i) \right] \quad (18)$$

where

$$P(D|H_i) = \int P(D|\mathbf{w}, H_i)\, P(\mathbf{w}|H_i)\, d\mathbf{w}. \quad (19)$$

Here, $H_i$ represents the models considered, $\mathbf{w}$ denotes a vector of parameters defining the models, $P(\mathbf{w}|H_i)$ embodies a "prior" distribution, which states which values the parameters of the model might credibly take. $P(D|\mathbf{w},H_i)$ signifies a likelihood, $P(H_i|D)$ represents posterior probability, $P(D|H_i)$ denotes "evidence" or marginal likelihood, and $P(H_i)$ embodies prior probabilities for the models. A model is defined in terms of an ensemble of probability distributions.

As noted in [67], the evidence can be approximated as the height of the apex of the integrand, $P(D|\mathbf{w},H_i)\,P(\mathbf{w}|H_i)$, times the width, $\sigma_{w|D}$:

$$P(D|H_i) \cong P(D|\mathbf{w}_{MP}, H_i) \times P(\mathbf{w}_{MP}|H_i)\,\sigma_{w|D} \qquad (20)$$

$$\text{Evidence} \cong \text{Best fit likelihood} \times \text{Occam factor}. \qquad (21)$$

Therefore, the evidence can be estimated as the multiple of the best fit likelihood, that the model can attain, and an 'Occam factor', which is a *penalty factor* with *magnitude less than one*, that penalizes $H_i$ for taking on the parameter, $\mathbf{w}$ [67]. In the context of AM process optimization, the 'Occam factor' can be viewed as a penalty term that penalizes complex models, that may yield a very good match to a training data set, but that may not yield that good predictions when compared to a testing data set (that may not generalize that well to the testing data set), in favor of simpler models, that may yield not quite as good fit to the training data set, but which may generalize better to the testing data set.

## 2.8 Scaling of the Inputs

Practical implementations of neural networks may assume that individual inputs are scaled to the same range. While this feature may not be needed in theory, this may be preferred in practice. Note that scaling of the inputs can be addressed separately from the specifics of the algorithm implementation (the neural network) at hand.

## 2.9 Training (in Context with the Systematic Framework for Model Selection)

The systematic framework for model selection can easily support a variety of training algorithms. In the case of a Matlab® implementation, the training algorithm can be specified as an input, when the neural network is created:

**input_data**     = [ in1_data, in2_data ]'; (22)

**output_data** = [ out_data' ]; (23)

**net**               = feedforwardnet ([10, 10], 'trainlm' ); (24)

**[ net, tr]**       = train (net, input_data, output data); (25)

The training algorithm can also be modified after the formation of the network:

**net.trainFcn** = 'trainlm'; (26)

In this paper, the authors assume training using the Levenberg-Marquardt algorithm ('trainlm') by default [68]. Matlab® supports many alternative training algorithms, summarized in [68].

*2.10 Guidelines for Selection of the Number of Layers and Nodes for the Neural Network*

Neural networks are capable of reconstructing functional dependence (surfaces) of arbitrary complexity, as noted above [see Eq. (3)]. But to our knowledge, there are no theoretical formulas available for determining the optimal number of layers or nodes for an actual NN providing such a reconstruction. For a formal overview on the representation through the dual (regularization) representation, principal component analysis, and corresponding sparse representations (decompositions) for function approximation, refer to [69]. In [69], the authors introduce a general representation for a function as a linear combination of local correlation kernels at optimal sparse locations (and at optimal scales). The authors explain the connection of this general representation to principal component analysis, regularization, sparsity principles, and to support vector machines (SVMs). As noted in [70], there is a trade-off between the selection of basis functions used in the functional decomposition and the size of the network. Per the Kolmogorov-Arnold Representation Theorem [Eq. (3)], the selection of the basis functions comes down to obtaining a sparse representation of the decomposition over the basis functions selected. If you select too few nodes or layers (basis functions), the network may not possess the ability to reconstruct the underlying functional dependence, e.g., to accurately model the functions $f_1(\cdot)$ and $f_2(\cdot)$ in Eqs. (4) and (5). But if there are too many nodes or layers, then the network may fit the training data to the model structure (over-fit), resulting in poor ability of the network to generalize, i.e., yield accurate function estimates for new inputs [per the 'Occam factor' in Eq. (21)]. Hence, the better one understands the underlying physics, the fewer kernel functions one may need to accurately model the functional dependance at hand. Understanding of the underlying physics can help improve understanding of what types of kernel functions are needed (suitable) for sparse decomposition, i.e., for accurate but sparse representation of the functional dependance at hand [70].

*2.11 Extension to Higher Dimensions*

In principle, our systematic framework lends itself very well to scaling to higher dimensions:

**input_data**    = [ **in1_data, in2_data, in3_data, …**]';                    (27)

**output_data**  = [ **out_data'** ];                                             (28)

**net**                = **feedforwardnet ([10, 10], 'trainlm' );**              (29)

**[ net, tr ]**     = **train (net, input_data, output_data );**                 (30)

But such an approach may rely on extensive input data to be available (may require significant data collection). Accounting for the extent of the input data needed, the approach may not scale well to the higher dimensions. A more efficient approach, based on sequential learning and a series of local searches, may be preferred in practice. For an example of the latter, refer to Section 4.3.

*2.12 Extent of Input Data Needed for Accurate Construction - Suitability for HTEM*

With specifics depending on the implementation (on the correlation of the input predictors [features] to the quantity being predicted as well as on the number of dimensions involved), significant input data may be needed in general for accurate NN construction of the multi-dimensional surfaces, as suggested by Figure 7. Here, the random variable $Q$, used to estimate the variance of a Gaussian distributed variables within 10% of the true value, is defined as [71]:

$$Q = \frac{(N-1)S^2}{\sigma^2} = \frac{1}{\sigma^2}\sum_{i=1}^{N}(X_i - \bar{X})^2. \tag{31}$$

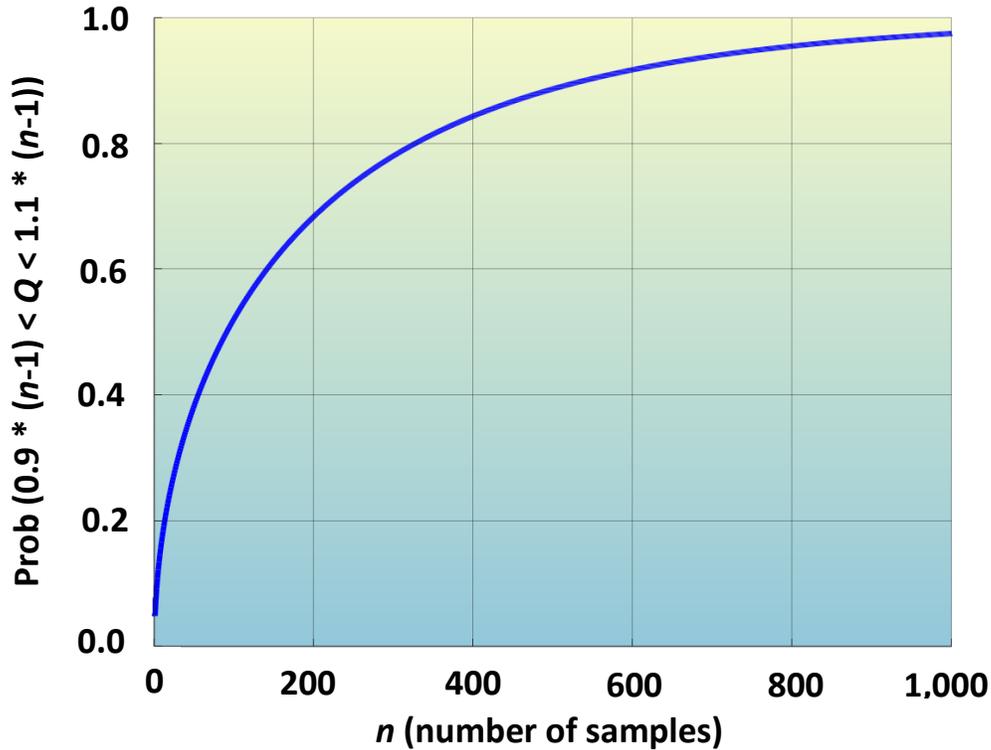

**Figure 7:** Assessment of the data needed to estimate the variance of Gaussian distributed variables within 10% of the true value.

If $X_1$, $X_2$, $X_3$, …, $X_N$ are assumed to represent samples from a normal distribution $N(\mu, \sigma^2)$. the random variable $Q$ happens to have a chi-squared distribution with $N-1$ degrees of freedom, i.e., [71]

$$Q \sim \chi^2(N-1). \tag{32}$$

Of course, in the extreme case of the input predictors (features) comprising of pure noise, i.e., with no correlation to the quantity being predicted, then no NN will generalize well, even when provided with arbitrary large volumes of input data. Our recommendation is to make the most of the HTEM data from each SLM build plate available, and carefully keep track of confidence or uncertainty intervals corresponding to quantities estimated. For additional information, refer to the Supplementary Note 3.

*2.13 Alternative (and Similar) Methods Considered*

Upon qualitatively assessing the input data in Figure 6, it appears that polynomial regression and Gaussian process regression may prove to be good modeling choices, for comparison, in this case. Hence, we compare and contrast the NN-based multi-dimensional functional construction to multi-variate linear regression [14], polynomial regression [14], and to GPR [72], [73], [74], [75], [76]. Multi-variate linear regression can be a suitable approach, in the event of a relatively small-to-modest number, or even in the event of a moderately large number, of HTEM measurements (observations). Here, the relationship between the input predictors and the observations is modeled as [14]

$$\tilde{y}_i = b_0 + \sum_{j=1}^{p} b_j \tilde{x}_{ij} + e_i, \qquad (33)$$

for $i \in \{1,\ldots,n\}$. In this context, $\tilde{y}_i \in \mathbb{R}$ represents the *i*-th observation, $b_0 \in \mathbb{R}$ is the unknown regression coefficient for the intercept, $b_j \in \mathbb{R}$ is the unknown regression coefficient for the *j*-th predictor, and $e_i \in \mathbb{R}$ represents error samples, which we assume to be independent and identically distributed (i.i.d.) with [14]

$$E[e_i] = 0, \qquad \mathrm{Var}(e_i) = \sigma^2, \qquad \mathrm{Cov}(e_i, e_j) = 0 \ \forall \ i \neq j. \qquad (34)$$

While the error samples may often be Gaussian distributed, we are not explicitly making that assumption. When given *n* independent observations, one can organize the observations (measurements) into a matrix form as follows [14]:

$$\tilde{\mathbf{y}} = \widetilde{\mathbf{X}} \mathbf{b} + \mathbf{e}. \qquad (35)$$

In this context, $\tilde{\mathbf{y}}$ represents a $\mathbb{R}^{n x 1}$ vector, $\widetilde{\mathbf{X}}$ a $\mathbb{R}^{n x (p+1)}$ matrix, **b** a $\mathbb{R}^{(p+1) x 1}$ a vector, and **e** a $\mathbb{R}^{n x 1}$ vector. The $n \times n$ covariance matrix for **e** is now given by $\mathrm{E}[\mathbf{e}\,\mathbf{e}^T] = \sigma^2 \, \mathbf{I}$. Analogously [14],

$$E[\tilde{\mathbf{y}}] = \tilde{\mathbf{X}}\,\mathbf{b}, \qquad \mathrm{Cov}(\tilde{\mathbf{y}}) = \sigma^2\,\mathbf{I}. \qquad (36)$$

The corresponding ordinary least-squares problem becomes [14]

$$\min_{\mathbf{b}\in R^{(p+1)\times 1}} \|\tilde{\mathbf{y}} - \tilde{\mathbf{X}}\mathbf{b}\|^2, \qquad (37)$$

where $\|\cdot\|$ represents the Frobenius norm [14]. In the case of continuous-valued inputs (predictors), the ordinary least-squares problem has the following, closed-form solution [14]:

$$\hat{\mathbf{b}} = (\tilde{\mathbf{X}}^T\tilde{\mathbf{X}})^{-1}\tilde{\mathbf{X}}^T\,\tilde{\mathbf{y}} \qquad (38)$$

In the case of Eq. (4), a linear regression model would take the form

Relative density $= b_0 + b_1 *$ laser power $+ b_2 *$ scan speed $+ b_3 *$ layer thickness $+ b_4 *$ hatch spacing $+ e_1$. (39)

In the event the data justifies the use of a model more sophisticated than linear regression, one can employ a NN approach or one can consider adding a quadratic term to the linear-regression model. The quadratic term corresponds to the 2$^{nd}$ term from the Taylor expansion of the system model from Eq. (4) or Eq. (5). In the absence of a major, discrete jump in the data, i.e., as long as the underlying physics does not produce a discrete jump in the data, such as caused by the on-set of super-conductivity, one can expect the system model to lend itself to the Taylor expansion. A least-squared solution to the quadratic model [14]

$$y = a\,x^2 + bx + c, \qquad (40)$$

where $a \neq 0$, exists in a closed form, and is given by [14]:

$$\begin{bmatrix} \sum x_i^4 & \sum x_i^3 & \sum x_i^2 \\ \sum x_i^3 & \sum x_i^2 & \sum x_i \\ \sum x_i^2 & \sum x_i & n \end{bmatrix} \begin{bmatrix} a \\ b \\ c \end{bmatrix} = \begin{bmatrix} \sum x_i^2 y_i \\ \sum x_i y_i \\ \sum y_i \end{bmatrix}. \qquad (41)$$

For an example of application of second order polynomial regression to the optimization of density in SLM-ed Ti-6Al-4V samples, refer to [49].

In context with the input data in Figure 6, we expect the polynomial regression to yield reasonably good performance, since the functions being modeled here look relatively simple. A strength of the polynomial regression relates to its simplicity. Keep in mind that constructing complex models, for cases involving noisy data offers little value. In such cases, the accuracy of any functional model constructed will be limited by the measurement noise. A primary limitation of polynomial regression may reveal itself when the underlying function drops to zero and stays at zero for a while. A secondary limitation of polynomial regression can show up in the event of very large and dense linear systems comprising of 50,000 parameters or more. Such systems can be solved but may necessitate more complicated methods. Standard methods for solving such large and dense systems may become computationally burdensome.

For background context on Gaussian process regression, refer to [72], [76], [77], [78] or to Supplementary Note 2. Suffice to say that GPR is a nonlinear, Bayesian, and nonparametric regression technique, which is useful for interpolating between data points scattered in a high-dimensional input space [72]. The non-parametric GPR regression does not assume a closed functional form, nor does it seek to explain the process underlying the data by drawing upon on theory [72]. As a Bayesian regression technique, the GPR, however, does offer a solution to the modeling problem such that the locality of the interpolation can be explicitly and quantitatively controlled, by encoding it in the a priori assumption of smoothness of the underlying function [72]. The NN approach for the multi-dimensional functional reconstruction exhibits significant resemblance with the GPR-based GAP potentials used for molecular dynamics and Monte Carlo simulations of material behavior [72], [73], [74], [75]. In both cases, one seeks to reconstruct a multi-dimensional function by interpolating (fitting) a smooth surface through a grid of reference points. In the case of AM, the grid of reference points may consist of HTEM measured density or hardness values, per Figure 1. But in the case of the MD, the grid of reference points is usually obtained from high-precision density functional theory (DFT) calculations. Given the extent of the resemblance, the Gaussian-process regression may be considered a similar (a related) method, more so than an alternative, to the NN approach for the multi-dimensional functional reconstruction. In context with the input data in Figure 6, the Gaussian-process regression comes across as quite suitable, despite its complexity exceeding that of polynomial regression. The Gaussian-process regression tends to be well suited for tracking movement in the presence of Gaussian noise. The measurement errors are often Gaussian. In addition, measurement tools are often affected by numerous factors and tend to report noiseless observations +/- a noisy value. For example, as the head of the laser moves, it tends to undergo small, noisy fluctuations, caused by heat, micro-vibrations, etc.

For information on additional ML models considered for melt pool classification and characterization of melt pool geometry, including the random forest algorithm, SVMs, ridge linear regression, Lasso linear regression, gradient boosting trees and extreme gradient boosting (XGBoost), refer to [43].

*2.14 More on Model Parameter Selection - Cross-Validation for Model Hyperparameter Optimization*

Choosing an optimal value for the set of hyperparameters used for training a ML model is of vital importance, since the hyperparameters may significantly affect the prediction performance of the ML model. As indicated above, there are no theoretical formulas available, to our knowledge, for determining the optimal number of layers or nodes for an actual NN providing a reconstruction of functional dependence of arbitrary complexity. In [43], Akbari et al. optimize two different hyperparameters for the NN models presented, the overall number of neurons and a regularization parameter $\alpha$. Akbari et al. rightfully note that a higher number of neurons increases the complexity of the ML model but may result in overfitting. The regularization parameter, $\alpha$, constrains an $L_2$ norm of the model weights, hence reducing model complexity and discouraging overfitting [43].

According to Eqs. (13) and (18), the Bayesian approach to model selection assumes the selection of the model with the highest posterior probability. In the case of equal prior probabilities, Eq. (18) can be rewritten in terms of maximization over the likelihoods $P(D|H_i)$:

$$\operatorname*{argmax}_{i} P(H_i|D) = \operatorname*{argmax}_{i} [P(D|H_i)] = \operatorname*{argmax}_{i} [\log(P(D|H_i))], \quad \text{if } P(H_i) \text{ is identical } \forall\, i\text{'s.} \quad (42)$$

Here, we are utilizing the fact that log() is a monotonically increasing function. The maximum likelihood estimator (MLE) estimates the hyperparameters from the data by maximizing the logarithm of the likelihood distribution. GPR models tend to be quite suitable for Bayesian inferencing and optimization, since the integrals over hyperparameters of the model can be derived analytically [49]. For the GPR models, the marginal log-likelihood function can be expressed as [49]

$$\begin{aligned} \log(P(\boldsymbol{y}|\boldsymbol{X},\beta,\boldsymbol{\theta},\sigma^2)) &= -\frac{1}{2}(\boldsymbol{y}-\boldsymbol{H}\beta)^T [K(\boldsymbol{X},\boldsymbol{X}'|\boldsymbol{\theta}) + \sigma^2 \boldsymbol{I}]^{-1} (\boldsymbol{y}-\boldsymbol{H}\beta) \\ &\quad -\frac{n}{2}\log(2\pi) - \frac{1}{2}\log|K(\boldsymbol{X},\boldsymbol{X}'|\boldsymbol{\theta}) + \sigma^2 \boldsymbol{I}| \end{aligned} \quad (43)$$

In this context, the MLE estimates of the hyperparameters can be represented as [49]

$$(\hat{\beta}, \hat{\boldsymbol{\theta}}, \widehat{\sigma^2})_{MLE} = \operatorname*{argmax}_{\beta,\boldsymbol{\theta},\sigma^2} \log(P(\boldsymbol{y}|\boldsymbol{X},\beta,\boldsymbol{\theta},\sigma^2)), \quad (44)$$

For information on hyperparameter optimization for random forest, gradient boosting algorithms or SVMs, refer to [43]. Hyperparameter optimization for GPR kernels is further addressed in Section 2.5 of Supplementary Note 2. When used, five-fold cross-validation was carried out in Matlab as [66], [79]

**Fold = 5;** (45)
**indices = crossvalind( 'Kfold', length(input_data_set (:,1)), Fold );** (46)

## 3. Results

*3.1 Qualification of the Input Data – Including the Associated Measurement Error*

Figure 1 shows a build plate of alloy samples fabricated using an EOS M290 powder bed SLM

printer and subjected to HTEM testing. Figure 6 visualizes the input data for the relative density and the Rockwell hardness obtained from HTEM measurements from one of two such build plates of Inconel 718 samples. The relative density was measured using the Archimedes principle, which is an effective method for density evaluation for objects with irregular shapes, whereas the hardness was measured through conventional indentation, using a calibrated indentation machine with a diamond tip (see [3]). The functional behavior of the density and hardness looks qualitatively similar, exhibiting a clear dip at the low laser power and high scan speed, but with the hardness data clearly being subjected to a greater amount of noise. To obtain the results in Figure 6, both the layer thickness and the hatch spacing were set to the values recommended by the vendor of the EOS M290 printer, listed in the Table 1. The laser power and scan speed were varied over a fairly large interval (approx. 2 $x$) roughly centered on the EOS recommended values ($x$). More specifically, in the case of Build Run 1, $N_{BR1} = 175$ data (grid) points were collected, with the laser power increasing from 210 W to 360 W in increments of usually 15 W, but with the scan speed increasing from 600 mm/sec up to 1500 mm/sec in increments of usually 50 mm/sec, 75 mm/sec, 100 mm/sec or 150 mm/sec. However, in the case of Build Run 2, $N_{BR2} = 90$ data (grid) points were collected, with the laser power increasing from 210 W to 360 W in increments of usually 15 W, but with the scan speed increasing from 600 mm/sec, 675 mm/sec, 750 mm/sec or 825 mm/sec up to 1500 mm/sec in increments of usually 50 mm/sec, 75 mm/sec, 100 mm/sec or 150 mm/sec.

**Table 1**: Key parameters recommended by EOS for fabricating IN 718 using the M290 printer.

| Parameter | Value | Unit |
| --- | --- | --- |
| Laser power | 285 | W |
| Scanning speed | 960 | mm / sec |
| Hatch spacing | 0.11 | mm |
| Layer thickness | 0.04 | mm |
| Laser spot radius | 50 | μm |
| Laser spot diameter | 100 | μm |
| Volumetric energy density | 67.5 | J / mm$^3$ |

Provided with the two build runs from the EOS M290 printer, the authors estimated the measurement error from HTEM testing, defined in Figure 4(a), as

$$\sigma_{meas,1} = 0.21\% \qquad \text{(estimated over 90 samples)} \qquad (47)$$

$$\sigma_{meas,2} = 1.42 \text{ HV} \qquad \text{(estimated over 73 samples)} \qquad (48)$$

Given the shape of the data set in Figure 6, the flatness, in case of the density, and the noisy behavior, in case of the hardness, combined with the measurement error from Eqs. (47) and (48), accurate maximization of the relative density or hardness may not be straight forward, since there are no clear peaks apparent for a ML algorithm to extract. But in the case of the AM process optimization, super-accurate estimates of process parameters may not be needed. Good approximations of process windows may suffice.

The qualitative similarity in the functional behavior between the density and the hardness from Figure 6 looks encouraging. Cognizant of the definition of the VED in Eq. (2), and our experimentally-verified processing maps for Inconel 718, overlaid on Figure 6 [3], the authors have a reason to believe that the dip in density and hardness, observed at the low laser power and high scan speed, is caused by the low VED and the lack-of-fusion defects.

*3.2. Qualification of the Three-Dimensional (3D) Surfaces Synthesized*

Figures 8 and 9 capture the 3D surfaces synthesized for the relative density from the input data in Figure 6. Given these surfaces, one can predict the relative density for any candidate combination of the laser power and scan speed. Table 2 and Supplementary Tables S3 – S7 summarize the qualification of the 3D surfaces.

To obtain the results in Figures 8 and 9, the authors employed the default configuration of the **feedforwardnet**() and **train**() functions from the Matlab® deep NN library. This approach

---

[3] These experimentally-verified processing maps reappear in Figures 10 - 12 and in Figure 16.

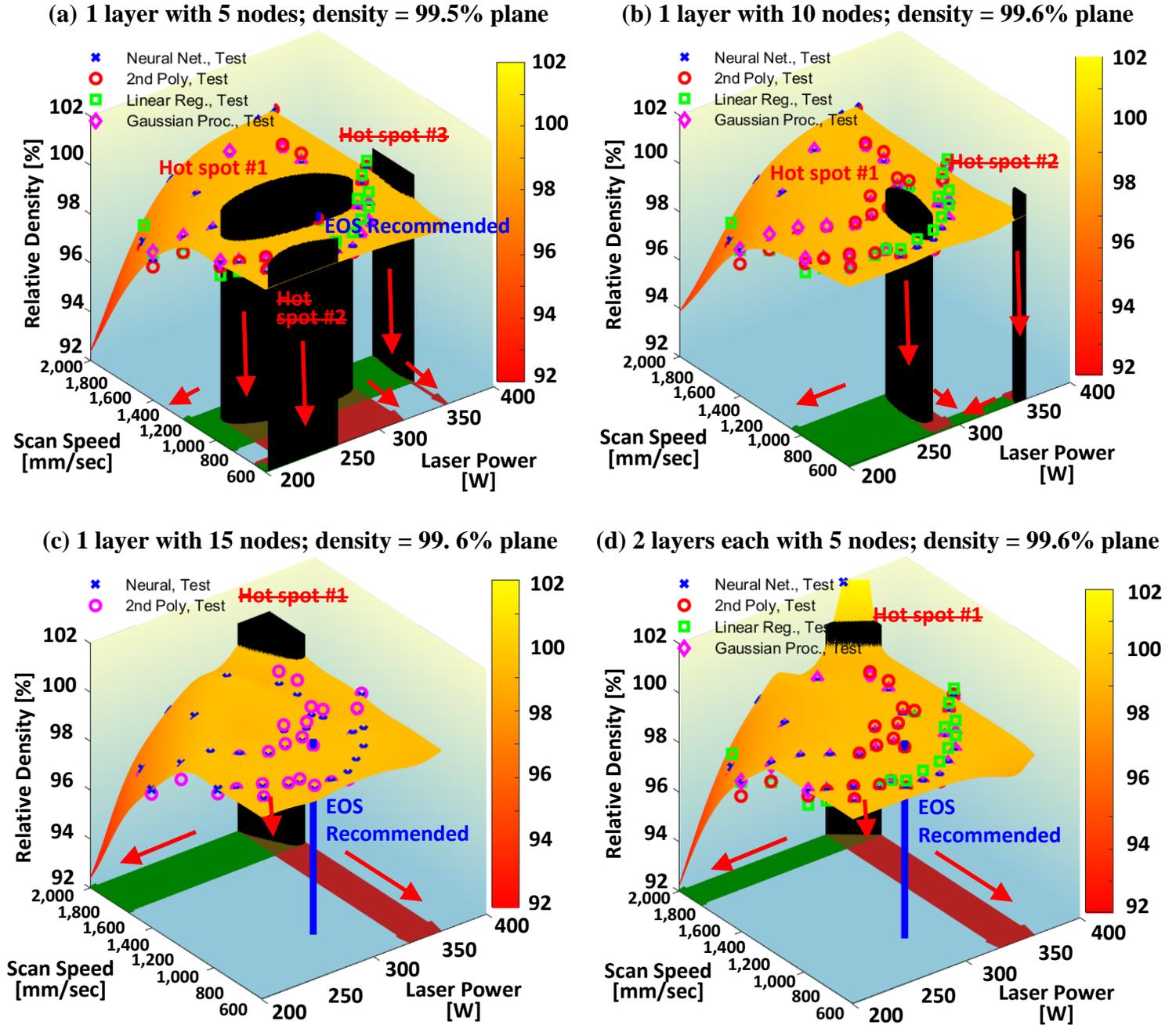

**Figure 8**: Reconstructed density mappings generated from the real AM data (Build Run 1) for IN 718 produced by the HT testing technique. Boundary conditions were excluded from the data set, which was split into a training and testing data, in accordance with the 80/20 rule of data science [65] - [67]. No cross-validation was needed to produce the Figure. The hot spots with strike through represent hot spots located along the boundary which the algorithm disregards. "Neural, Test" refers to predictions obtained from the test set for the NN approach. Similarly, "2nd Poly, Test" refers to predictions obtained from this same test set for the 2nd-order polynomial regression, "Linear Reg. Test" refers to predictions obtained from this test set for the multi-variate linear regression, and "Gauss Proc., Test" refers to predictions obtained by applying Gaussian process regression to this same test set. For corresponding 2D contour maps, refer to Supplementary Figures S1 and S2.

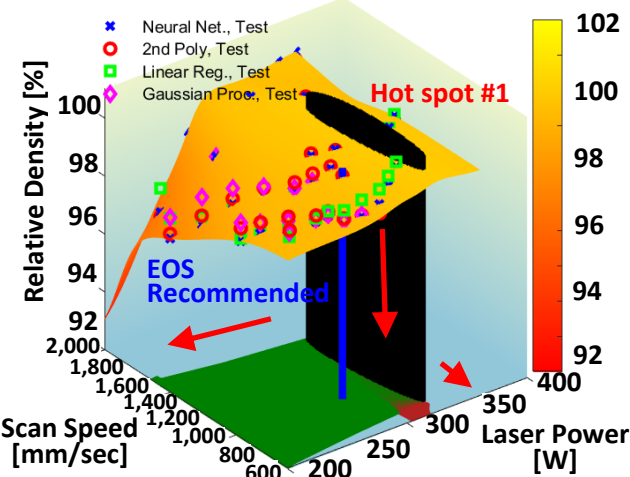
(a) 5 nodes in layer 1, and 10 in layer 2; density = 100% plane

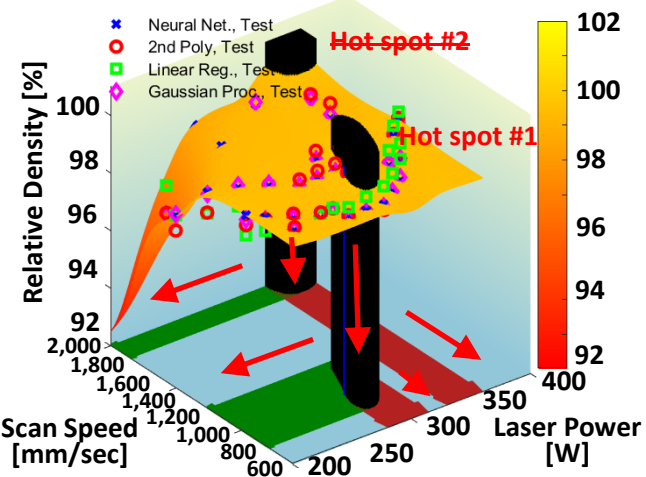
(b) 10 nodes in layer 1, and 5 nodes in layer 2; density = 99.6% plane

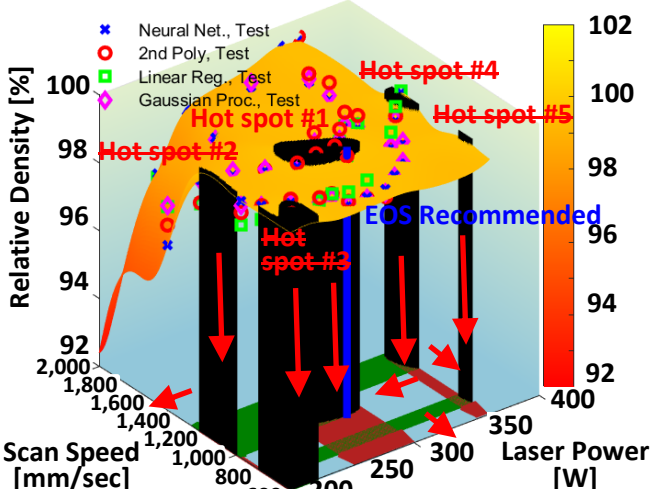
(c) 2 layers with 10 nodes each; density = 99.6% plane

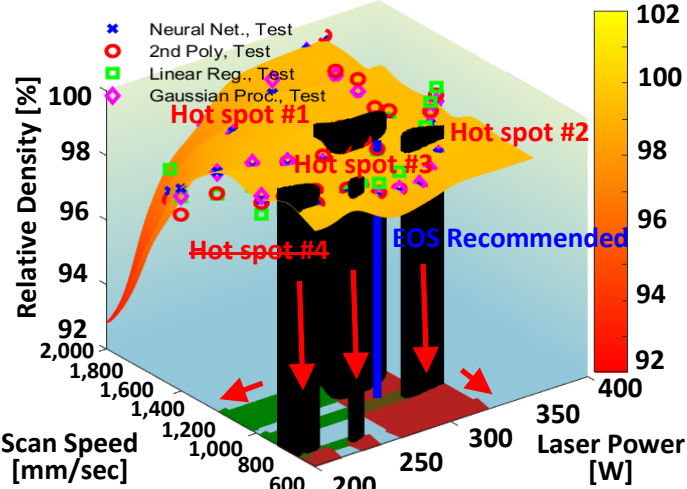
(d) 15 nodes in layer 1, and 10 in layer 2; density = 99.6% plane

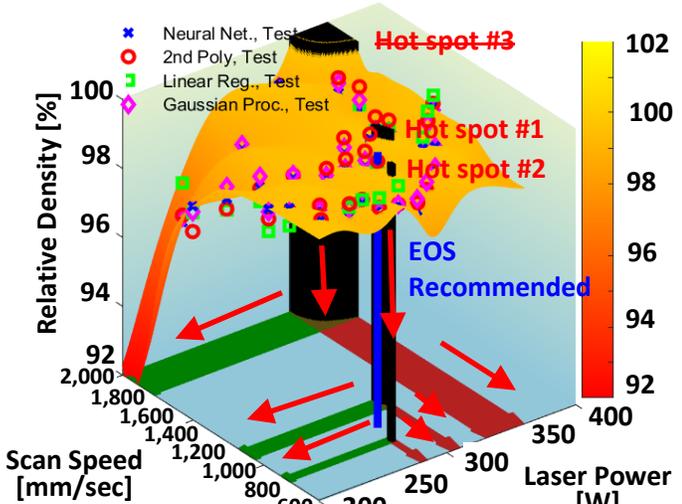
(e) 10 nodes in layer 1 and 15 in layer 2; density = 99.7% plane

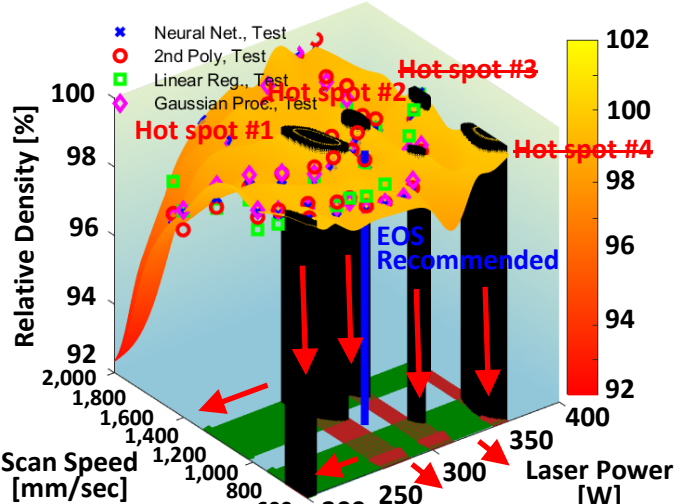
(f) 15 nodes in layer 1 and 15 in layer 2; density = 99.8% plane

**Figure 9**: Additional reconstructed density mappings generated from the real AM data (Build Run 1) for IN 718 produced by the HT-testing technique. Boundary conditions were excluded from the data set, which was split into training and testing data using the 80/20 rule (no cross-validation) [65] - [67].

**Table 2:** Characterization of the quality of fit of the 3D surfaces for the density and hardness synthesized relative to the original input data shown in Figure 6. Here, no boundary conditions have been applied. The quality of fit was estimated exclusively from testing portion of Build Run 1 ($N_{test,BR1}$ = 35 samples for each randomized split). $P(M|D)$ represents the posterior probability of the model listed given the data, whereas $\sigma_{test}$ and $\sigma_{meas}$ appear in Eq. (14). The specific values for $\sigma_{meas,1}$ and $\sigma_{meas,2}$ used are listed in Eqs. (47) and (48). $N_{param}$ represents the number of parameters in the model involved, but $N_{train,BR1}$ the number of samples in the training set from Build Run 1. In this case, $N_{train,BR1}$ = 140 samples. For the NNs, using the syntax of Eq. (24), $N_{param}$ was estimated in Matlab® as **length(getwb(net))**. But for the Gaussian process regression, $N_{param}$ was estimated in Matlab® as **length(gaussianProcessModel.KernelInformation.KernelParameters)**. Similarly, in the case of the polynomial regression, the model size as estimated as **length(regPoly2.Coefficients)**. One run of five-fold cross-validation was performed to obtain the results in the Table. For similar results, obtained for the cases of no cross-validation and ten-fold cross-validation, refer to Supplementary Tables S3 and S5-S7.

| Method | Density | | | | Hardness | | | | Size |
|---|---|---|---|---|---|---|---|---|---|
| | | | | No Boundary Conditions Included | | | | | |
| | $R^2$ | $P(M\|D)$ | $\sigma_{test,1}$ [%] | $\sigma_{test,1}$ / $\sigma_{meas,1}$ | $R^2$ | $P(M\|D)$ | $\sigma_{test,2}$ [HV] | $\sigma_{test,2}$ / $\sigma_{meas,2}$ | $N_{train,BR1}$ / $N_{param,BR1}$ |
| 1-layer NN: 5 nodes | 0.834 | 0.1077 | 0.250 | 1.179 | 0.054 | 0.0759 | 2.627 | 1.850 | 6.67 |
| 1-layer NN: 10 nodes | 0.746 | 0.1054 | 0.347 | 1.637 | 0.092 | 0.0825 | 2.630 | 1.852 | 3.41 |
| 1-layer NN: 15 nodes | 0.733 | 0.0967 | 0.386 | 1.821 | 0.063 | 0.0904 | 2.914 | 2.052 | 1.54 |
| 2-layer NN: 5 + 5 nodes | 0.821 | 0.1069 | 0.325 | 1.533 | 0.037 | 0.1064 | 2.857 | 2.012 | 2.75 |
| 2-layer NN: 5 + 10 nodes | 0.760 | 0.1025 | 0.280 | 1.321 | 0.129 | 0.1073 | 2.667 | 1.878 | 1.62 |
| 2-layer NN: 10 + 5 nodes | 0.783 | 0.1070 | 0.243 | 1.146 | 0.035 | 0.0931 | 2.797 | 1.970 | 1.54 |
| 2-layer NN: 10 + 10 nodes | 0.732 | 0.0919 | 0.352 | 1.660 | 0.032 | 0.1336 | 2.875 | 2.025 | 0.93 |
| 2-layer NN: 10 + 15 nodes | 0.780 | 0.0965 | 0.358 | 1.689 | 0.055 | 0.1105 | 3.114 | 2.193 | 0.66 |
| 2-layer NN: 15 + 10 nodes | 0.687 | 0.0887 | 0.361 | 1.703 | 0.049 | 0.1025 | 2.785 | 1.961 | 0.65 |
| 2-layer NN: 15 + 15 nodes | 0.725 | 0.0968 | 0.291 | 1.373 | 0.162 | 0.0979 | 2.671 | 1.881 | 0.47 |
| 2nd-order polynomial reg. | 0.668 | 1.0 | 0.510 | 2.406 | 0.056 | 1.0 | 2.591 | 1.825 | 23.33 |
| 1st-order linear regression | 0.490 | 1.0 | 0.738 | 3.481 | 0.055 | 1.0 | 2.604 | 1.834 | 46.67 |
| Gaussian process regression | 0.756 | 1.0 | 0.278 | 1.311 | 0.081 | 1.0 | 2.690 | 1.894 | 70 |

included specifying the Levenberg-Marquardt algorithm as the training algorithm ('trainlm') together with random data division ('dividerand'). For the five-fold cross-validation in Table 2, the results were similarly obtained, by using 80% of the data from Build Run 1 for training, but 20% for testing, for each randomized split of the input data set, in accordance with the classical 80 / 20 rule [65], [67]. In case of no cross-validation in Supplementary Tables S6 and S7, the results were also obtained, by using 80% of the input data set from Build Run 1 for training, and 20% for testing. But in case of the ten-fold cross-validation in Supplementary Tables S3 and S5, only 10% of the input data set was used for testing, for each of the ten randomized splits of the input data set. Note that Build Run 1 featured $N_{BR1} = 140 + 35 = 175$ samples, with $N_{train,BR1} = 140$ samples and $N_{test,BR1} = 35$ samples (in case of five-fold cross-validation or no cross-validation). Build Run 2 only featured $N_{BR2} = 90$ samples in total. Initially, the authors refrained from splitting Build Run 2 into a training and a testing set, due to small size of the testing set, and the resulting lack of statistical relevance anticipated. Supplementary Figures S3 – S4 capture 3D surfaces synthesized for the relative density, obtained by using all $N_{BR2} = 90$ samples from Build Run 2 for training, whereas Supplementary Figures S5 – S6 contain corresponding 2D contour maps. Build Run 2 was then treated in analogous fashion as Build Run 1, in accordance with the classical 80/20 rule of data science [65] - [67], in an effort to develop insights into consistency of the methods considered across build runs (see Figures S7 – S8 in the Supplementary Manuscript).

In Figures 8 and 9 [4], the portions of the 3D surfaces exceeding the intersecting density plane (threshold) specified, referred to as hot spots, were determined as follows:

1. For each of the network structures, the authors started with a density plane (threshold) set at 100.0%.

---

[4] For corresponding 2D contour figures, refer to the Supplementary Manuscript.

2. The authors then lowered the density threshold, approx. in increments of 0.1%, and determined intersections with the 3D surface, until the authors identified the hot spots shown (illustrated in form of the black vertical lines in Figures 8 and 9 and corresponding to the 3D surface exceeding the intersecting the density threshold specified).

3. As far as the parameter optimization was concerned, the authors ignored (crossed out), but still presented in Figures 8 and 9, the hot spots located along the boundary of the region covered by the training samples.

To estimate the Bayesian posterior probability for different NN models, $P(M|D)$, listed in Table 2 and in Supplementary Tables S3 and S6, the authors employ an extension of a simple example outlined in [80], for the Bayesian model selection:

$$M_i: \text{NN Model } i: \qquad p(D/\mathbf{x};M_i) = p(\mathbf{y}_{\text{test}}/\mathbf{x};M_i) = N(\mathbf{y}_{\text{test}}; \mu = y_{\text{train},i}(\mathbf{x}), \sigma^2_{\text{meas}}) \qquad (49)$$

Here, $\mathbf{y}_{\text{test}}$ represents a vector containing the measured density or the hardness values corresponding to the testing set, respectively. Similarly, $\sigma_{\text{meas}}$ denotes the standard deviation of the measurement error. Furthermore, the vector $\mathbf{y}_{\text{train},i}(\mathbf{x})$ contains the measured density or the hardness values predicted by the NN Model, $M_i$, for the input, $\mathbf{x}$, based on the training set. In Table 2 and Supplementary Tables S3 – S6, the authors obtain the coefficient of determination, $R^2$, from the cross-correlation coefficient, $\rho(\cdot,\cdot)$, as follows:

$$R^2 = [\rho(\mathbf{y}_{\text{test}}, \mathbf{y}_{\text{train},i}(\mathbf{x}))]^2. \qquad (50)$$

Supplementary Table S7 presents the corresponding comparison of model structures, in terms of the mean squared error (MSE), which is estimated by utilizing the Froebenius norm [81].

According to Table 2 and Supplementary Tables S3 – S6, the NN structures, which yield the highest coefficient of determination, $R^2$, for the relative density, high model posterior probability, $P(M|D)$, and the lowest normalized standard deviation for the testing error, $\sigma_{test}$ [defined in Figure 4(b)], generally consist of

1. A single-layer neural network with five (5) nodes;

2. A single-layer neural network with ten (10) nodes.

These observations are consistent with the 'Occam factor' for model selection, i.e., the term that penalizes complex NNs yielding good fit to the training data but with limited ability to accurately predict new testing data [see Eq. (21)]. Comparison of Table 2, Supplementary Table S3 and Supplementary Table S6 suggests qualitative similarity in observations obtained from five-fold cross-validation, ten-fold cross-validation and from no cross-validation, but considerable difference in the absolute prediction performance observed. Supplementary Tables S4 and S5 further reveal considerable differences in the absolute prediction performance observed between successive runs of randomized splitting of the input data set, for a given cross-validation method (for the five-fold cross-validation or for the ten-fold cross-validation). While $R^2$ for the 1-layer NN with 5 or 10 nodes tends to be quite similar to that for the Gaussian process regression, and while the prediction performance observed for different methods tends to be qualitatively preserved across successive test runs on average, both for the five-fold cross-validation and for the ten-fold cross-validation, there is quite significant difference in the absolute prediction performance observed across successive test runs, even of the same cross-validation method, according to Supplementary Tables S4 and S5. The randomized splitting seems to have quite significant impacts on the absolute prediction performance observed.

Supplementary Tables S7 is intended to offer insight into the effect of including the boundary conditions from Eqs. (9) – (10) on the quality of the multi-dimensional surfaces constructed. Simply put, including the boundary conditions from Eq. (9) – (10) in the training set does not help in terms of improving the overall quality of the multi-dimensional function synthesized. As Supplementary Figures S9 and S10 illustrate, the boundary conditions significantly impacted the overall shape of the multi-dimensional surfaces over and beyond what the measured HTEM data points accounted for. While a physics-assisted approach to multi-dimensional optimization of AM process parameters likely does make sense, a milder approach for accounting for the boundary conditions may be warranted. The authors did not extend Supplementary Table S7, such as to account for cross-validation, because the authors did not want the boundary conditions in the extended input data set to be assigned to the testing data set, through randomized splitting. Furthermore, comparison between Figure 6 and Figures 8 - 9, on one hand, and Figures S3 – S8, on the other hand, reveal that *there are significant variations in the location of the hot spots between Build Runs 1 and 2*. Such behavior may be inherent to the AM fabrication process and the HTEM measurement process employed [may suggest large $\sigma_{meas}$, per Figure 4(a)]. Note that the authors are shifting the density plane by 0.1% (or even less), to determine the hot spots, while the standard deviation in the density measurements for the input data is $\sigma_{meas,1} = 0.21\%$, per Eq. (47). With this being the case, the NN approach may offer good guidance towards optimized AM process parameters (good estimates of process windows), but may not be expected to offer super-accurate estimates, at least not in case of the relative densities from Build Runs 1 and 2. Note also that additional build runs may yield more accurate estimates of $\sigma_{meas}$, per Figures 4 - 5, but may not decrease the level of variations actually observed, i.e., not decrease the value of $\sigma_{meas}$ observed, provided the variations are inherent to the AM fabrication process and the HTEM measurement process employed. In this case, a density threshold of 99.6% looks reasonable, for the purpose of generating hot spots, for many of the NN implementations considered. Moreover, in Figure 8 - 9, the predictions from the test sets looked qualitatively reasonable not only for the NN implementations, but also for the comparative methods considered (the multi-variate linear regression, the 2nd-order polynomial regression and for the Gaussian process regression), which are further covered in Section 4.4 below. In all, these results highlight the importance of incorporating good estimation and approximation analysis into the optimization framework.

## 4. Discussion

*4.1 Preference to Simple, Single-Layer Networks with 5 – 10 Nodes, as Far as Functional Reconstruction Is Concerned - Especially. in Light of the Limited Training Data*

The first take-away from Figures 8 and 9, and Table 2 pertains to the general ability of the framework for multi-dimensional functional synthesis to reconstruct the 3D-density surfaces with a relatively-high degree of accuracy. As far as the AM parameter optimization is concerned, there may exist more than one solution. In other words, optimized parameters for the laser power and scan speed may be derived from more than one NN structure. The optimization framework may produce a few candidate sets of optimized AM process parameters.

Another take-away from Table 2 relates to the initial preference to simple, single-layer NNs with 5 or 10 nodes, as noted above, on basis of the highest coefficient of determination, $R^2$, high model posterior probability, $P(M|D)$, and the lowest testing error standard deviation, $\sigma_{test}$. This observation is consistent with the theme of the 'Occam factor' in Eq. (21) [14], [67], [70]. Figure 6 and Figures 8 - 9 intuitively show the principles behind the 'Occam factor' (also referred to in the literature as Occam's razor) in action. The simple, single-layer neural networks produce 3D surfaces with relatively smooth and simple shapes and without extraneous wrinkles. Per the qualification condition in Eq. (14), these models are also considered good enough.

Note the observation here regarding the initial preference to simple, single-layer network is specific to the data set being analyzed, and its relatively simple (flat) functional behavior. In the case of higher dimensions (the optimization of a data set obtained from a larger number of AM process input parameters) or other alloys, the same conclusion may or may not be fully held, depending on the extent and nature of the data analyzed. The fundamental principles behind the 'Occam factor' will though still hold [67].

*4.2 Assessment of Ability to Estimate Two-Dimensional (2D) Hot Spots – In Context with Experimentally Verified Process Windows*

Despite the relatively simple functional behavior, the simple, single-layer neural networks with 5 - 10 nodes are still able to estimate 2D hot spots, as Figures 10 - 11 show. Each of the structures in Figures 10 - 11 produces a Hot Spot #1 that roughly aligns with the (laser power, scan speed) combination recommended by the vendor of the M290 printer, EOS, and that produces a narrowed-down search space, to which a new dimension (e.g., layer thickness or hatch spacing) can be appended, in preparation for the AM parameter optimization in a higher dimension. The 2-layer network with 15 nodes in the first layer and 10 in the second layer provides two additional Hot Spots, labeled as #3 and #4, within the region covered by the training samples. But Hot Spot #2 falls along the boundary of that region and is excluded from consideration. There certainly are variations in the hot spot location between the three structures in Figure 10. Moreover, a comparison between Figures 8 - 9 and Supplementary Figures S3 – S4 shows that there are even larger variations in the hot spot locations between Build Runs 1 and 2. Comparison between the corresponding 2D contour maps (Supplementary Figures S1 – S2 and S5 – S6) suggests the same. The message from Figure 10 and Table 2, in terms of selecting a NN structure and estimating the 2D hot spots, can be summarized as follows:

1. Start out by estimating the measurement error, $\sigma_{meas}$.
2. Select neural network structure(s) yielding multi-dimensional functional reconstruction with

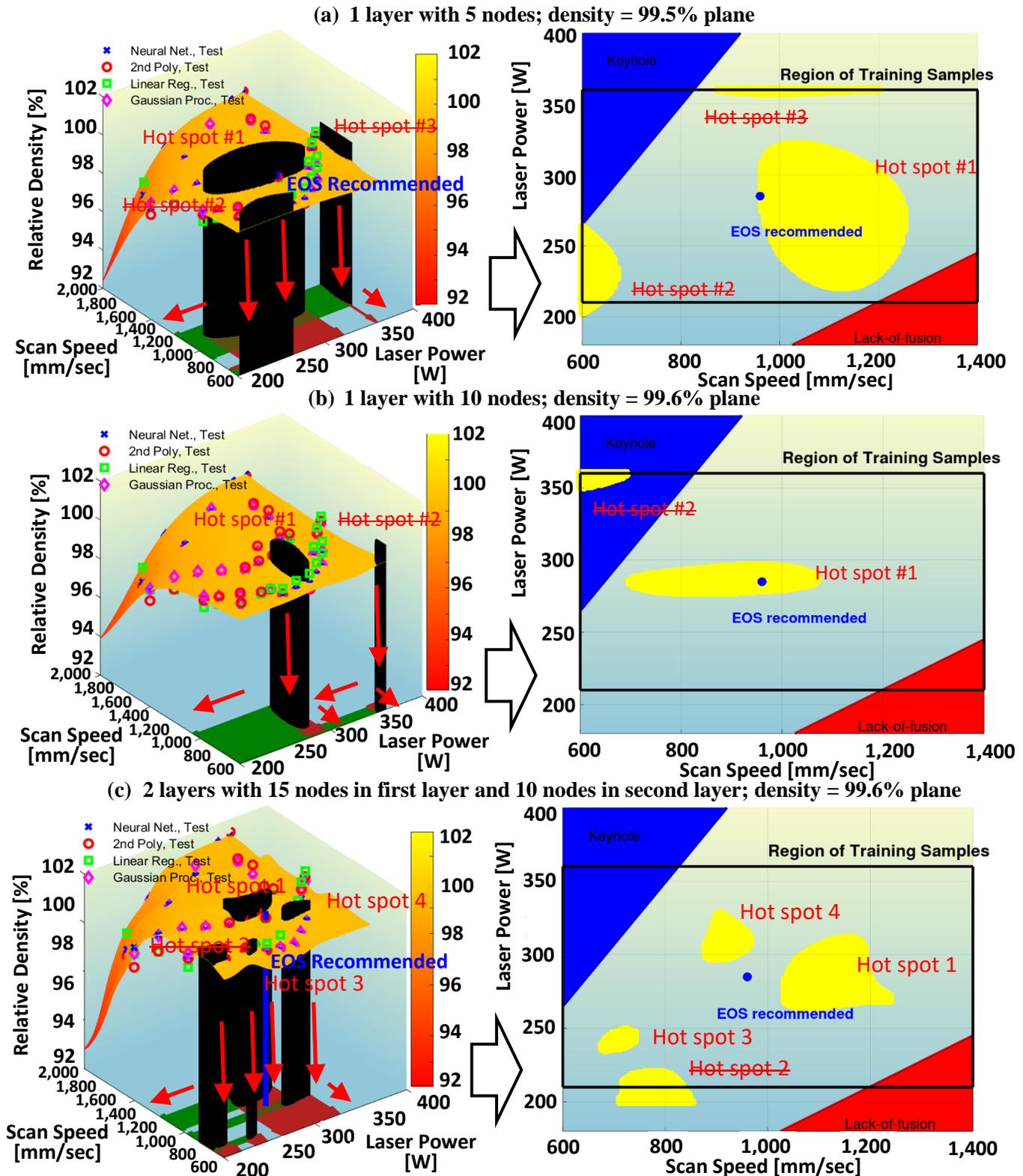

**Figure 10:** Key ingredients of a practical approach for extending the AM parameter optimization from Figures 8 - 9 (Build Run 1) to higher dimensions. The experimentally-verified processing map for keyholing is defined in terms of VED > 100 J/mm$^3$. The experimentally-verified processing map for lack-of-fusion is, correspondingly, defined in terms of VED < 40 J/mm$^3$.

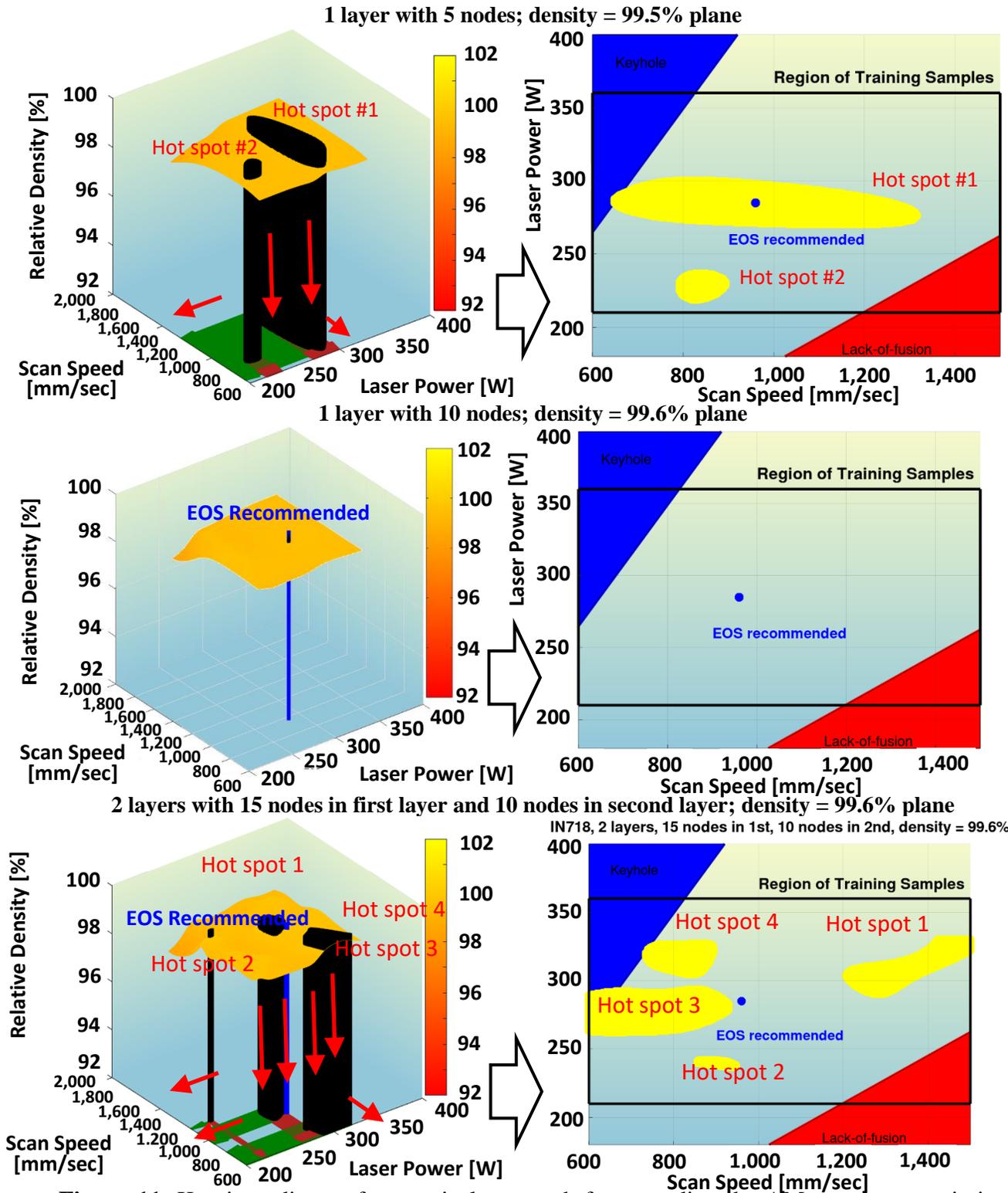

**Figure 11**: Key ingredients of a practical approach for extending the AM parameter optimization from Figures S3 - S4 (Build Run 2) to higher dimensions. Here, all $N_{BR2} = 90$ samples have been used for training. The experimentally verified processing map for keyholing is defined in terms of VED > 100 J/mm$^3$. The experimentally verified processing map for lack-of-fusion is, correspondingly, defined in terms of VED < 40 J/mm$^3$. No cross-validation was needed to produce the Figure.

the highest $R^2$, high model posterior probability, $P(M|D)$, and the lowest testing error, $\sigma_{\text{test}}$.

3. Qualify the trained neural network(s) by applying the condition from Eq. (14). Verify that the neural network(s) produce meaningful 2D hot spots, i.e., hot spots not aligning with the boundary values from the training set.

4. If more than one network fulfills qualification Steps 2 – 4 above, then select the network(s) whose parameters have been properly trained, given the input data available, i.e., are supported by the figure for the data points in the training set, $N_{\text{train,BR1}}$, exceeding the number of model parameters, $N_{\text{param}}$, per Table 2 ($N_{\text{train,BR1}} / N_{\text{param}} > 1$).

The network selection procedure summarized in Steps 2 – 4 above may not produce a unique solution. It is possible that more than one network structure qualifies. Step 5 also embodies the impact of the 'Occam factor' from Eq. (21).

Figures 12 - 17 describe the low-throughput microstructure characterization of as-fabricated Inconel 718 samples that inspires the estimation of the experimentally verified process windows shown in Figure 10. Figure 16 specifically captures the process windows verified through the microstructure analysis. The dip in the density and hardness in Figure 6 at low VED (for the lack-of-fusion side) suggests that similar process bounds can be estimated using an ML approach. But to estimate a similar process boundary for the keyhole side, additional training data may be needed to be sampled, for purpose of producing a similar dip at high VED, as observed at the low VED.

*4.3 Extension to Four and Five Dimensions – Incorporation of Layer Thickness and Hatch Spacing*

Table 3 presents a parameter configuration for a HTEM test aimed at generating training data enabling extension of the AM parameter optimization to four dimensions (4D), where one optimizes over the hatch spacing in addition to the laser power and the scan speed. Build Run 3 assumes reduced sample space for the laser power and scan speed, compared to Build Runs 1 and 2, but smaller hatch spacing. Build Run 4 similarly assumes reduced sample space for the laser power and scan speed, compared to Build Runs 1 and 2, but larger hatch spacing. Build runs 3 and 4 have, further, been selected such as to fit on a build plate capable of accommodating up to 100 samples. And the grid comprising the laser power and scan speed has been selected such as to overlap with the sampling grid from Build Runs 1 and 2, for ease of comparison. Build Run 3 extends to slightly higher VED than Build Runs 1 and 2, but not by much.

Table 4 captures parameter settings recommended for optimization of SLM processing parameters for Inconel 718 in 4D and 5D. The parameter setting was derived based on ML analysis of the 3D density surfaces in Figures 6 - 8 of the main manuscript combined with expert insight. The 300 samples recommended for 5D analysis fit on three (3) build plates. Build Run 5 may be preferred over Build Runs 3 and 4, in part because it is more extensively sampling the high VED region, despite a sparser sampling grid, and one that does not fully align with the sampling grid from Build Runs 1 and 2.

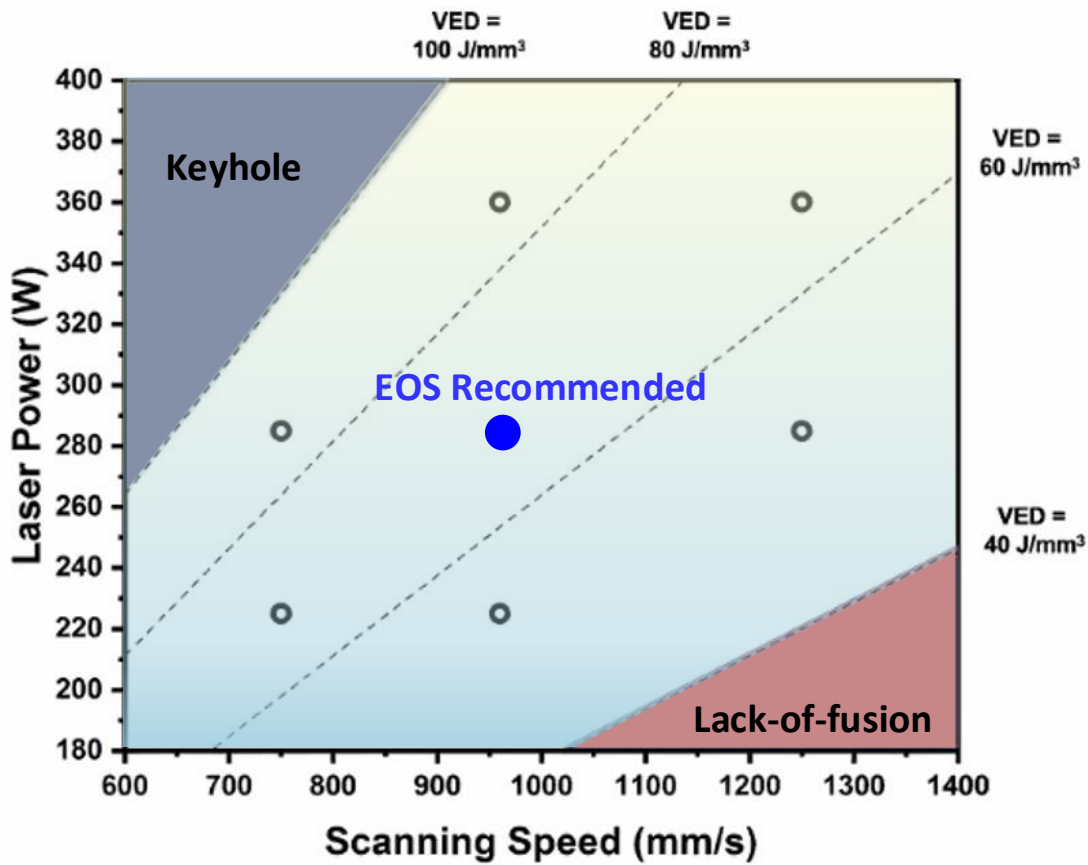

**Figure 12**: Properties (laser power and scan speed) of seven (7) samples that have been selected for low-throughput characterization (for correlation with microstructure properties). The intent is to select a minimum number of samples for which one can assess the impact of laser power, the scanning speed and the VED. The sample corresponding to the EOS recommended settings (VED = 67.5 J/mm$^3$) is shown as a blue, filled circle, but the other six (6) samples are shown as dark, unfilled circles.

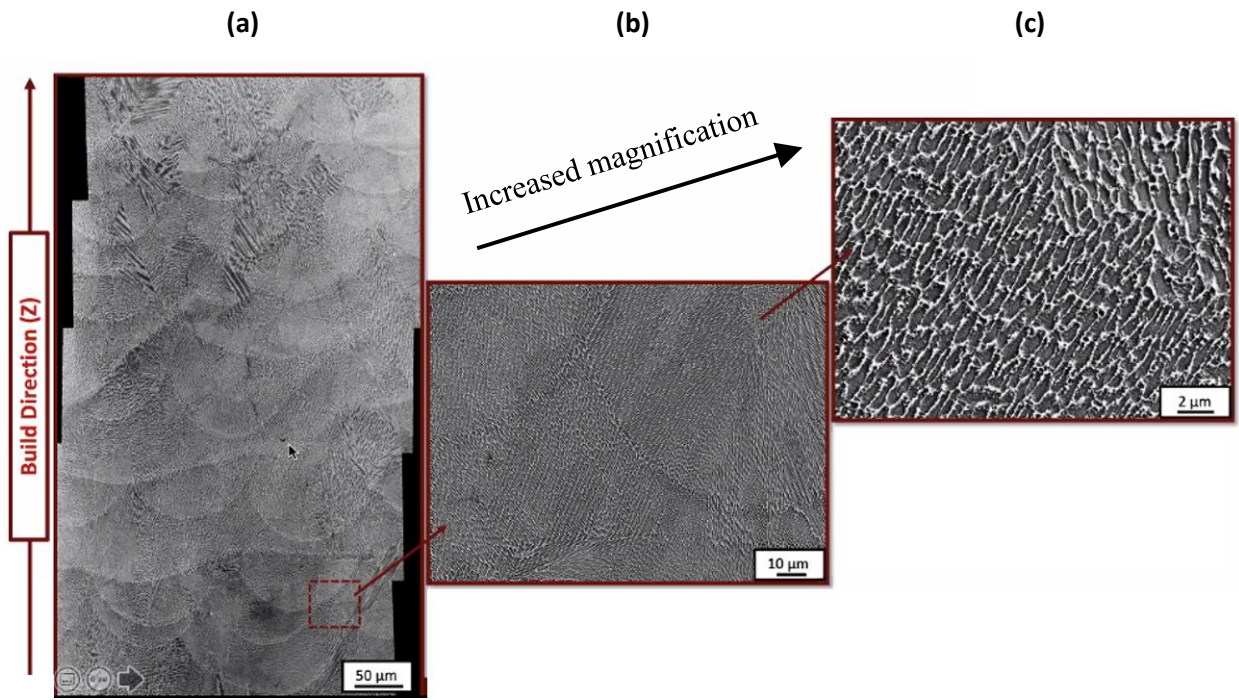

**Figure 13**: SEM images showing cross sectional views of the sample from Figure 12 corresponding to the EOS recommended parameters at different magnification. We carefully controlled the magnification for the purpose of identifying the region from where we can analyze the dendritic arm spacing. We also looked to avoid the area which exhibited overlap with the melt pool. Figure 13(c) suggests columnar grains as well as possibility of a second phase. Figure 13(c), furthermore, corresponds to high calibration allowing calibration of the dendritic arm spacing.

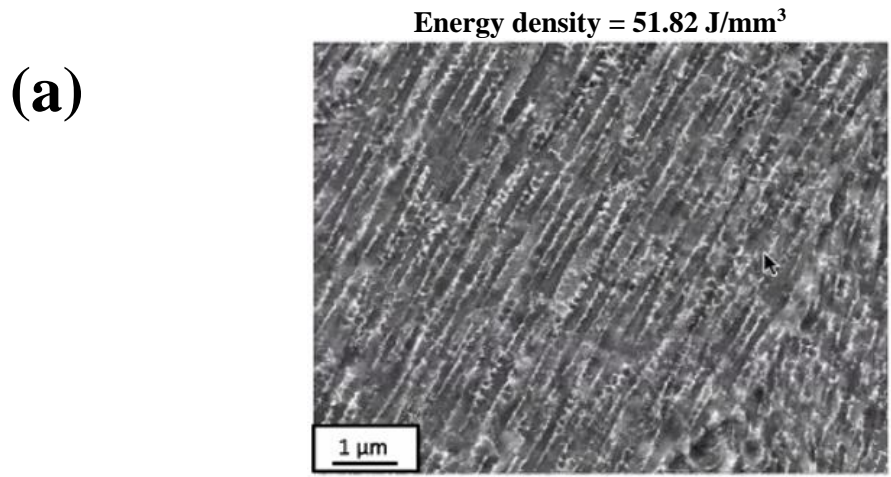

**(a)** Energy density = 51.82 J/mm³

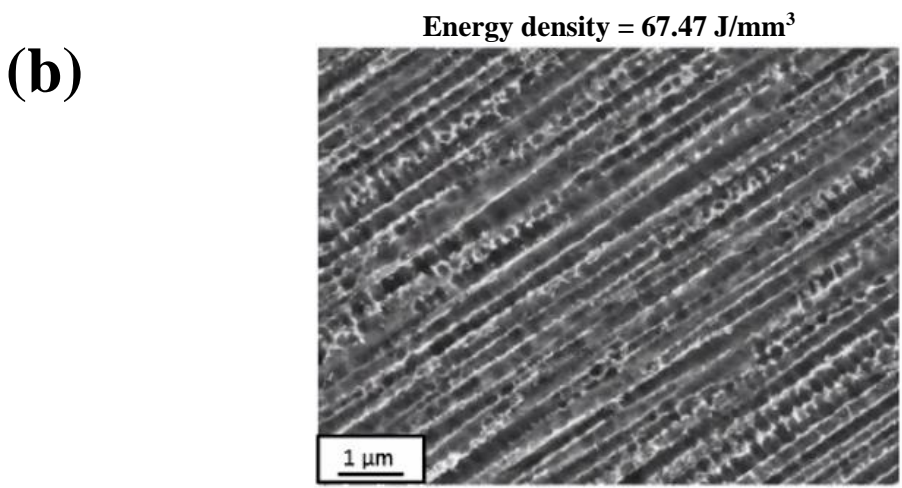

**(b)** Energy density = 67.47 J/mm³

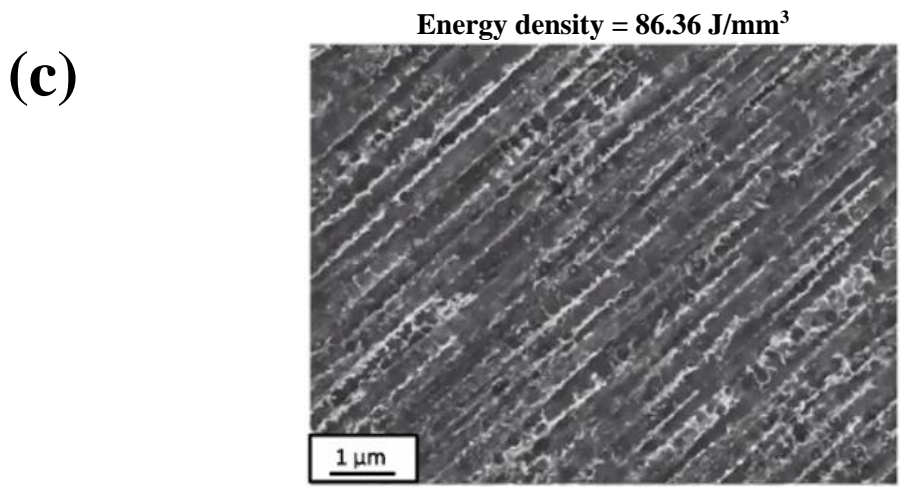

**(c)** Energy density = 86.36 J/mm³

**Figure 14**: Primary dendritic arm spacing for three (3) samples selected for low-throughput characterization. The range of VED, for which we can fabricate dense components for Inconel 718, is between 40 and 100 J/mm³, as indicated in Figure 6. At the lower energy density, the value for the dendritic arm spacing is around 1 – 4 μm. As we increase the energy density, the dendritic arm spacing comes to be about .58 μm. The dendritic arm spacing seems strongly dependent on the VED.

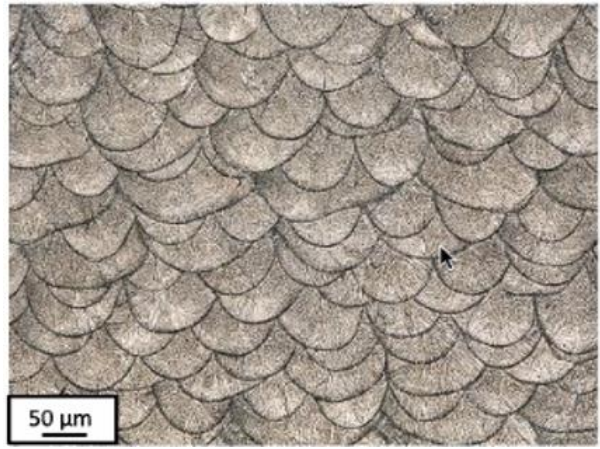

**Energy density = 51.82 J/mm³**

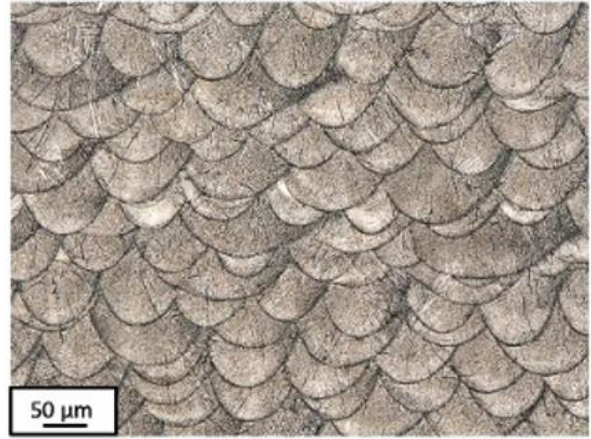

**Energy density = 67.47 J/mm³**

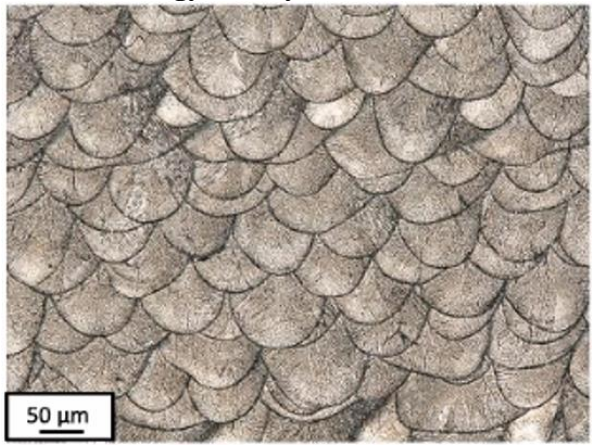

**Energy density = 86.36 J/mm³**

**Figure 15**: Melt pool characteristics for the three samples from Figure 14. At the lower energy density, the melt pool generically is semi-spherical and shallow. And then, when one increases the energy density, the melt pool becomes deeper The AM process parameters indeed seem to affect the dendritic arm spacing and the melt pool dimensions (along with the grain size, aspect ratios, solidification conditions and the mechanical properties).

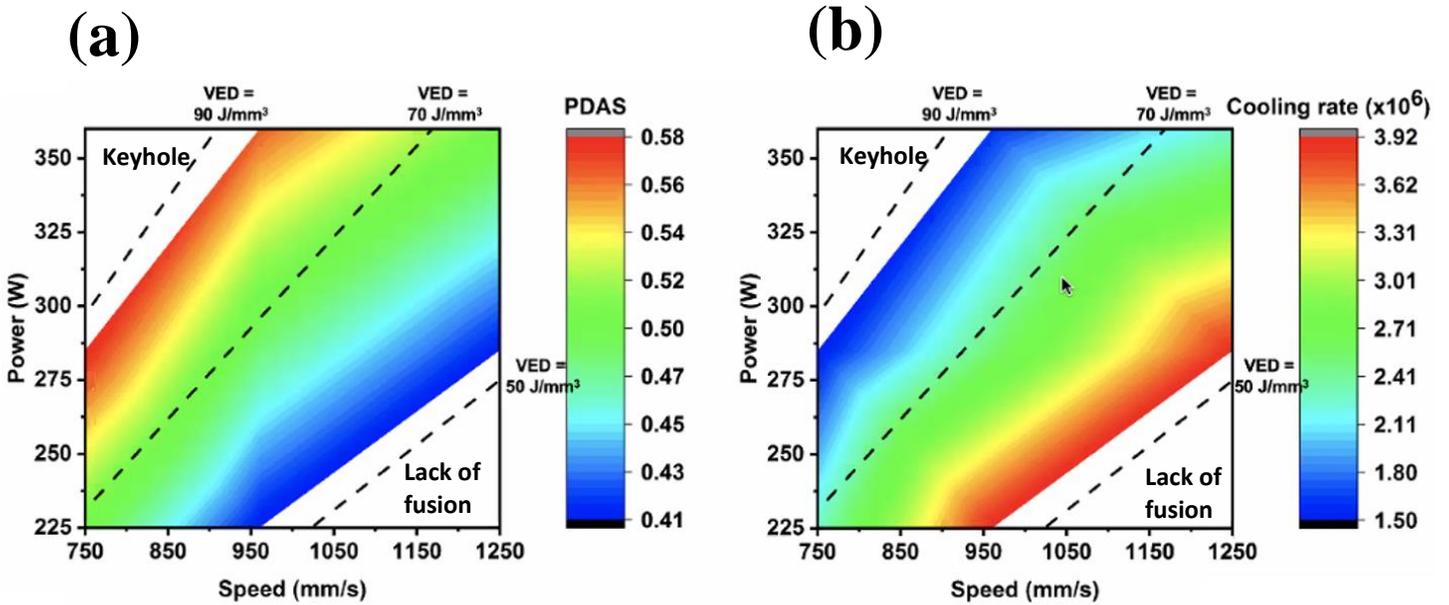

**Figure 16**: Process windows for Inconel 718 verified experimentally through microstructure analysis. The figure to the left characterizes the primary dendritic arm spacing, but the one on the right, the cooling rate. At the lower energy density, the cooling rate is clearly higher than at the higher energy densities. This is consistent with analytical models. The cooling rate impacts thermal gradients and local solidification conditions.

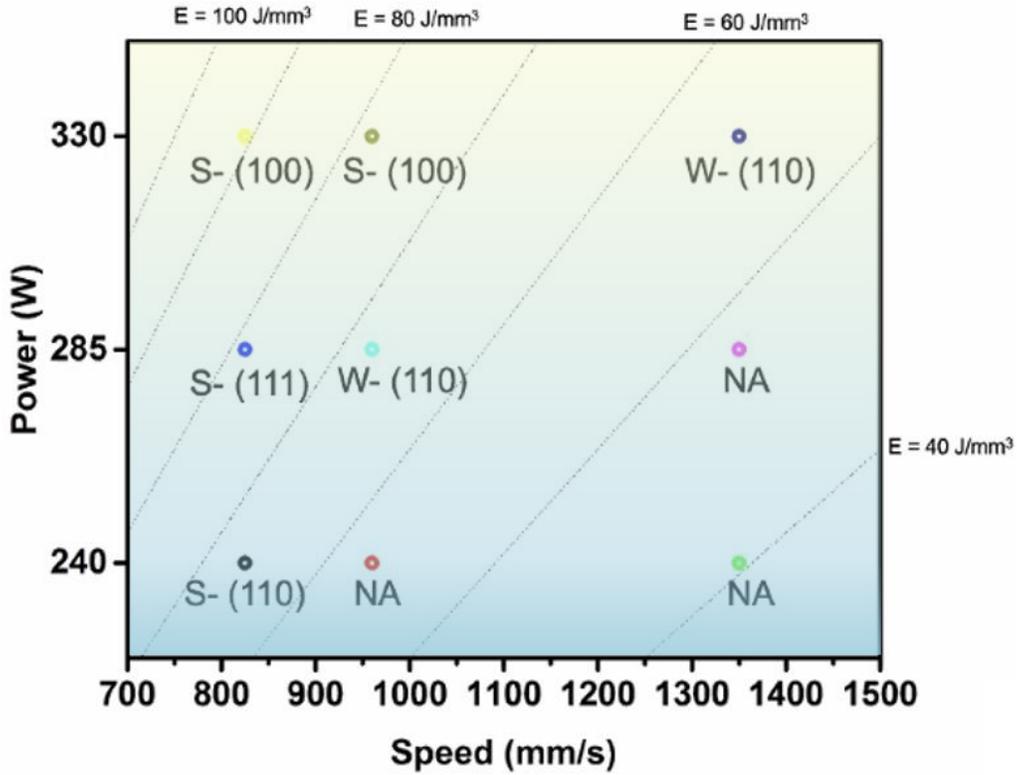
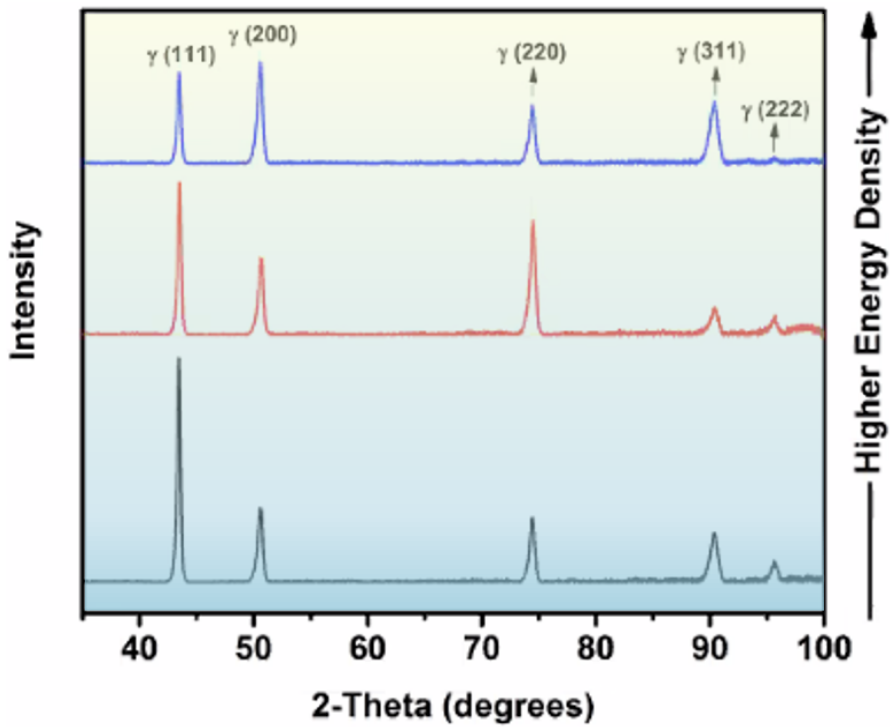

**Figure 17**: Analysis of texture properties, i.e., grain orientation relative to the build direction, from XRD. As the energy density increases, we observe transition from a random texture first to 110 planes and then – at a very high energy density – to 100 planes. At the low energy density, one tends to observe more random texture, smaller dendritic arm spacing and semi-spherical melt pool than at the high energy density.

| Parameter | | Build Runs 1 and 2 | Build Run 3 | Build Run 4 |
|---|---|---|---|---|
| **Laser Power** | Min. [W] | 210 | 240 | 240 |
| | Max. [W] | 360 | 330 | 330 |
| | Δ [W] / No. samples | 15 / **11** | 15 / **7** | 15 / **7** |
| **Scan Speed** | Min. [mm/sec] | 600 | 825 | 825 |
| | Max. [mm/sec] | 1500 | 1275 | 1275 |
| | Δ / No. samples | 75, 100 or 150 mm/sec / **11** | 75 mm/sec / **7** | 75 mm/sec / **7** |
| **Hatch Spacing** | Min. [mm] | 0.11 | 0.07 | 0.13 |
| | Max. [mm] | 0.11 | 0.09 | 0.15 |
| | Δ [mm]/ No. samples | 0.0 / **1** | 0.02 / **2** | 0.02 / **2** |
| **Layer Thickness** | Min. [mm] | 0.04 | 0.04 | 0.04 |
| | Max. [mm] | 0.04 | 0.04 | 0.04 |
| | Δ [mm]/ No. samples | 0.0 / **1** | 0.0 / **1** | 0.0 / **1** |
| **VED** | Min. [J/mm$^3$] | 31.8 | 52.3 | 31.4 |
| | Max. [J/mm$^3$] | 136.4 | 142.9 | 76.9 |
| **Overall** | No. samples | **11 * 11 * 1 = 121** | **7 * 7 * 2 = 98** | **7 * 7 * 2 = 98** |

**Table 3:** Illustration of parameter selection for a HTEM test run applied to SLM fabricated Inconel 718 samples from a build plate capable of accommodating up to 100 samples for 4D analysis. Here the grid comprising the laser power and scan speed for Build Runs 3 and 4 has been selected such as to overlap with the grid from Build Runs 1 and 2, for ease of comparison.

| Parameter | | *Build Run 5* |
|---|---|---|
| **Laser Power** | $LP_1$ [W] | 200 |
| | $LP_2$ [W] | 240 |
| | $LP_3$ [W] | 280 |
| | $LP_4$ [W] | 320 |
| | $LP_5$ [W] | 360 |
| **Scan Speed** | $SS_1$ [mm/sec] | 600 |
| | $SS_2$ [mm/sec] | 800 |
| | $SS_3$ [mm/sec] | 1,000 |
| | $SS_4$ [mm/sec] | 1,250 |
| | $SS_5$ [mm/sec] | 1,500 |
| **Hatch Spacing** | $HS_1$ [mm] | 0.09 |
| | $HS_2$ [mm] | 0.11 |
| | $HS_3$ [mm] | 0.13 |
| | $HS_4$ [mm] | 0.15 |
| **Layer Thickness** | $LT_1$ [mm] | 0.02 |
| | $LT_2$ [mm] | 0.04 |
| | $LT_3$ [mm] | 0.06 |
| **VED** | Min. [J/mm$^3$] | 14.8 |
| | Max. [J/mm$^3$] | 333.3 |
| **Overall** | No. samples | **5 * 5 * 4 * 3 = 300** |

**Table 4:** Parameter setting recommended for optimization of SLM processing parameters for Inconel 718 in 4D and 5D. The parameter setting was derived based on ML analysis of the 3D density surfaces in Figures 8 - 11 combined with expert insight. The 300 samples recommended for 5D analysis fit on three (3) build plates.

*4.4 More on the Comparison with Alternative Methods*

Multivariate linear and 2nd order polynomial regression was implemented using Ver. 1.4.0.0 of the Matlab® library provided by Cecen [82]. The Gaussian process regression was implemented using the Matlab® function **fitrgp()**, from the Statistics and Machine Learning Toolbox, with default settings, along with the **predict()** function, from the same toolbox:

**gaussianProcessModel = fitrgp( input_data_train, noisy_output_train );**

**noisy_output_GaussianProcess_modeled_test = predict( gaussianProcessModel, data_test );**

For further specifics of our implementation of the Gaussian process regression, and on Bayesian regression in general, refer to Supplementary Note 2.

In the case of no cross-validation, the multivariate linear regression model, trained on 80% of the data from Build Run 1 ($N_{\text{train,BR1}} = 140$ samples), has the structure:

$$\text{Relative Density [\%]} = 98.7003 + 0.00872753 * (\text{Laser power}) - 0.00161589 * (\text{Scan speed}), \quad (51)$$

whereas the 2nd-order polynomial regression model, trained on the same data and with no cross-validation, has the following structure:

$$\begin{aligned}\text{Relative Density [\%]} = &\ 97.6825 + 0.0264329 * (\text{Laser power}) - 0.00339034 * (\text{Scan speed}) \\ &+ 3.24226 * 10^{-5} * (\text{Laser power}) * (\text{Scan speed}) - 0.000107201 * \\ &(\text{Laser power})^2 - 2.59929 * 10^{-6} * (\text{Scan speed})^2.\end{aligned} \quad (52)$$

Figure 18 compares the prediction accuracy of the neural network solutions to that of multi-variate linear regression, 2nd-order polynomial regression and Gaussian process regression for Build Run 1 and the case of five-fold cross-validation. In terms of the coefficient of determination, $R^2$, Figure 18 indicates that the relatively simple single-layer NN with 5 or 10 nodes outperforms multi-variate linear regression, 2nd-order polynomial regression and may slightly outperform GPR for Build Run 1. Supplementary Figures S11 and S12 confirm similar observations for the cases of ten-fold cross-validation and no cross-validation, respectively. These observations are consistent with the 'Occam factor' in Eq. (21) [67]. As noted above, the 'Occam factor' can be viewed as a penalty term that penalizes complex models, which may yield a very good match to a training data set, but that may not yield that good predictions when compared to a testing data set (that may not the generalize that well to the testing data set, due to over-fitting), in favor of simpler models, that may yield not quite as good fit to the training data set, but which may generalize better to the testing data set [67].

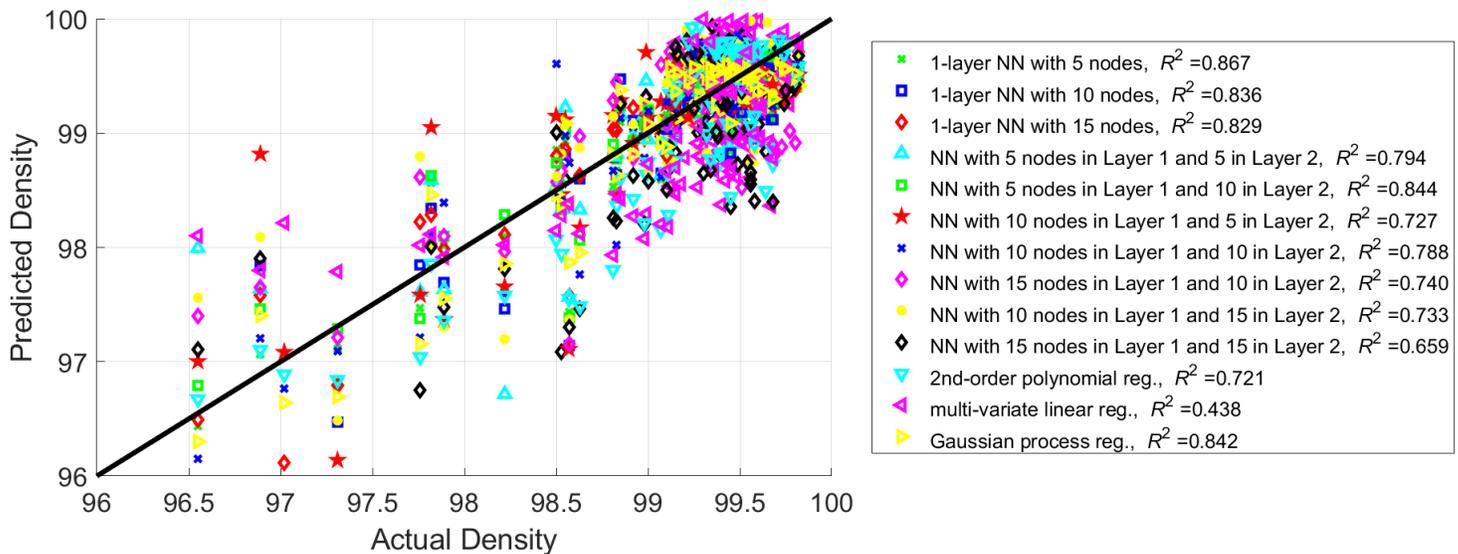

**Figure 18**: Prediction accuracy of the neural network solutions compared to that of the multi-variate linear regression, 2$^{nd}$-order polynomial regression, and Gaussian process regression for the test set (Inconel 718, Build Run 1, five-fold cross-validation, boundary conditions excluded in the neural network solutions). For similar results, obtained for the cases of no cross-validation and ten-fold cross-validation, refer to Supplementary Figures S11 and S12.

Note again that the NN approach offers significant resemblance with GPR. Both in the case of the NNs applied to AM and in the case of GPR-based GAP potentials for MD or Monte Carlo simulations, one seeks to reconstruct a multi-dimensional function by interpolating (fitting) a smooth surface through a grid of reference points. In the case of AM, the grid of reference points may consist of HTEM measured density or hardness values, per Figure 1. But in the case of the MD, the grid of reference points is usually obtained from high-precision DFT calculations. Given the extent of the resemblance, the Gaussian-process regression may be considered a similar (a related) method, more so than an alternative, to the NN approach for the multi-dimensional functional reconstruction. Given this inherent resemblance, it does not come as a surprise that the Gaussian process regression yields similar performance as, or slightly underperforms, the 1-layer NN's with 5 or 10 nodes, in terms of the $R^2$, according to Figure 18, Supplementary Figures S11 and S12, Table 2 and Supplementary Tables S3 – S6. As suggested by Maitra et al. [49], there certainly may exist other applications, not involving fitting through a grid of reference points, but instead noisy and random input data set, such as stream-water-temperature prediction [83] or prediction of $CO_2$ emission in the agriculture section of Iran [84], where Gaussian noises present in a GPR model can be accurately quantified, and where properly tuned GPR may outperform NNs. As noted above, the observation here regarding the preference to simple, single-layer NNs is specific to the data set being analyzed, and its relatively simple (flat) functional behavior. In the case of higher dimensions (the optimization of a data set obtained from a larger number of AM process input parameters) or other alloys, the same conclusion may or may not be fully held, depending on the extent and nature of the data analyzed.

According to Figure 11, the variable signal component is relatively small compared to the constant background, in the case of Build Run 2. Therefore, relatively small noise in the HTEM measurements of the relative density, in case of a testing set generated from Build Run 2, can result in test points exceeding the relative flat signal blanket, resulting in a hot spot. In the case of a training set generated from Build Run 2, large models may attempt to fit to noise in the training set, resulting in relatively poor ability to generalize to new data, again consistent with the 'Occam factor' [67]. For additional information, refer to Figures S3 – S9 in the Supplementary Manuscript.

## 5. Conclusions

In this paper, the authors presented a generic NN (ML) framework for the AM parameter optimization, a framework for the multi-dimensional functional construction, one summarized in the Figure 2, and one tailored towards HTEM. The authors showed how the framework could be applied to the optimization of the most essential (Tier I) processing parameters for Inconel 718 superalloy, namely laser power and scan speed.

Figures 8 - 9 illustrated how the NN framework could give rise to multi-dimensional mappings from the AM input parameters to the relative density, which characterized the absence of defects (voids or porosity), or to the material hardness, which represented a mechanical property. Kolmogorov's theorem and the Universal Approximation Theorem provided the theoretical basis for Tier 1 of the AM parameter optimization framework, i.e., the multi-dimensional functional construction, presented. The authors showed how to optimize the multi-dimensional functional construction such as obtaining the highest quality of fit. The authors introduced the objective criterion for qualifying the fit of a reconstructed function and put the model qualification in context with the Bayesian model selection. The authors also demonstrated how the AM parameter optimization could be extended to higher dimensions, by combining the NN prediction of process windows (hot spots) from the lower dimensions with practical insight. In addition, the authors introduced experimentally verified processing windows for Inconel 718 (in Figure 12 and Figure 16). In terms of the coefficient of determination, $R^2$, for the relative density, and consistent with the 'Occam factor' in Eq. (21) [67], a relatively simple single-layer NN with 5 or 10 nodes tends to outperform multi-variate linear regression, $2^{nd}$-order polynomial regression, and even slightly outperform GPR (i.e., the default configuration), despite inherent structural similarity of GPR with the NN-approach, for the primary Inconel 718 HTEM data set studied, both with and without cross-

validation. The good performance of the simple NNs observed (the comparatively high $R^2$ values), i.e., the good ability to generalize, as well as the propensity of large NNs to overfitting, can both be explained in terms of the 'Occam factor' [see Eq. (21)] [67]. In terms of future research, the authors advocate for looking at high throughput in context with the associated measurement accuracy and with what one is seeking to measure or detect (but not in isolation). The novelty of the research work entails the versatile and scalable NN framework presented, suitable for use in conjunction with HTEM, for the AM parameter optimization of superalloys. The impact entails broad applications of technology within AM and beyond.

## 6. Data Availability

The data presented in this paper can be requested by contacting the corresponding author (baldur@imagars.com, baldur.steingrimsson@oregonstate.edu, baldur.steingrimsson@oit.edu).

## 7. Acknowledgements

XF and PKL very much appreciates the support of (1) the U.S. Army Research Office Project (W911NF-13-1-0438 and W911NF-19-2-0049) with the program managers, Drs. J. C. Marx, A. D. Brown, M. P. Bakas, S. N. Mathaudhu, and D. M. Stepp, (2) the National Science Foundation (DMR-1611180, 1809640, and 2226508) with the program directors, Drs. J. Madison, J. Yang, G. Shiflet, and D. Farkas, (3) the Department of Energy (DOE DE-EE0011185) with the program director of J. R. Terneus, and (4) the Air Force Office of Scientific Research (AF AFOSR-FA9550-23-1-0503) with the program manager of Dr. D. P. Cole. XF and PKL also express gratitude for the support from the Bunch Fellowship. In addition, XF and PKL would like to acknowledge funding from the State of Tennessee and Tennessee Higher Education Commission (THEC) through their support of the Center for Materials Processing (CMP) with Prof. C. Rawn and Prof.

P. Rack as directors. BS very acknowledges support from the National Science Foundation (IIP-1632408 and IIP-1447395), with the program directors, Drs. R. Mehta and G. Larsen, from the U.S. Navy (N6833521C0420), with Drs. D. Shifler and J. Wolk as the program managers, and from the U.S. Air Force (FA864921P0754), with J. Evans as the program manager.

The authors also want to thank Dr. E. Huang for sharing the content related to an alternative solution approach aimed at strengthening the overall framework for the AM parameter optimization. Dr. J. Wang is thanked for offering overall comments and assessment of the manuscript package, just prior to submission. Dr. M. Kisialiou is thanked for offering guidance throughout the development of the ML (NN) portion. Dr. J. Shi is thanked for providing access to the as-built SLM-ed Ti-6Al-4V density data, for reference.

# References

[1] W. Frazier, "Metal Additive Manufacturing: A Review," *J. Mater. Eng. Perform.,* vol. 23, p. 1917–1928, 2014.

[2] S. Chen, Y. Tong, and P.K. Liaw, "Additive Manufacturing of High-Entropy Alloys: A Review," *Entropy,* vol. 20, no. 12, p. 937, 2018.

[3] A.K. Agrawal, G.M. de Bellefon, and D. Thoma, "High-throughput experimentation for microstructural design in additively manufactured 316L stainless steel," *Materials Science & Engineering A,* vol. 793, p. 139841, 2020.

[4] R. Jiang, R. Kleer, and F.T. Piller, "Predicting the future of additive manufacturing: a Delphi study on economic and societal implications of 3D printing for 2030," *Technological Forecasting and Social Change,* vol. 117, no. https://doi.org/10.1016/j.techfore.2017.01.006, pp. 84 - 97, 2017.

[5] T. Wohlers, Wohlers report 2019 : 3D printing and additive manufacturing state of the industry, Wohlers Associates, 2019.

[6] C.Y. Yap, C.K. Chua, Z.L. Dong, Z.H. Liu, D.Q. Zhang, L.E. Loh and S.L. Sing, "Review of selective laser melting: materials and applications," *Applied Physics Review,* vol. 2, p. 041101, 2015.

[7] B. Rankouhi, A.K. Agrawal, F.E. Pfefferkorn, and D.J. Thoma, "A dimensionless number for predicting universal processing parameter," *Manufacturing Letters,* vol. 27, pp. 13 - 17, 2021.

[8] D. Zhang, S. Sun, D. Qiu, M.A. Gibson, M.S. Dargusch, M. Brandt, M. Qian, and M. Easton, "Metal alloys for fusion-based additive manufacturing," *Advanced Engineering Materials,* vol. 20, pp. 1-20, 2018.

[9] J.H. Martin, B.D. Yahata, J.M. Hundley, J.A. Mayer, T.A. Schaedler, and T.M. Pollock, "3D Printing of High-Strength Aluminium Alloys," *Nature,* vol. 549, p. 365–369, 2017.

[10] S. Guan, J. Ren, S. Mooraj, Y. Liu, S. Feng, S. Zhang, J. Liu, X. Fan, P.K. Liaw, and W. Chen, "Additive Manufacturing of High-Entropy Alloys: Microstructural Metastability and Mechanical Behavior," *Journal of Phase Equilibria and Diffusion,* vol. 42, pp. 748 - 771, 2021.

[11] C. Zhao, N.D. Parab, X. Li, K. Fezzaa, W.Tan, A.D. Rollett, and T. Sun, "Critical instability at moving keyhole tip generates porosity in laser melting," *Science,* vol. 370, no. 6520, pp. 1080 - 1086, 2020.

[12] C. Zhao, K. Fezzaa, R.W. Cunningham, H. Wen, F. De Carlo, L. Chen, A.D. Rollett, and T. Sun, "Real-time monitoring of laser powder bed fusion process using high-speed X-ray imaging and diffraction," *Scientific Reports,* vol. 7, no. 3602, 2017.

[13] R. Cunningham, C. Zhao, N. Parab, C. Kantzos, J. Pauza, K. Fezzaa, T. Sun, and A.D. Rollett, "Keyhole threshold and morphology in laser melting revealed by ultrahigh-speed x-ray imaging," *Science,* vol. 363, no. 6429, pp. 849 - 852, 2019.

[14] B. Steingrimsson, X. Fan, A. Kulkarni, M.C. Gao, and P.K. Liaw, "Machine Learning and Data Analytics for Design and Manufacturing of High-Entropy Materials Exhibiting Mechanical or Fatigue Properties of Interest," in *Fundamental Studies in High-Entropy Materials, eds. Dr. James Brechtl and Dr. Peter K. Liaw*, https://arxiv.org/abs/2012.07583, Springer, 2021.

[15] A. Caggiano, J. Zhang, V. Alfieri, F. Caiazzo, R. Gao, and R. Teti, "Machine learning based image processing for on-line defect recognition in additive manufacturing," *CIRP Annals,* vol. 68, pp. 451-454, 2019.

[16] C.Y. Lin, T. Wirtz, F. LaMarca, and S.J. Hollister, "Structural and mechanical evaluations of a topology optimized titanium interbody fusion cage fabricated by selective laser melting process," *Journal of Biomedical Materials Research. Part A,* vol. 83A, no. 2, pp. 272-279, 2007.

[17] A. Bandyopadhyay and K.D. Traxel, "Invited review article: Metal-additive manufacturing—Modeling strategies for application-optimized designs," *Additive Manufacturing,* vol. 22, pp. 758-774, 2018.

[18] I.E. Anderson, E.M.H. White, and R. Dehoff, "Feedstock powder processing research needs for additive manufacturing development," *Curr. Opin. Solid State Mater. Sci.,* vol. 22, pp. 8 - 15, 2018.

[19] B. Zhang, R. Seede, L. Xue, K.C. Atli, C. Zhang, A. Whitt, I. Karaman, R. Arroyave, and A. Elwany, "An efficient framework for printability assessment in Laser Powder Bed Fusion metal additive manufacturing," *Journal of Additive Manufacturing,* vol. 46, no. 102018, 2021.

[20] T. DebRoy, T. Mukherjee, J.O. Milewski, J.W. Elmer, B. Ribic, J.J. Blecher, and W. Zhang, "Scientific, technological and economic issues in metal printing and their solutions," *Nature Materials,* vol. 18, no. https://doi.org/10.1038/s41563-019-0408-2, p. 1026–1032, 2019.

[21] H.D. Carlton, A. Haboub, G.F. Gallegos, D.Y. Parkinson, and A.A. MacDowell, "Damage evolution and failure mechanisms in additively manufactured stainless steel," *Mater Sci Eng A,* vol. 651, no. https://doi.org/10.1016/j.msea.2015.10.073, p. 406–14, 2016.

[22] T. DebRoy, H.L. Wei, J.S. Zuback, T. Mukherjee, J.W. Elmer, J.O. Milewski, A.M. Beese, A. Wilson-Heid, A. De, and W. Zhang, "Additive manufacturing of metallic components – process, structure and properties," *Progress in Materials Science,* vol. 92, no. https://doi.org/10.1016/j.pmatsci.2017.10.001, p. 112–224, 2018.

[23] R. Li, J. Liu, Y. Shi, L. Wang, and W. Jiang, "Balling behavior of stainless steel and nickel powder during selective laser melting process," *Int. J. Adv. Manuf. Technol.,* vol. 59, p. 1025–1035, 2012.

[24] A.A. Martin, N.P. Calta, S.A. Khairallah, J. Wang, P.J. Depond, A.Y. Fong, V. Thampy, G.M. Guss, A.M. Kiss, K.H. Stone, C.J. Tassone, J. Nelson Weker, M.F. Toney, T. van Buuren, and M.J. Matthews, "Dynamics of pore formation during laser powder bed fusion additive manufacturing," *Nat. Commun.,* vol. 10, p. 1987, 2019.

[25] S.A. Khairallah, A.T. Anderson, A.M. Rubenchik, and W.E. King, "Laser powder-bed fusion additive manufacturing: physics of complex melt flow and formation mechanisms of pores, spatter, and denudation zones," *Addit. Manuf. Handb. Prod. Dev. Def. Ind.,* vol. 108, p. 613–628, 2017.

[26] J.C. Tseng, W.C. Huang, W .Chang, A. Jeromin, T.F. Keller, J. Shen, A.C. Chuang, C.C. Wang, B.H. Lin, L. Amalia, N.T. Tsou, S.J. Shih, and E.W. Huang, "Deformations of Ti-6Al-4V additive-manufacturing-induced isotropic and anisotropic columnar structures: Insitu measurements and underlying mechanisms," *Additive Manufacturing,* vol. 35, October, 2020.

[27] N.M. Heckman, T.A. Ivanoff, A.M. Roach, B.H. Jared, D.J. Tung, H.J. Brown-Shaklee, T. Huber, D.J. Saiz, J.R. Koepke, J.M. Rodelas, J.D. Madison, B.C. Salzbrenner, L.P. Swiler, R.E. Jones, and B.L. Boyce, "Automated high-throughput tensile testing reveals stochastic process parameter sensitivity," *Mater. Sci. Eng.,* vol. 772, p. 138632, 2020.

[28] B.C. Salzbrenner, J.M. Rodelas, J.D. Madison, B.H. Jared, L.P. Swiler, Y.L. Shen, and B. L. Boyce, "High-throughput stochastic tensile performance of additively manufactured stainless steel," *J. Mater. Process. Technol.,* vol. 241, pp. 1 - 12, 2017.

[29] H. Chae, E.W. Huang, W. Woo, S.H. Kang, J. Jain, K. An, and S.Y. Lee, "Unravelling thermal history during additive manufacturing of martensitic stainless steel," *Journal of Alloys and Compounds,* vol. 857, March 15, 2021.

[30] R. Jafari-Marandi, M. Khanzadeh, W. Tian, B. Smith, and L. Bian, "From in-situ monitoring toward high-throughput process control: cost-driven decision-making framework for laser-based additive manufacturing," *J. Manuf. Syst.,* vol. 51, p. 29–41, 2019.

[31] F. Ren, L. Ward, T. Williams, K.J. Laws, C. Wolverton, J. Hattrick-Simpers, and A. Mehta, "Accelerated discovery of metallic glasses through iteration of machine learning and high-throughput experiments," *Sci. Adv.,* vol. 4, p. 1566, 2018.

[32] D.R. Clymer, J. Cagan, and J. Beuth, "Power–velocity process design charts for powder bed additive manufacturing," *Journal of Mechanical Design,* vol. 139, p. 100907, 2017.

[33] M. Thomas, G.J. Baxter, and I. Todd, "Normalised model-based processing diagrams for additive layer manufacture of engineering alloys," *Acta Materialia,* vol. 108, pp. 26 - 35, 2016.

[34] L. Johnson, M. Mahmoudi, B. Zhang, R. Seede, J.T. Maier, H.J. Maier, I. Karaman, and A. Elwany, "Assessing printability maps in additive manufacturing of metal alloys," *Acta Materialia,* vol. 176, pp. 1 - 25, 2019.

[35] M. van Elsen, F. Al-Bender, and J.P. Kruth, "Application of dimensional analysis to selective laser melting," *Rapid Prototyp J,* vol. 14, no. 1, pp. 15-22, 2008.

[36] Z. Wang and M. Liu, "Dimensionless analysis on selective laser melting to predict porosity and track morphology," *J Mater Process Technol,* vol. 273, no. https://doi.org/10.1016/j.jmatprotec.2019.05.019, p. 116238, 2019.

[37] T. Mukherjee, V. Manvatkar, A. De, and T. DebRoy, "Dimensionless numbers in additive manufacturing," *J Appl Phys,* vol. 121, no. 6, p. 064904, 2017.

[38] A. Raghavan, H.L. Wei, T.A. Palmer, and T. DebRoy, "Heat transfer and fluid flow in additive manufacturing," *J Laser Appl,* vol. 25, no. 5, p. 052006, 2013.

[39] T. Mukherjee and T. DebRoy, "Mitigation of lack of fusion defects in powder bed fusion additive manufacturing," *J Manuf Processes,* vol. 36, no. https://doi.org/10.1016/j.jmapro.2018.10.028, p. 442–9, 2018.

[40] A.M. Rubenchik, W.E. King, and S.S. Wu, "Scaling laws for the additive manufacturing," *J. Mater Process Technol,* vol. 257, no. https://doi.org/10.1016/j.jmatprotec.2018.02.034, p. 234–43, 2018.

[41] W. Fu, C. Haberland, E.V. Klapdor, D. Rule, and S. Piegert, "Streamlined Frameworks for Advancing Metal Based Additive Manufacturing Technologies in Gas Turbine Industry," in *Proceedings of the 1st Global Power and Propulsion Forum (GPPF 2017)*, Zurich, Switzerland, Jan. 16-18, 2017.

[42] N. Johnson, P. Vulimiri, A. To, X. Zhang, C. Brice, B. Kappes, and A. Stebner, "Invited review: Machine learning for materials developments in metals additive manufacturing," *Journal of Additive Manufacturing,* vol. 36, p. 101641, 2020.

[43] P. Akbari, F. Ogoke, N.Y. Kao, K. Meidani, C.Y. Yeh, W. Lee, and A.B. Farimani, "MeltpoolNet: Melt pool characteristic prediction in Metal Additive Manufacturing using machine learning," *Additive Manufacturing,* vol. 55, p. 102817, 2022.

[44] L. Scime and J. Beuth, "Using machine learning to identify in-situ melt pool signatures indicative of flaw formation in a laser powder bed fusion additive manufacturing process," *Journal of Additive Manufacturing,* vol. 25, pp. 151-165, 2019.

[45] F. Ogoke and A.B. Farimani, "Thermal control of laser powder bed fusion using deep reinforcement learning," *Journal of Additive Manufacturing,* vol. 46, p. 102033, 2021.

[46] S. Lee, J. Peng, D. Shin, and Y.S. Choi, "Data analytics approach for melt-pool geometries in metal additive manufacturing," *Sci. Technol. Adv. Mater.,* vol. 20, no. 1, p. 972–978, 2019.

[47] T.L. Hong, P.C. Lin, J.-Z. Chen, T.D.Q. Pham, and X.V. Tran, "Data-driven models for predictions of geometric characteristics of bead fabricated by selective laser melting," *Journal of Intelligent Manufacturing,* vol. 34, p. 1241–1257, 2023.

[48] W. Mycroft, M. Katzman, S.T.-Williams, and E.H. Nava, "A data-driven approach for predicting printability inmetal additive manufacturing processes," *Journal of Intelligent Manufacturing,* vol. 31, p. 1769–1781, 2020.

[49] V. Maitra, J. Shi, and C. Lu, "Robust prediction and validation of as-built density of Ti-6Al-4V parts manufactured via selective laser melting using a machine learning approach," *Journal of Manufacturing Processes,* vol. 78, pp. 183 - 201, 2022.

[50] M. Zou, W.G. Jiang, Q.H. Qin, Y.C. Liu, and M.L. Li, "Optimized XGBoost Model with Small Dataset for Predicting Relative Density of Ti-6Al-4V Parts Manufactured by Selective Laser Melting," *Materials,* vol. 15, p. 5298, 2022.

[51] G. Tapia, S. Khairallah, M. Matthews, W.E. King, and A. Elwany, "Gaussian process based surrogate modeling frame work for process planning in laser powder bed fusion additive manufacturing of 316L stainless steel," *International Journal of Advanced Manufacturing Technology,* vol. 94, no. 9-12, p. 3591–603, 2018.

[52] B. Yuan, G.M. Guss, A.C. Wilson, S.P. Hau-Riege, P.J. DePond, S. McMains, M.J. Matthews, and B. Giera, "Machine-learning-based monitoring of laser powder bed fusion," *Sci. Technol. Adv. Mater.,* vol. 3, no. 12, p. 1800136, 2018.

[53] A. Gaikwad, B. Giera, G.M. Guss, J.-B. Forien, M.J. Matthews, and P. Rao, "Heterogeneous sensing and scientific machine learning for quality assurance in laser powder bed fusion - A single-track study," *Additive Manufacturing,* vol. 36, p. 101659, 2020.

[54] G.O. Barrionuevo, P.M. Sequeira-Almeida, S. Ríos, J.A.R. Grez, and S.W. Williams, "A machine learning approach for the prediction of melting efficiency in wire arc additive manufacturing," *The International Journal of Advanced Manufacturing Technology,* vol. 120, p. 3123–3133, 2022.

[55] Z. Ren, L. Gao, S.J. Clark, K. Fezzaa, P. Shevchenko, A. Choi, W. Everhart, A.D. Rollett, L. Chen, and T. Sun, "Machine learning-aided real-time detection of keyhole pore generation in laser powder bed fusion," *Science,* vol. 379, pp. 89-94, 2023.

[56] S. Li, G. Wang, Y. Di, L. Wang, H. Wang, and Q. Zhou, "A physics-informed neural network framework to predict 3D temperature field without labeled data in process of laser metal deposition," *Engineering Applications of Artificial Intelligence,* vol. 120, p. 105908, 2023.

[57] J. Horňas, J. Běhal, P. Homola, R. Doubrava, M. Holzleitner, and S. Senck, "A machine learning based approach with an augmented dataset for fatigue life prediction of additively manufactured Ti-6Al-4V samples," *Engineering Fracture Mechanics,* vol. 293, p. 109709, 2023.

[58] B. Steingrimsson, X. Fan, A. Kulkarni, D. Kim, and P.K. Liaw, "Machine Learning to Accelerate Alloy Design". US Patent 11,915,105, https://patents.google.com/patent/US11915105B2/, 27 February 2024.

[59] A. Kulkarni, *Personal communications about Siemens' experience with additive manufacturing,* 2022.

[60] T.G. Spears and S.A. Gold, "In-Process Sensing in Selective Laser Melting (SLM) Additive Manufacturing," *Integrating Materials and Manufacturing Innovation,* vol. 5, no. 1, pp. 16-40, 2016.

[61] B. Steingrimsson, R. Jones, M. Kisialiou, A. Kulkarni, and K. Yi, "Decisions with Big Data". Utility Patent Application Patent 16/182,389, 6 November 2018.

[62] A. Kolmogorov, "On the representation of continuous functions of several variables by superpositions of continuous functions of a smaller number of variables," *Proceedings of the USSR Academy of Sciences,* vol. 108, pp. 179 - 182, 1956.

[63] V. Arnold, "On functions of three variables," *Proceedings of the USSR Academy of Sciences,* vol. 114, no. English translation: Amer. Math. Soc. Transl., 28 (1663), pp. 51 - 54, pp. 679 - 681, 1957.

[64] M. Hassoun, Fundamentals of Artificial Neural Networks, MIT Press, 1995.

[65] I. Guyon, "A scaling law for the validation-set training-set size ratio," AT&T Bell Laboratories, http://citeseerx.ist.psu.edu/viewdoc/download?doi=10.1.1.33.1337&rep=rep1&type=pdf, 1997.

[66] scikit-learn, "Cross-validation: evaluating estimator performance," 2024. [Online]. Available: https://scikit-learn.org/stable/modules/cross_validation.html. [Accessed 7 October 2024].

[67] D. MacKay, Bayesian Methods for Neural Networks: Theory and Applications, Cavendish Laboratory: doi=10.1.1.47.6409, 1995.

[68] Matlab Help Center, "feedforwardnet - Generate feedforward neural network," [Online]. Available: https://www.mathworks.com/help/deeplearning/ref/feedforwardnet.html. [Accessed 13 July 2021].

[69] T. Poggio and F. Girosi, "A Sparse Representation for Function Approximation," *Neural Computation,* vol. 10, no. 6, p. 1445–1454, 1998.

[70] B. Steingrimsson, X. Fan, X. Yang, M.C. Gao, Y. Zhang, and P. K. Liaw, "Predicting Temperature-Dependent Ultimate Strengths of Body-Centered Cubic (BCC) High-Entropy Alloys," *npj Computational Materials,* vol. 7, no. 152, pp. 1-10, September 24, 2021.

[71] R.E. Walpole, R.H. Myers, S.L. Myers, and K. Ye, Probability and Statistics for Engineers and Scientists, 9th ed., Boston, MA: Prentice Hall, 2012.

[72] V.L. Deringer, A.P. Bartok, N. Bernstein, D.M. Wilkins, M. Ceriotti, and G. Csanyi, "Gaussian Process Regression for Materials and Molecules," *Chemical Reviews,* vol. 121, no. 16, pp. 10073-10141, 2021.

[73] J. Byggmästar, A. Hamedani, K. Nordlund, and F. Djurabekova, "Machine Learning Interatomic Potential for Radiation Damage and Defects in Tungsten," *Physical Review B,* vol. 100, no. 14, p. 144105, 2019.

[74] J. Byggmästar, K. Nordlund, and F. Djurabekova, "Modeling refractory high-entropy alloys with efficient machine-learned interatomic potentials: Defects and segregation," *Physical Review B,* vol. 104, no. 10, p. 104101, 2021.

[75] M. Koskenniemi, J. Byggmästar, K. Nordlund, and F. Djurabekova, "Efficient atomistic simulations of radiation damage in W and W-Mo using machine-learning potentials," *Journal of Nuclear Materials,* vol. 577, p. 154325, 2023.

[76] J.Q. Shi and T. Choi, Gaussian Process Regression Analysis for Functional Data, Boca Raton, FL 33487-2742: CRC Press, Taylor & Francis Group, 2011.

[77] E. Schulz, M. Speekenbrink, and A. Krause, "A tutorial on Gaussian process regression: Modelling, exploring, and exploiting functions," *Journal of Mathematical Psychology,* vol. 85, pp. 1-16, 2018.

[78] C.E. Rasmussen and C.K.I. Williams, "Gaussian Processes for Machine Learning," in *Adaptive Computation and Machine Learning*, The MIT Press, 2006.

[79] Matlab Help Center, "cvpartition - Partition data for cross-validation," 2024. [Online]. Available: https://www.mathworks.com/help/stats/cvpartition.html. [Accessed 7 October 2024].

[80] K. P. Murphy, "Bayesian model selection/ Frequentist parameter estimation," University of British Columbia, Lecture notes for CS340 (Machine learning), Fall 2007.

[81] G. Golub and C.F. Van Loan, Matrix Computations - Third Edition, Baltimore: The Johns Hopkins University Press, 1996.

[82] A. Cecen, "Multivariate Polynomial Regression," GitHub, 2024. [Online]. Available: https://github.com/ahmetcecen/MultiPolyRegress-MatlabCentral. [Accessed 26 May 2024].

[83] R. Grbic, D. Kurtagic, and D. Sliskovic, "Stream water temperature prediction based on Gaussian process regression," *Expert Systems with Applications,* vol. 40, p. 7407–14, 2013.

[84] E. Shabani, B. Hayati, E. Pishbahar, M.AP. Ghorbani, and M. Ghahremanzadeh, "A novel approach to predict $CO_2$ emission in the agriculture sector of Iran based on inclusive multiple model," *Journal of Cleaner Production,* vol. 279, p. 123708, 2021.

## 8. Author Contributions

B.S. devised the framework for multi-dimensional synthesis, created the associated Matlab® scripts, crafted the theoretical ML analysis, characterized the quality of fit, and optimized the structures. A.A. conducted the SLM fabrication, and HTEM characterization and provided to B.S. A.A. also ran the Matlab® scripts, prepared by B.S. and returned the outputs. X.F. ran selected Matlab® scripts and reviewed the analysis algorithms for accuracy. A.K. contributed important insights on aspects related to practicality and impact. D.T. proposed the idea of applying machine learning for extending the 2D analysis to higher dimensions. P.K.L. played an instrumental role in formulating the scope of the manuscript. All authors contributed to the manuscript, including editing and proof reading, but with B.S. primarily responsible for assembly and revisions of the manuscript.

## 9. Competing Interests

The authors declare no competing interests.

## Supplementary Tables

**Table S1**: Summary of key process parameters of SLM and SLS [1, 2]. Tier I is assumed to consist only of the VED.

| No. | Parameter | Description | Tier |
|---|---|---|---|
| Laser and scanning parameters | | | |
| 1 | Average power | Measure of total energy output of a laser | Tier III |
| 2 | Mode | Continuous wave or pulsed | Tier III |
| 3 | Peak power | Maximum powder in a laser purse | Tier II |
| 4 | Pulse width | Length of a laser pulse when operating in pulsed mode | Tier II |
| 5 | Frequency | Pulses per unit time | Tier III |
| 6 | Wavelength | Distance between crests in laser electromagnetic waves | Tier III |
| 7 | Polarization | Orientation of electromagnetic waves in laser beam | Tier III |
| 8 | Beam quality | Related to intensity profile and used to predict how well beam can be focused and determine min. theoretical spot size | Tier III |
| 9 | Intensity profile | Determines how much energy added at a specific location | Tier III |
| 10 | Spot size | Length and width of elliptical spot | Tier II |
| 11 | Scan velocity | Velocity at which laser moves across build surface | Tier II |
| 12 | Scan spacing | Distance between neighboring laser passes | Tier II |
| 13 | Scan strategy | Pattern in which the laser is scanned across the build surface (hatches, zig-zags, spirals, etc.) and associated parameters | Tier III |
| Powder material properties | | | |
| 14 | Bulk density | Material density, limits maximum density of final component | Tier III |
| 15 | Thermal conductivity | Measure of material's ability to conduct heat | Tier III |
| 16 | Heat capacity | Measure of energy required to raise temp. of material | Tier III |
| 17 | Latent heat fusion | Energy required for solid-liquid and liquid-solid phase change | Tier III |
| 18 | Melting temperature | Temperature at which material melts | Tier III |
| 19 | Boiling temperature | Temperature at which material vaporizes | Tier III |
| 20 | Melt pool viscosity | Measure of resistance of melt to flow | Tier III |
| 21 | Coef. Of thermal expansion | Measure of volume change of material on heating or cooling | Tier III |
| 22 | Surface free energy | Free energy required to form new unit area on heating or cooling | Tier III |
| 23 | Vapor pressure | Measure of the tendency of material to vaporize | Tier III |
| 24 | Heat (enthalpy) of reaction | Energy associated with a chemical reaction of material | Tier III |
| 25 | Material absorptivity | Measure of laser energy absorbed by the material | Tier III |
| 26 | Diffusivity | Important for solid state sintering, not critical for melting | Tier III |
| 27 | Solubility | Solubility of solid material in liquid melt | Tier III |
| 28 | Particle morphology | Measures of shape of individual particles and their distribution | Tier III |
| 29 | Surface roughness | Arithmetic mean of the surface profile | Tier III |
| 30 | Particle size distribution | Distribution of particle sizes, usually diameter, in a powder sample | Tier III |
| 31 | Pollution | Ill-defined factor describing change in properties of powder due to reuse | Tier III |
| Powder bed and recoat parameters | | | |
| 32 | Density | Measure of packing density of powder particles, influence heat balance | Tier III |
| 33 | Thermal conductivity | Measure of powder bed's ability to conduct heat | Tier III |
| 34 | Heat capacity | Measure of energy required to raise temp. of powder bed | Tier III |
| 35 | Absorpivity | Measure of laser energy absorbed | Tier III |
| 36 | Emissivity | Ratio of energy radiated to that of black body | Tier III |
| 37 | Deposition system param's | Recoater velocity, pressure, recoater type, dosing | Tier III |
| 38 | Layer thickness | Height of a single powder layer | Tier II |
| 39 | Powder bed temp. | Bulk temperature of the powder bed | Tier III |
| Build environment parameters | | | |
| 40 | Shield gas | Usually Ar or $N_2$, but may also be He or something else | Tier IV |
| 41 | Oxygen level | Probably most important environment parameter; oxygen can lead to oxide formation in metal, change wettability, energy required for welding, etc. | Tier IV |
| 42 | Shield gas mol.weight | Influences heat balance, diffusivity into and out of part | Tier IV |
| 43 | Shield gas viscosity | May influence free surface activity of melt pool, convective heat balance | Tier IV |
| 44 | Thermal conductivity | Term in heat balance | Tier IV |
| 45 | Heat capacity of gas | Term in heat balance | Tier IV |
| 46 | Pressure | Influences vaporization of metal as well as oxygen content | Tier IV |
| 47 | Gas flow velocity | Influences convective cooling, removal of condensate | Tier IV |
| 48 | Convective heat transfer coef. | Convective cooling of just melted part by gas flowing over the surface | Tier IV |
| 49 | Ambient temperature | Appears in heat balance, may impact powder preheat and residual stress | Tier IV |
| 50 | Surface free energy | Between liquid and surround gas influence melt pool shape | Tier IV |

**Table S2**: Key Technical Terms and Concepts Related to Gaussian process regression [3].

| Term | Meaning |
| --- | --- |
| Covariance | The covariance is a measure of the strength of statistical correlation between two data values, $y(\mathbf{x})$ and $y(\mathbf{x'})$, usually expressed as a function of the distance between $\mathbf{x}$ and $\mathbf{x'}$. The covariance evaluates to zero for uncorrelated data. |
| Descriptor | In the context of regression, descriptors, sometimes referred to as "features", encode independent variables into a vector, $\mathbf{x}$, on which the modeled variable, $y$, depends. |
| Hyperparameter | A hyperparameter is a global parameter of an ML model that controls the behavior of the fit. The hyperparameters can be distinct from the potentially very large number of "free parameters" that are determined, when the model is fitted to the data. The hyperparameters are usually estimated from experience or iteratively optimized using data. |
| Kernel | A kernel captures a similarity measure between two data points, x and x'. This similarity measure is usually denoted as $k(\mathbf{x},\mathbf{x'})$. The kernels can be used to construct models of covariance. |
| Overfitting | Overfitting refers to a fit that yields high accuracy for the input (training) data, but produces much larger errors elsewhere (for new testing data). This may be cause by absence of proper regularization. |
| Prior | A prior refers to a formal quantification, as a probability distribution, of our initial knowledge or assumption about the behavior of a model, before the model gets fitted to any data. |
| Regularity | A function is considered to be regular, if all of its derivatives are confined to moderate bounds. The term "regularity" may be considered loosely interchangeable with the term "degree of smoothness". |
| Regularization | Regularization refers to techniques used to enforce the regularity of fitted functions. In the case of Gaussian process regression, this is achieved by penalizing solutions which have large values for basis coefficients. The extent of regularization may be viewed as corresponding to the "expected error" of the fit. |
| Sparsity | A sparse model, in the context of Gaussian process regression, is one where there are far fewer kernel basis functions than input data points. Note that the locations of these basis functions, which we also refer to as the representative set, may not coincide with the input data locations. |
| Underfitting | Underfitting refers to cases, where the accuracy achievable by a better choice of hyperparameters is not reached, neither for the training nor the testing data. |

**Table S3:** Characterization of the quality of fit of the 3D surfaces for the density and hardness synthesized relative to the original input data shown in Figure 6. Here, no boundary conditions have been applied. The quality of fit was estimated exclusively from testing portion of Build Run 1 ($N_{test,BR1}$ = 35 samples). $P(M|D)$ represents the posterior probability of the model listed given the data, whereas $\sigma_{test}$ and $\sigma_{meas}$ appear in Eq. (14). The specific values for $\sigma_{meas,1}$ and $\sigma_{meas,2}$ used are listed in Eqs. (47) and (48) in the main manuscript. $N_{param}$ represents the number of parameters in the model involved, but $N_{train,BR1}$ the number of samples in the training set from Build Run 1. In this case, $N_{train,BR1}$ = 140 samples. In case of the neural networks (NNs) and using the syntax from Eq. (24) in the main manuscript, $N_{param}$ was estimated in Matlab® as **length (getwb(net))**. But for the Gaussian process regression, $N_{param}$ was estimated in Matlab® as **length (gaussianProcessModel.KernelInformation.KernelParameters)**. Similarly, in the case of the polynomial regression, the model size as estimated as **length (regPoly2.Coefficients)**. One run of ten-fold cross-validation was performed to obtain the results in the Table.

| Method | Density | | | | Hardness | | | | Size |
| --- | --- | --- | --- | --- | --- | --- | --- | --- | --- |
| | | | | No Boundary Conditions Included | | | | | |
| | $R^2$ | $P(M|D)$ | $\sigma_{test,1}$ [%] | $\sigma_{test,1} / \sigma_{meas,1}$ | $R^2$ | $P(M|D)$ | $\sigma_{test,2}$ [HV] | $\sigma_{test,2} / \sigma_{meas,2}$ | $N_{train,BR1} / N_{param,BR1}$ |
| 1-layer NN: 5 nodes | 0.910 | 0.1091 | 0.222 | 1.047 | 0.113 | 0.0863 | 2.721 | 1.916 | 6.67 |
| 1-layer NN: 10 nodes | 0.898 | 0.1060 | 0.281 | 1.326 | 0.122 | 0.1296 | 2.547 | 1.794 | 3.41 |
| 1-layer NN: 15 nodes | 0.837 | 0.1009 | 0.282 | 1.330 | 0.128 | 0.1109 | 2.615 | 1.841 | 1.54 |
| 2-layer NN: 5 + 5 nodes | 0.879 | 0.1024 | 0.306 | 1.443 | 0.181 | 0.1015 | 2.383 | 1.678 | 2.75 |
| 2-layer NN: 5 + 10 nodes | 0.883 | 0.1020 | 0.256 | 1.208 | 0.102 | 0.0924 | 2.604 | 1.834 | 1.62 |
| 2-layer NN: 10 + 5 nodes | 0.860 | 0.1024 | 0.271 | 1.278 | 0.125 | 0.0918 | 2.600 | 1.831 | 1.54 |
| 2-layer NN: 10 + 10 nodes | 0.840 | 0.0992 | 0.291 | 1.373 | 0.147 | 0.1035 | 2.529 | 1.781 | 0.93 |
| 2-layer NN: 10 + 15 nodes | 0.809 | 0.0955 | 0.308 | 1.453 | 0.053 | 0.0717 | 2.951 | 2.078 | 0.66 |
| 2-layer NN: 15 + 10 nodes | 0.858 | 0.0902 | 0.313 | 1.476 | 0.085 | 0.1014 | 2.826 | 1.990 | 0.65 |
| 2-layer NN: 15 + 15 nodes | 0.833 | 0.0923 | 1.444 | 6.812 | 0.117 | 0.1108 | 2.786 | 1.962 | 0.47 |
| 2nd-order polynomial reg. | 0.807 | 1.0 | 0.461 | 2.175 | 0.077 | 1.0 | 2.540 | 1.789 | 23.33 |
| 1st-order linear regression | 0.455 | 1.0 | 0.742 | 3.500 | 0.090 | 1.0 | 2.493 | 1.756 | 46.67 |
| Gaussian-process regression | 0.895 | 1.0 | 0.245 | 1.156 | 0.140 | 1.0 | 2.393 | 1.685 | 70 |

**Table S4:** Characterization of variations observed across different runs of the five-fold cross-validation. The variations reported are caused by randomized splitting into the training and testing data sets. For each run, the quantity reported is average of $R^2[i]$ averaged across the five folds (splits). Here, $R^2[i]$ represents the coefficient of determination for the relative density between the training data set and the testing data set for fold (split), $i$.

| Method | Coefficient of Determination, $R^2$, for the Relative Density — No Boundary Conditions Included ||||||||||| |
|---|---|---|---|---|---|---|---|---|---|---|---|
| | Run 1 | Run 2 | Run 3 | Run 4 | Run 5 | Run 6 | Run 7 | Run 8 | Run 9 | Run 10 | Avg. |
| 1-layer NN: 5 nodes | 0.948 | 0.831 | 0.923 | 0.951 | 0.863 | 0.768 | 0.773 | 0.919 | 0.825 | 0.897 | 0.870 |
| 1-layer NN: 10 nodes | 0.899 | 0.913 | 0.901 | 0.937 | 0.936 | 0.849 | 0.834 | 0.928 | 0.821 | 0.880 | 0.890 |
| 1-layer NN: 15 nodes | 0.886 | 0.771 | 0.909 | 0.911 | 0.944 | 0.832 | 0.672 | 0.925 | 0.862 | 0.773 | 0.849 |
| 2-layer NN: 5 + 5 nodes | 0.914 | 0.770 | 0.918 | 0.934 | 0.915 | 0.794 | 0.781 | 0.932 | 0.849 | 0.881 | 0.869 |
| 2-layer NN: 5 + 10 nodes | 0.932 | 0.888 | 0.884 | 0.912 | 0.938 | 0.824 | 0.779 | 0.917 | 0.823 | 0.881 | 0.878 |
| 2-layer NN: 10 + 5 nodes | 0.932 | 0.881 | 0.896 | 0.850 | 0.938 | 0.799 | 0.799 | 0.842 | 0.730 | 0.887 | 0.855 |
| 2-layer NN: 10 + 10 nodes | 0.917 | 0.787 | 0.854 | 0.904 | 0.881 | 0.805 | 0.742 | 0.905 | 0.775 | 0.871 | 0.844 |
| 2-layer NN: 10 + 15 nodes | 0.876 | 0.858 | 0.791 | 0.891 | 0.880 | 0.782 | 0.777 | 0.852 | 0.811 | 0.805 | 0.832 |
| 2-layer NN: 15 + 10 nodes | 0.913 | 0.873 | 0.852 | 0.891 | 0.755 | 0.768 | 0.747 | 0.835 | 0.758 | 0.847 | 0.824 |
| 2-layer NN: 15 + 15 nodes | 0.693 | 0.735 | 0.847 | 0.911 | 0.875 | 0.803 | 0.609 | 0.867 | 0.764 | 0.816 | 0.792 |
| 2nd-order polynomial reg. | 0.807 | 0.789 | 0.815 | 0.806 | 0.781 | 0.741 | 0.737 | 0.799 | 0.726 | 0.772 | 0.777 |
| 1st-order linear regression | 0.456 | 0.444 | 0.474 | 0.434 | 0.413 | 0.404 | 0.420 | 0.456 | 0.403 | 0.452 | 0.436 |
| Gaussian process regression | 0.928 | 0.916 | 0.921 | 0.943 | 0.943 | 0.828 | 0.823 | 0.916 | 0.849 | 0.882 | 0.895 |

**Table S5:** Characterization of variations observed across different runs of the ten-fold cross-validation. The variations reported are caused by randomized splitting into the training and testing data sets. For each run, the quantity reported is the average of $R^2[i]$ averaged across the ten folds (splits). Here, $R^2[i]$ represents the coefficient of determination for the relative density between the training data set and the testing data set for fold (split), $i$.

| Method | Coefficient of Determination, $R^2$, for the Relative Density No Boundary Conditions Included | | | | | | | | | | |
|---|---|---|---|---|---|---|---|---|---|---|---|
| | Run 1 | Run 2 | Run 3 | Run 4 | Run 5 | Run 6 | Run 7 | Run 8 | Run 9 | Run 10 | Avg. |
| 1-layer NN: 5 nodes | 0.918 | 0.807 | 0.881 | 0.798 | 0.801 | 0.846 | 0.857 | 0.756 | 0.779 | 0.789 | 0.823 |
| 1-layer NN: 10 nodes | 0.879 | 0.830 | 0.880 | 0.713 | 0.749 | 0.853 | 0.839 | 0.787 | 0.760 | 0.788 | 0.808 |
| 1-layer NN: 15 nodes | 0.865 | 0.811 | 0.836 | 0.771 | 0.746 | 0.816 | 0.822 | 0.735 | 0.717 | 0.749 | 0.787 |
| 2-layer NN: 5 + 5 nodes | 0.815 | 0.800 | 0.849 | 0.751 | 0.748 | 0.836 | 0.865 | 0.765 | 0.770 | 0.782 | 0.798 |
| 2-layer NN: 5 + 10 nodes | 0.831 | 0.823 | 0.867 | 0.783 | 0.716 | 0.803 | 0.845 | 0.762 | 0.754 | 0.731 | 0.792 |
| 2-layer NN: 10 + 5 nodes | 0.815 | 0.837 | 0.755 | 0.689 | 0.689 | 0.833 | 0.818 | 0.748 | 0.703 | 0.765 | 0.765 |
| 2-layer NN: 10 + 10 nodes | 0.819 | 0.832 | 0.781 | 0.745 | 0.696 | 0.824 | 0.809 | 0.718 | 0.741 | 0.782 | 0.775 |
| 2-layer NN: 10 + 15 nodes | 0.875 | 0.779 | 0.800 | 0.776 | 0.721 | 0.759 | 0.773 | 0.741 | 0.645 | 0.790 | 0.766 |
| 2-layer NN: 15 + 10 nodes | 0.800 | 0.718 | 0.737 | 0.727 | 0.733 | 0.803 | 0.777 | 0.681 | 0.627 | 0.794 | 0.740 |
| 2-layer NN: 15 + 15 nodes | 0.779 | 0.669 | 0.760 | 0.666 | 0.576 | 0.792 | 0.790 | 0.660 | 0.651 | 0.758 | 0.710 |
| 2nd-order polynomial reg. | 0.779 | 0.734 | 0.770 | 0.754 | 0.680 | 0.793 | 0.771 | 0.680 | 0.677 | 0.733 | 0.737 |
| 1st-order linear regression | 0.505 | 0.464 | 0.494 | 0.452 | 0.417 | 0.503 | 0.496 | 0.444 | 0.453 | 0.490 | 0.472 |
| Gaussian process regression | 0.862 | 0.822 | 0.844 | 0.762 | 0.743 | 0.858 | 0.846 | 0.742 | 0.770 | 0.801 | 0.805 |

**Table S6:** Characterization of the quality of fit of the 3D surfaces for the density and hardness synthesized relative to the original input data shown in Figure 6. Here, no boundary conditions have been applied. The quality of fit was estimated exclusively from testing portion of Build Run 1 ($N_{test,BR1}$ = 35 samples). $P(M|D)$ represents the posterior probability of the model listed given the data, whereas $\sigma_{test}$ and $\sigma_{meas}$ appear in Eq. (14). The specific values for $\sigma_{meas,1}$ and $\sigma_{meas,2}$ used are listed in Eqs. (47) and (48) in the main manuscript. $N_{param}$ represents the number of parameters in the neural network, but $N_{train,BR1}$ the number of samples in the training set from Build Run 1. In this case, $N_{train,BR1}$ = 140 samples. In case of the NNs and using the syntax from Eq. (24) in the main manuscript, $N_{param}$ was estimated in Matlab® as **length(getwb(net))**. No cross-validation was performed in this case.

| Network Structure | Density | | | | Hardness (No Boundary Conditions Included) | | | | Size |
|---|---|---|---|---|---|---|---|---|---|
| | $R^2$ | $P(M|D)$ | $\sigma_{test,1}$ [%] | $\sigma_{test,1} / \sigma_{meas,1}$ | $R^2$ | $P(M|D)$ | $\sigma_{test,2}$ [HV] | $\sigma_{test,2} / \sigma_{meas,2}$ | $N_{train,BR1} / N_{param,BR1}$ |
| *1 layer: 5 nodes* | 0.848 | 0.110 | 0.205 | 0.967 | 0.138 | 0.0990881 | 2.515 | 1.771 | 6.67 |
| 1 layer: 10 nodes | 0.841 | 0.104 | 0.211 | 0.995 | 0.037 | 0.104127 | 2.837 | 1.998 | 3.41 |
| 1 layer: 15 nodes | 0.703 | 0.102 | 0.319 | 1.502 | 0.005 | 0.0748633 | 2.917 | 2.054 | 1.54 |
| 2 layers: 5 + 5 nodes | 0.225 | 0.111 | 0.862 | 4.061 | 0.021 | 0.096526 | 2.827 | 1.991 | 2.75 |
| 2 layers: 5 + 10 nodes | 0.639 | 0.058 | 0.445 | 2.095 | 0.005 | 0.091 | 2.813 | 1.981 | 1.62 |
| 2 layers: 10 + 5 nodes | 0.786 | 0.103 | 0.293 | 1.383 | 0.251 | 0.115831 | 2.284 | 1.609 | 1.54 |
| 2 layers: 10 + 10 nodes | 0.625 | 0.117 | 0.336 | 1.584 | 0.060 | 0.010 | 2.631 | 1.853 | 0.93 |
| 2 layers: 10 + 15 nodes | 0.570 | 0.095 | 0.465 | 2.192 | 0.178 | 0.117196 | 2.445 | 1.722 | 0.66 |
| 2 layers: 15 + 10 nodes | 0.846 | 0.110 | 0.203 | 0.955 | 0.107 | 0.096874 | 2.681 | 1.888 | 0.65 |
| 2 layers: 15 + 15 nodes | 0.824 | 0.090 | 0.268 | 1.261 | 0.006 | 0.105221 | 3.162 | 2.226 | 0.47 |

**Table S7:** Estimation of the accuracy of fit of the 3D surfaces for the density and hardness synthesized relative to the original input data shown in Figure 6 (Build Run 1). The accuracy of the fit is estimated with or without boundary conditions imposed using the testing data exclusively. The mean squared error (MSE) is estimated using the Frobenius norm. No cross-validation was performed in this case.

| Network Structure | Density | | | | Hardness | | | |
|---|---|---|---|---|---|---|---|---|
| | *MSE* | | std. dev. | | *MSE* | | std. dev. | |
| | No Bound Cond. | W/ Bound Cond. | No Bound Cond. | W/ Bound Cond. | No Bound Cond. | W/ Bound Cond. | No Bound Cond. | W/ Bound Cond. |
| 1 layer: 5 nodes | 1.207 | 170.49 | 0.193 | 21.57 | 15.18 | 144.67 | 2.596 | 2.637 |
| 1 layer: 10 nodes | 1.619 | 49.80 | 0.277 | 8.27 | 16.74 | 31.76 | 2.865 | 5.352 |
| 1 layer: 15 nodes | 1.444 | 73.50 | 0.241 | 12.46 | 16.68 | 23.66 | 2.861 | 4.040 |
| 2 layers: 5 + 5 nodes | 1.462 | 91.85 | 0.246 | 9.27 | 15.06 | 54.31 | 2.572 | 9.129 |
| 2 layers: 5 + 10 nodes | 1.386 | 109.32 | 0.227 | 7.78 | 13.64 | 15.38 | 2.339 | 2.589 |
| 2 layers: 10 + 5 nodes | 2.483 | 185.40 | 0.403 | 11.82 | 13.88 | 58.39 | 2.352 | 6.109 |
| 2 layers: 10 + 10 nodes | 1.220 | 146.07 | 0.206 | 14.72 | 16.54 | 15.46 | 2.833 | 2.633 |
| 2 layers: 10 + 15 nodes | 2.278 | 85.59 | 0.388 | 9.09 | 15.39 | 46.89 | 2.638 | 7.837 |
| 2 layers: 15 + 10 nodes | 3.543 | 61.67 | 0.556 | 9.80 | 15.33 | 53.40 | 2.627 | 4.179 |
| 2 layers: 15 + 15 nodes | 2.054 | 69.49 | 0.348 | 11.82 | 17.51 | 15.86 | 2.945 | 2.717 |

**Supplementary Figures**

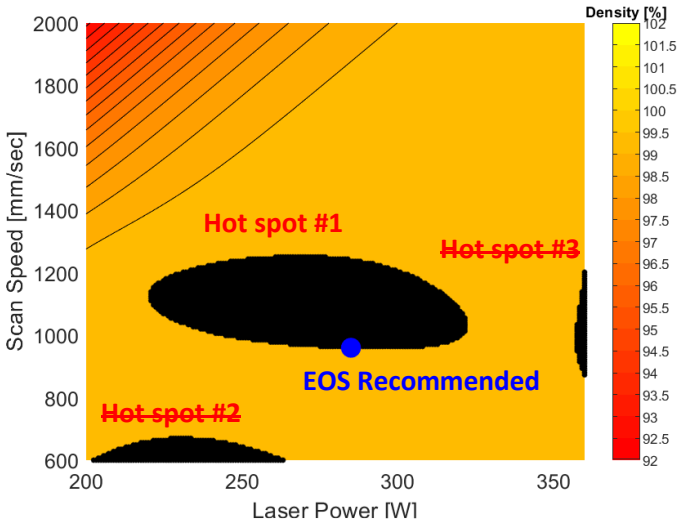

(a) 1 layer with 5 nodes; density = 99.5% plane

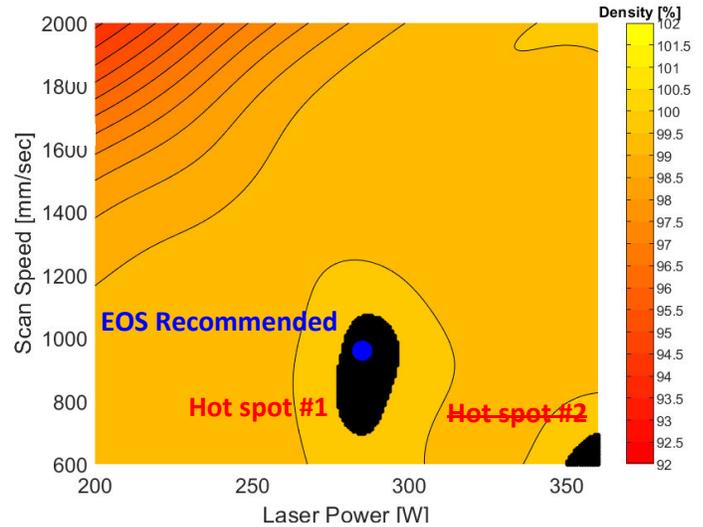

(b) 1 layer with 10 nodes; density = 99.6% plane

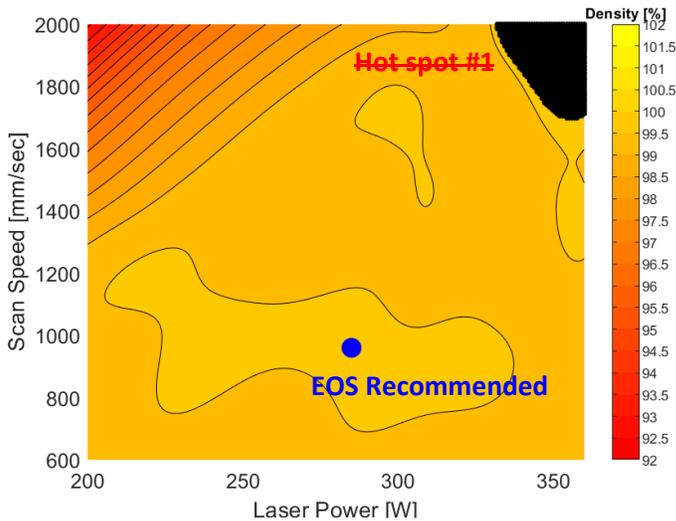

(c) 1 layer with 15 nodes; density = 99.6% plane

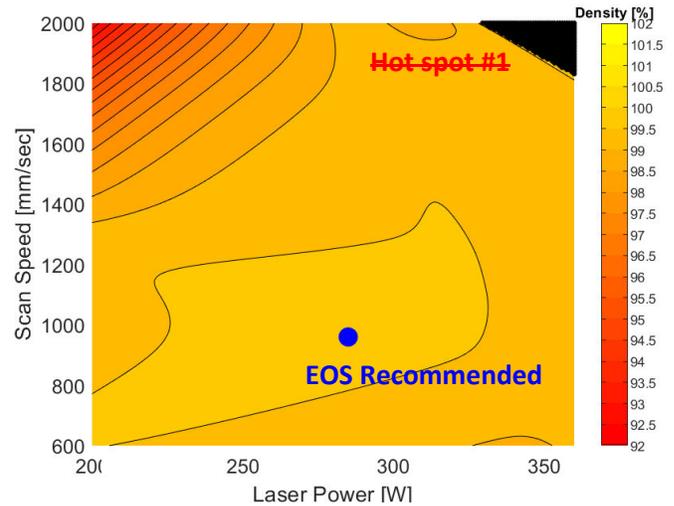

(d) 2 layers each with 5 nodes; density = 99.6% plane

**Figure S1**: 2D contour versions of reconstructed density mappings generated from the real AM data (Build Run 1) for IN 718 produced by the HT-testing technique. Boundary conditions were excluded from the data set, which comprised exclusively the training data. The hot spots with strike through represent hot spots located along the boundary which the algorithm disregards. No cross-validation was needed to produce the Figure.

**(a) 5 nodes in layer 1, and 10 in layer 2; density = 100% plane**

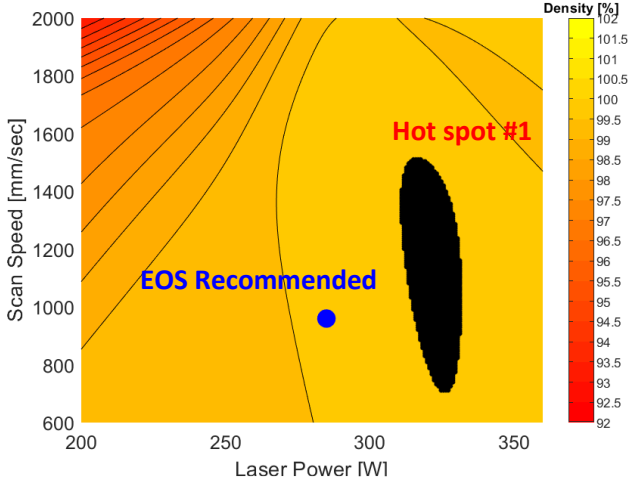

**(b) 10 nodes in layer 1, and 5 nodes in layer 2; density = 99.6% plane**

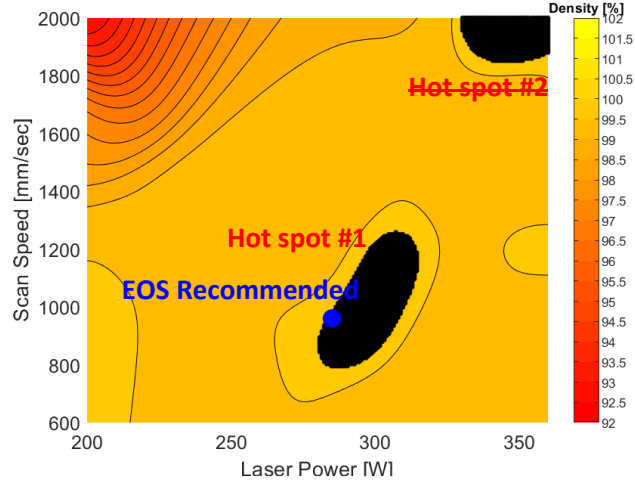

**(c) 2 layers with 10 nodes each; density = 99.6% plane**

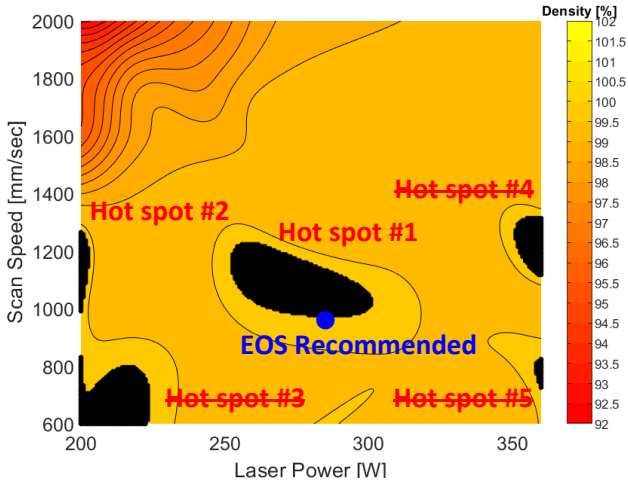

**(d) 15 nodes in layer 1, and 10 in layer 2; density = 99. 6% plane**

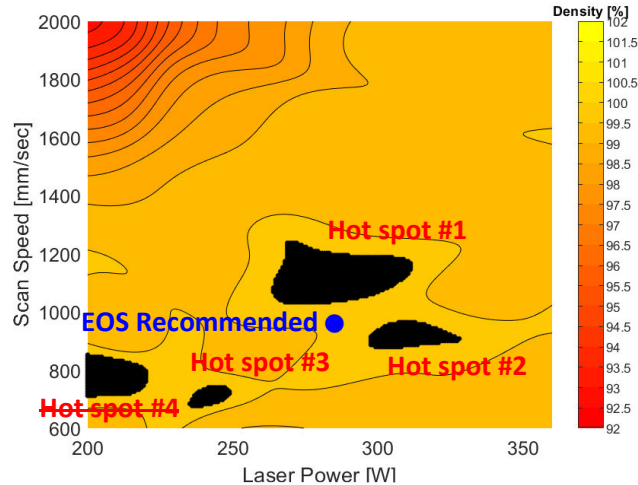

**(e) 10 nodes in layer 1 and 15 in layer 2; density = 99. 7% plane**

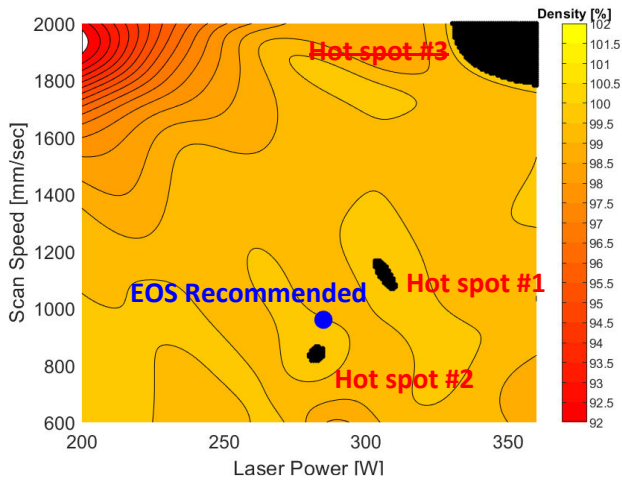

**(f) 15 nodes in layer 1 and 15 in layer 2; density = 99. 8% plane**

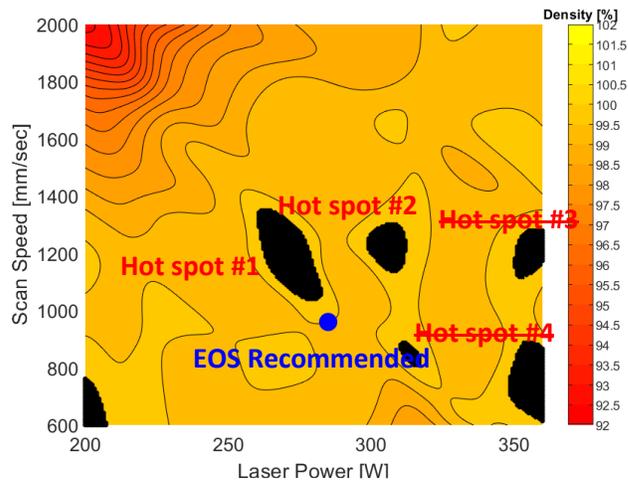

**Figure S2**: 2D contour versions of additional reconstructed density mappings generated from the real AM data (Build Run 1) for IN 718 produced by the HT-testing technique. Boundary conditions were excluded from the data set, which comprised exclusively the training data). No cross-validation was needed to produce the Figure.

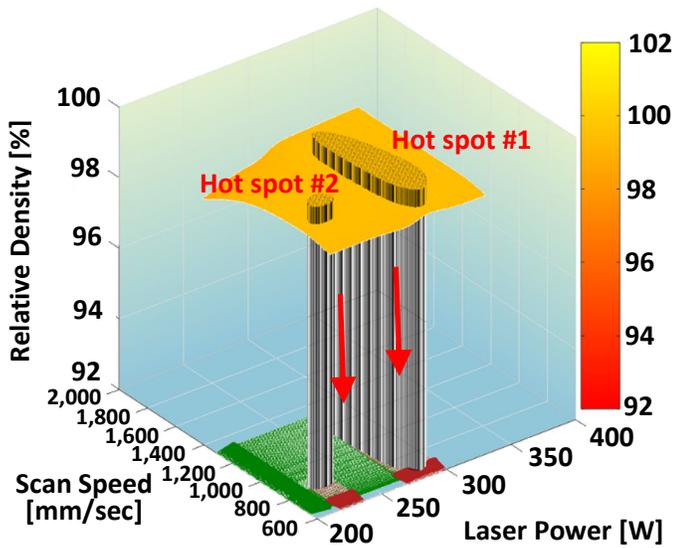
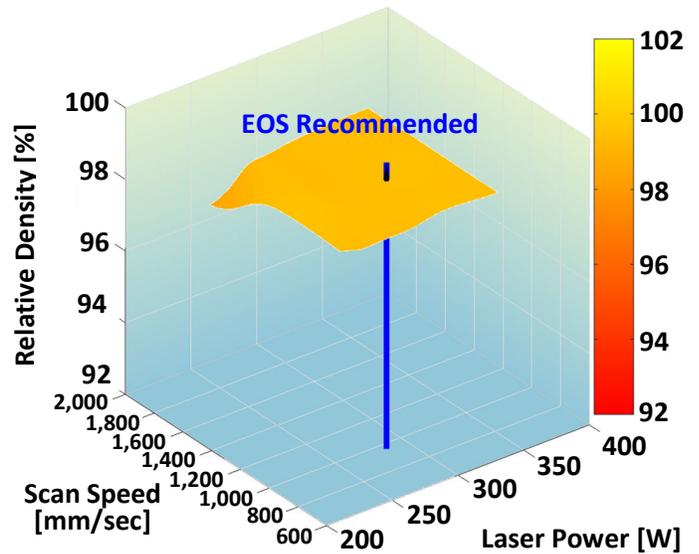
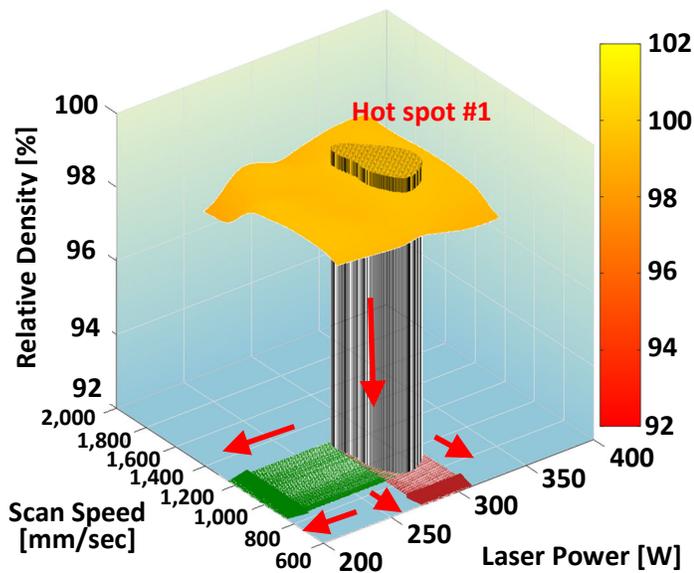
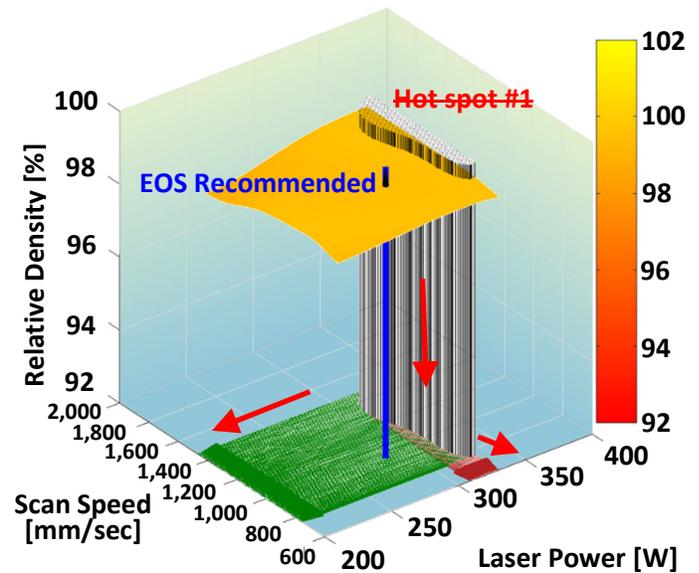

**Figure S3**: Reconstructed density mappings generated from real AM data (Build Run 2) for IN 718 produced by the HT-testing technique. Boundary conditions were excluded from the data set, which comprised exclusively of the training data. The hot spots with strike through represent hot spots located along the boundary, which the algorithm disregard. No cross-validation was needed to produce the Figure.

**(a)** 5 nodes in layer 1 and 10 in layer 2; density = 100% plane   **(b)** 10 nodes in layer 1 and 5 nodes in layer 2; density = 99.6% plane

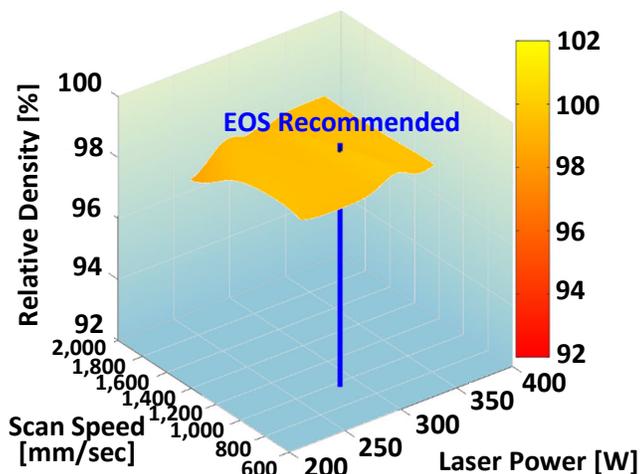
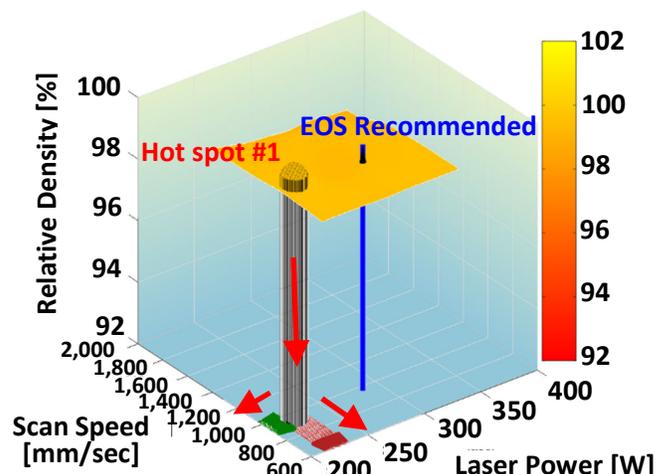

**(c)** 2 layers with 10 nodes each; density = 99.6% plane   **(d)** 15 nodes in layer 1 and 10 in layer 2; density = 99.6% plane

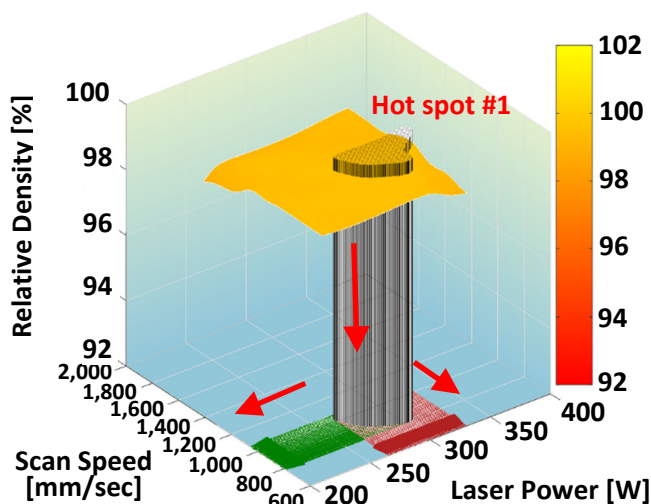
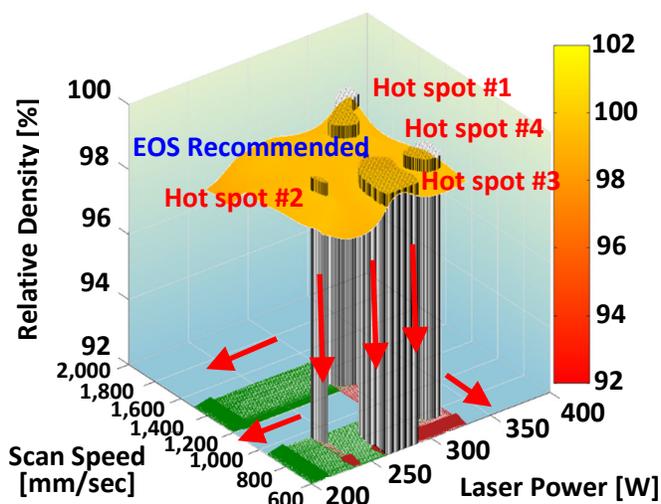

**(e)** 10 nodes in layer 1 and 15 in layer 2; density = 99.7% plane   **(f)** 15 nodes in layer 1 and 15 in layer 2; density = 99.8% plane

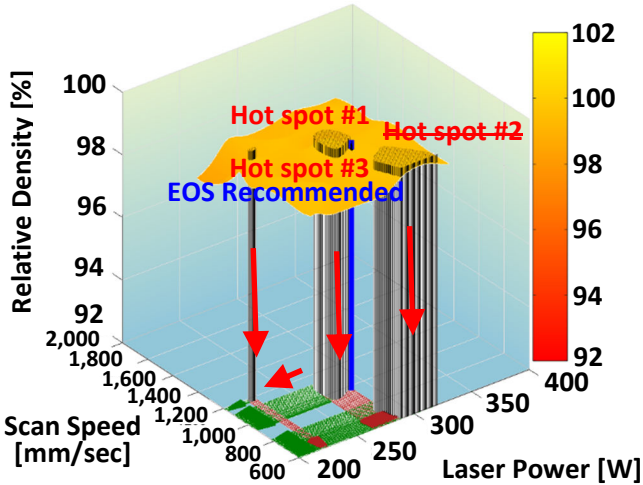
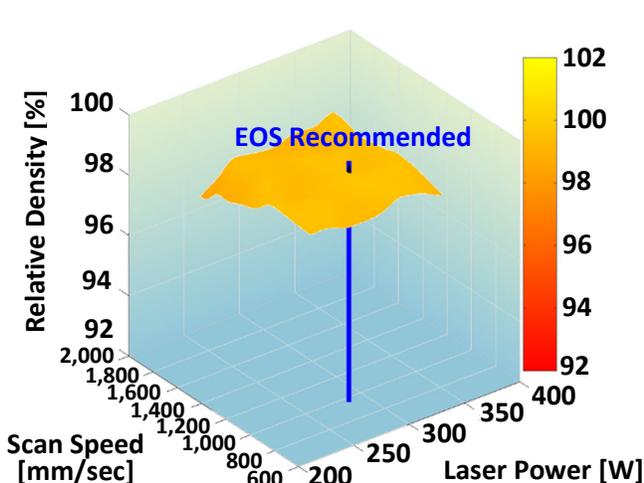

**Figure S4:** Additional reconstructed density mappings generated from real AM data (Build Run 2) for IN 718 produced by the HT-testing technique. Boundary conditions were excluded from the data set, which comprised exclusively of the training data. No cross-validation was needed to produce the Figure.

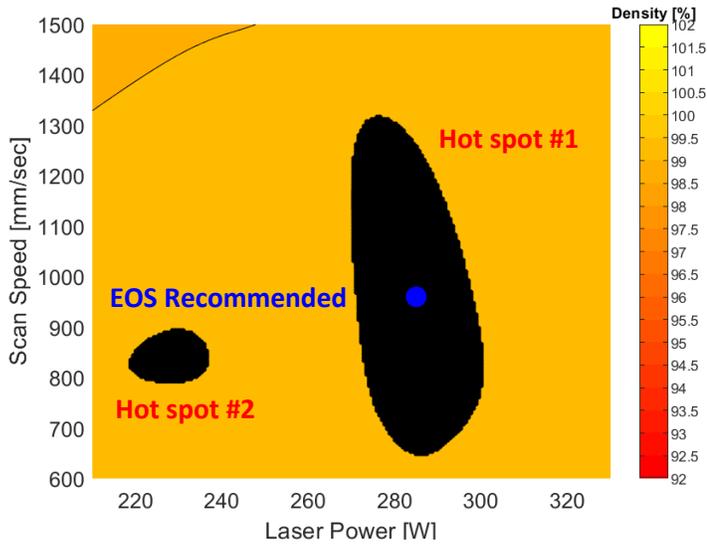
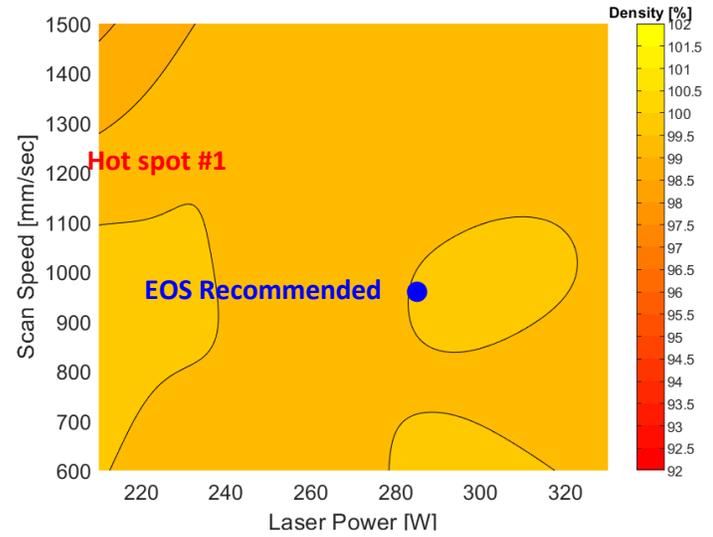
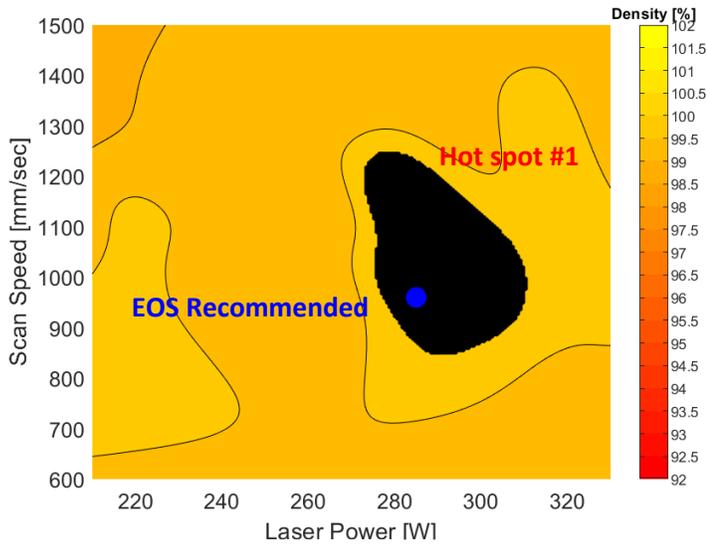
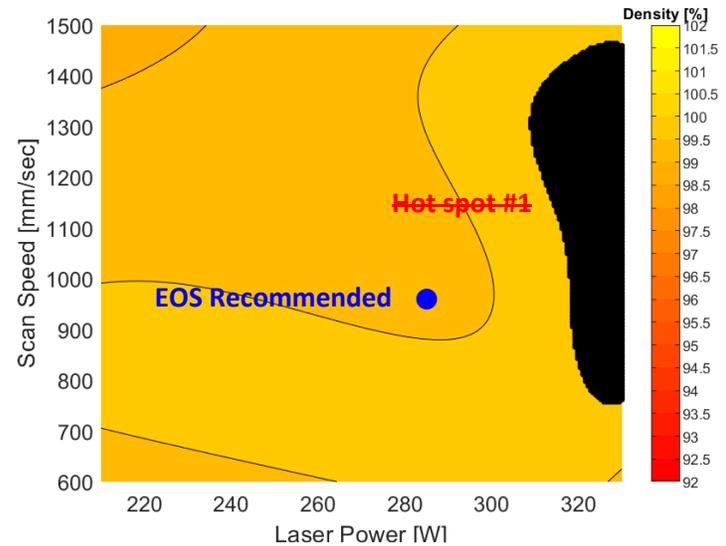

**Figure S5**: 2D contour versions of reconstructed density mappings generated from real AM data (Build Run 2) for IN 718 produced by the HT-testing technique. Boundary conditions were excluded from the data set, which comprised exclusively of the training data. The hot spots with strike through represent hot spots located along the boundary, which the algorithm disregard. No cross-validation was needed to produce the Figure.

**(a) 5 nodes in layer 1 and 10 in layer 2; density = 100% plane**  **(b) 10 nodes in layer 1 and 5 nodes in layer 2; density = 99.6% plane**

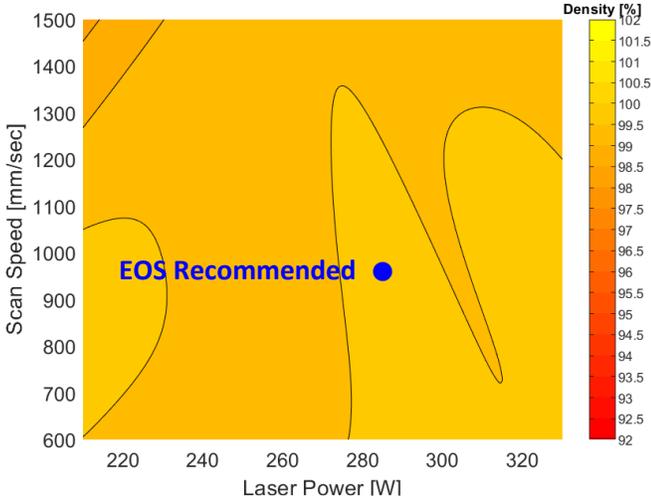 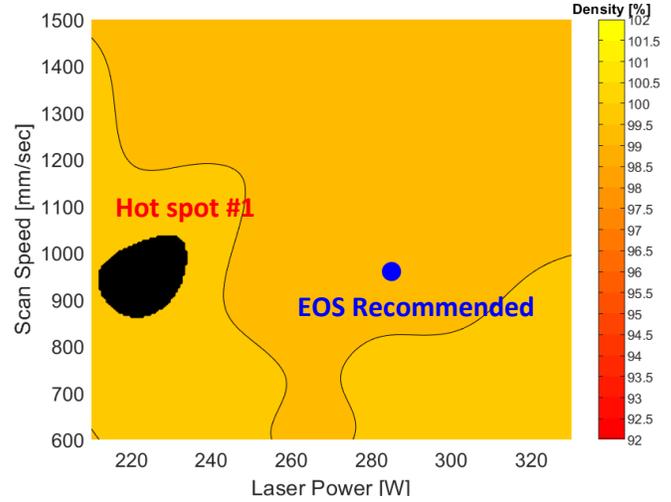

**(c) 2 layers with 10 nodes each; density = 99.6% plane**  **(d) 15 nodes in layer 1 and 10 in layer 2; density = 99.6% plane**

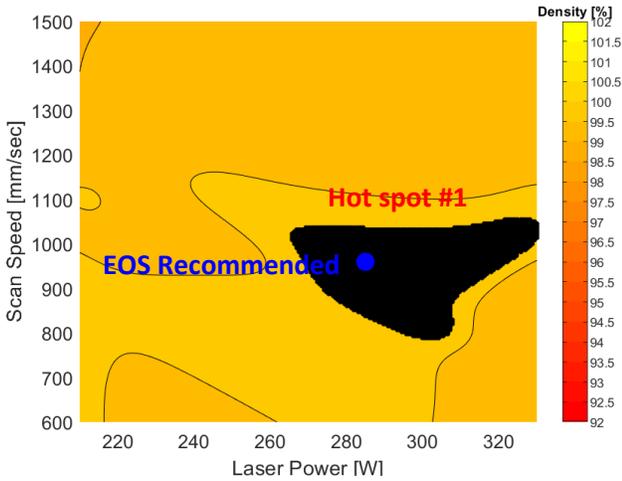 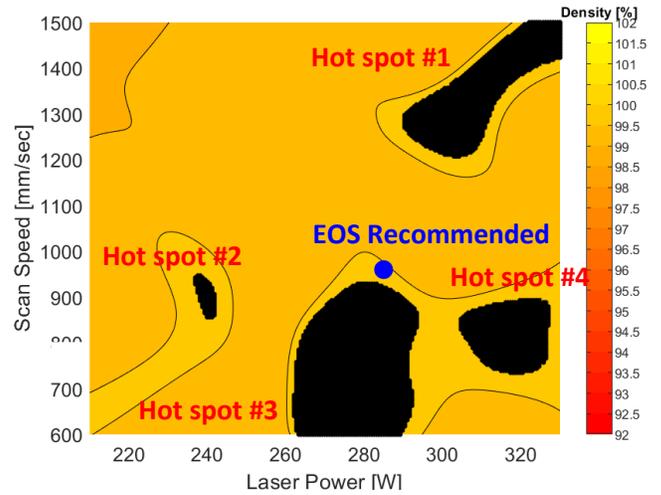

**(e) 10 nodes in layer 1 and 15 in layer 2; density = 99.7% plane**  **(f) 15 nodes in layer 1 and 15 in layer 2; density = 99.8% plane**

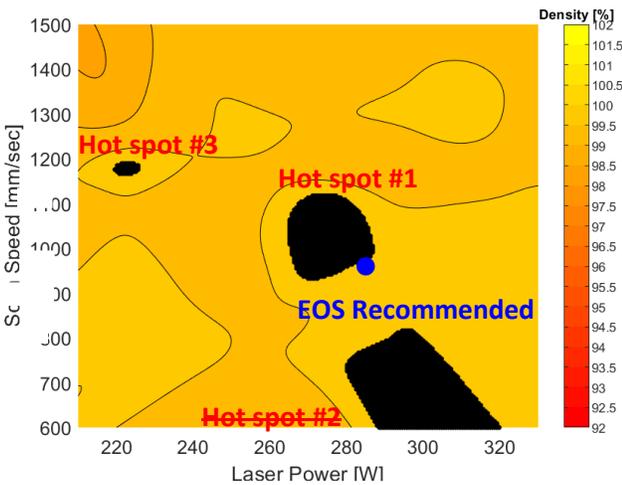 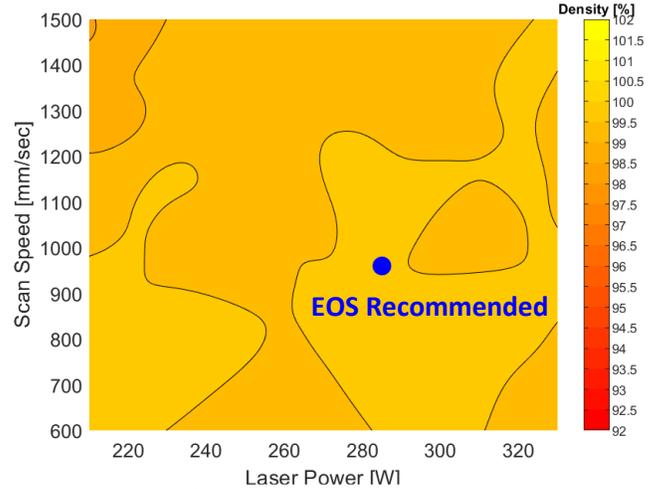

**Figure S6:** 2D contour versions of additional reconstructed density mappings generated from real AM data (Build Run 2) for IN 718 produced by the HT-testing technique. Boundary conditions were excluded from the data set, which comprised exclusively of the training data. No cross-validation was needed to produce the Figure.

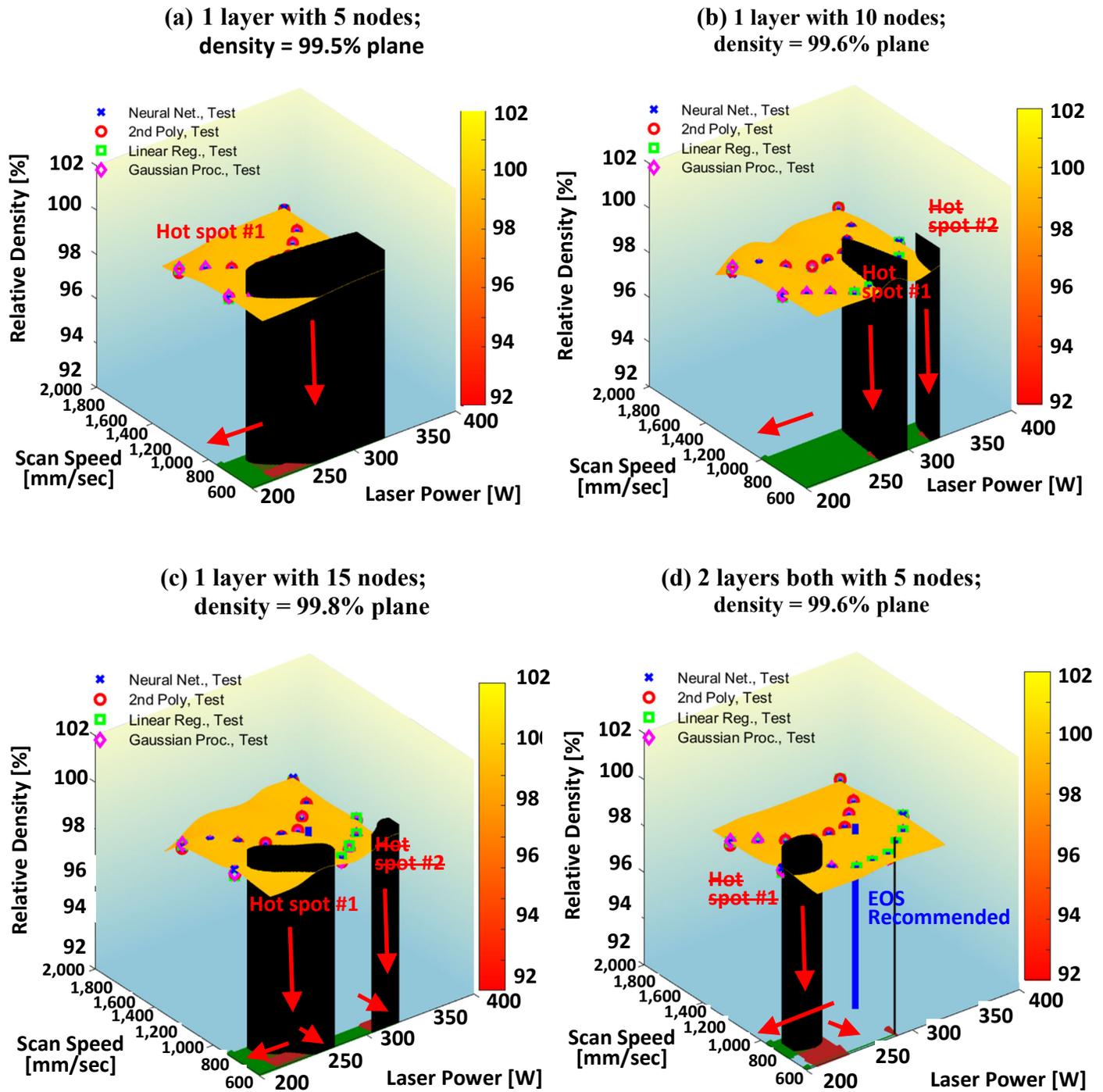

**Figure S7**: Reconstructed density mappings generated from real AM data (Build Run 2) for IN 718 produced by the HT-testing technique. Boundary conditions were excluded from the data set, which had been split into training data and testing data sets according to the 80/20 rule, but with no cross-validation needed to produce the Figure. The hot spots with strike through represent hot spots located along the boundary, which the algorithm disregard. Here, $N_{\text{train,BR2}}=72$ and $N_{\text{test,BR2}}=18$.

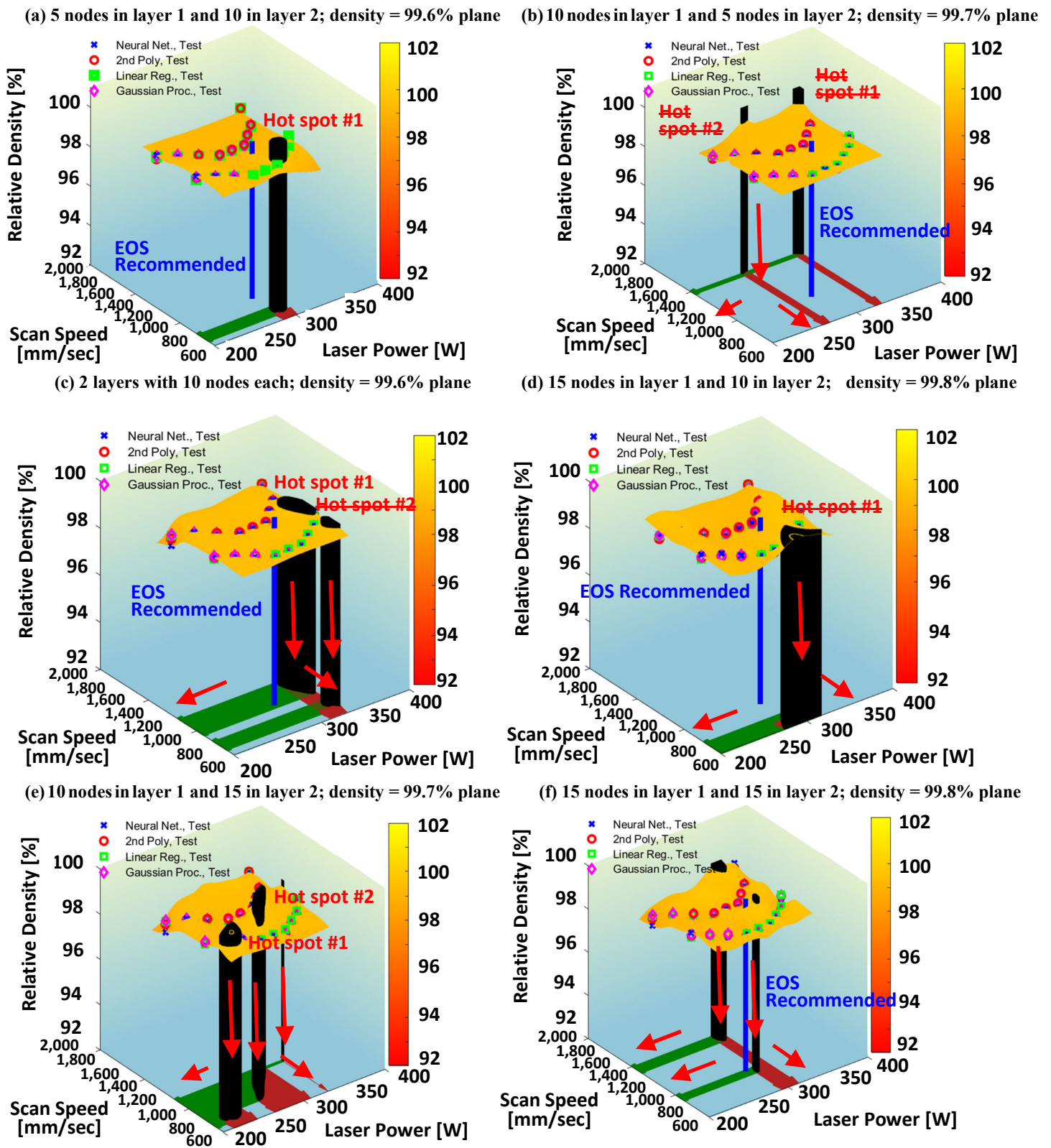

**Figure S8:** Additional reconstructed density mappings generated from real AM data (Build Run 2) for IN 718 produced by the HT-testing technique. Boundary conditions were excluded from the data set, which had been split into training data and testing data sets according to the 80/20 rule, but with no cross-validation needed to produce the Figure.

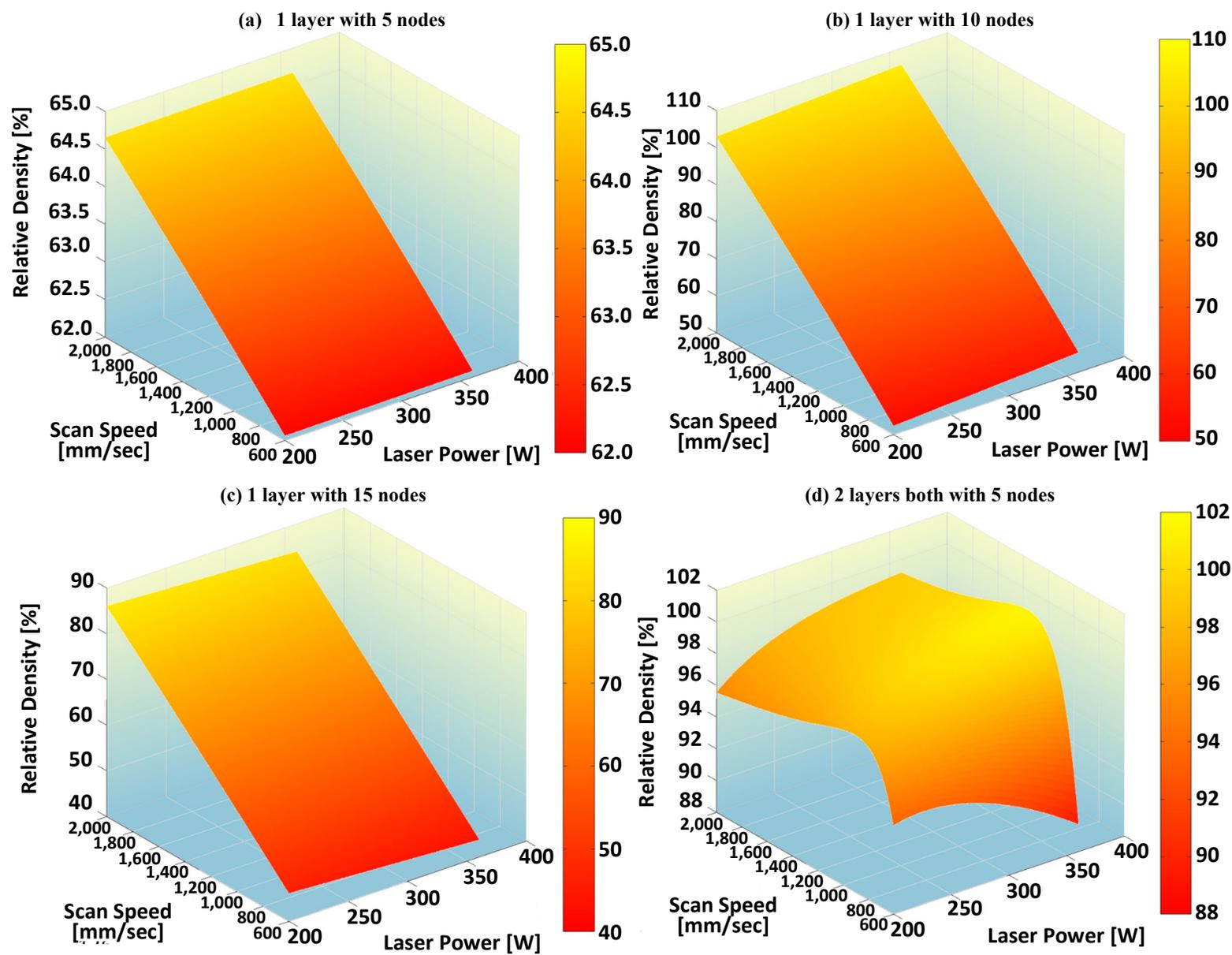

**Figure S9:** Reconstructed density as a function of laser power and scan speed with boundary conditions included. The input data set comprised of the training data exclusively.

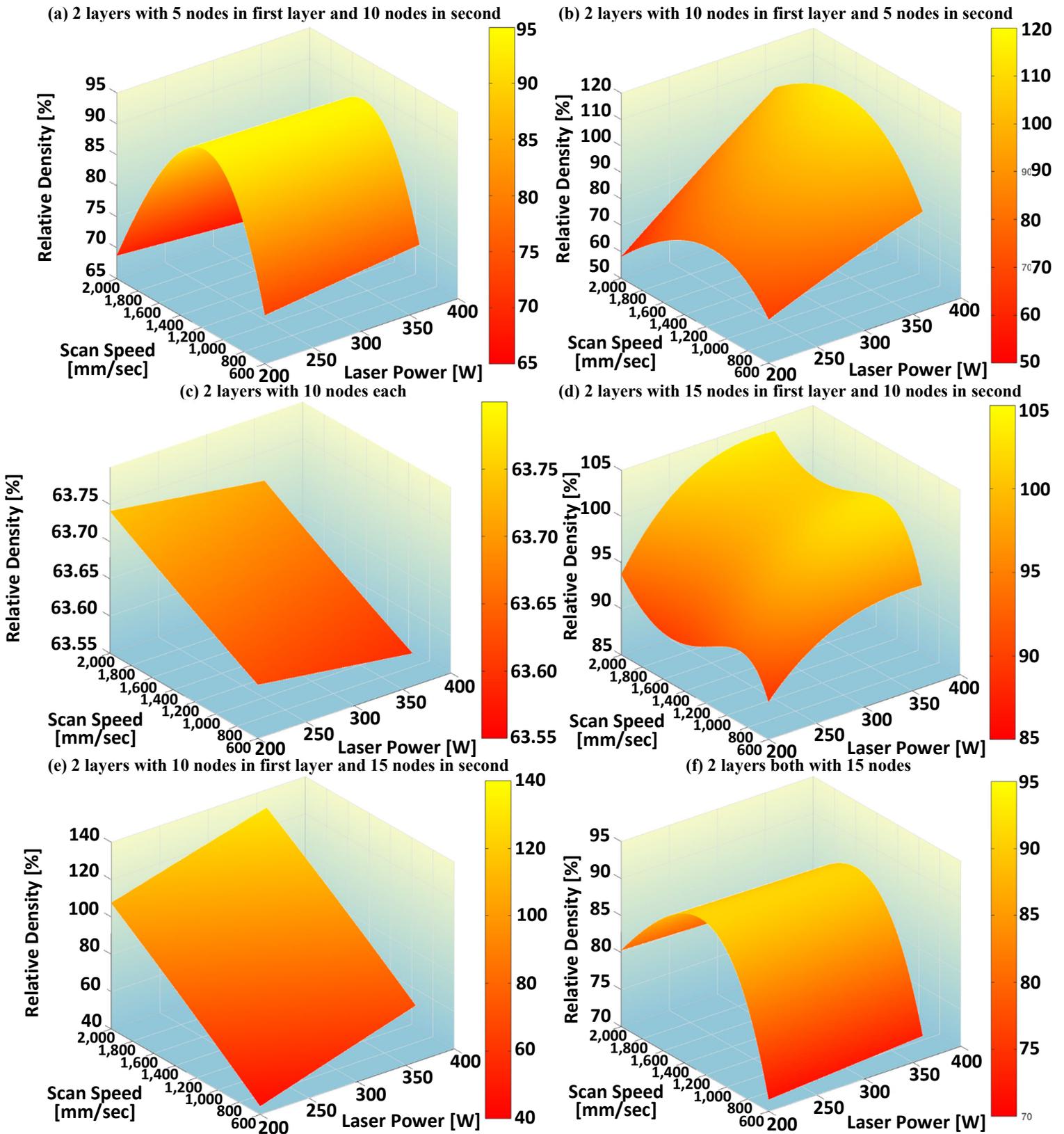

**Figure S10:** Reconstructed density as a function of laser power and scan speed with boundary conditions included. The input data set comprised of the training data exclusively.

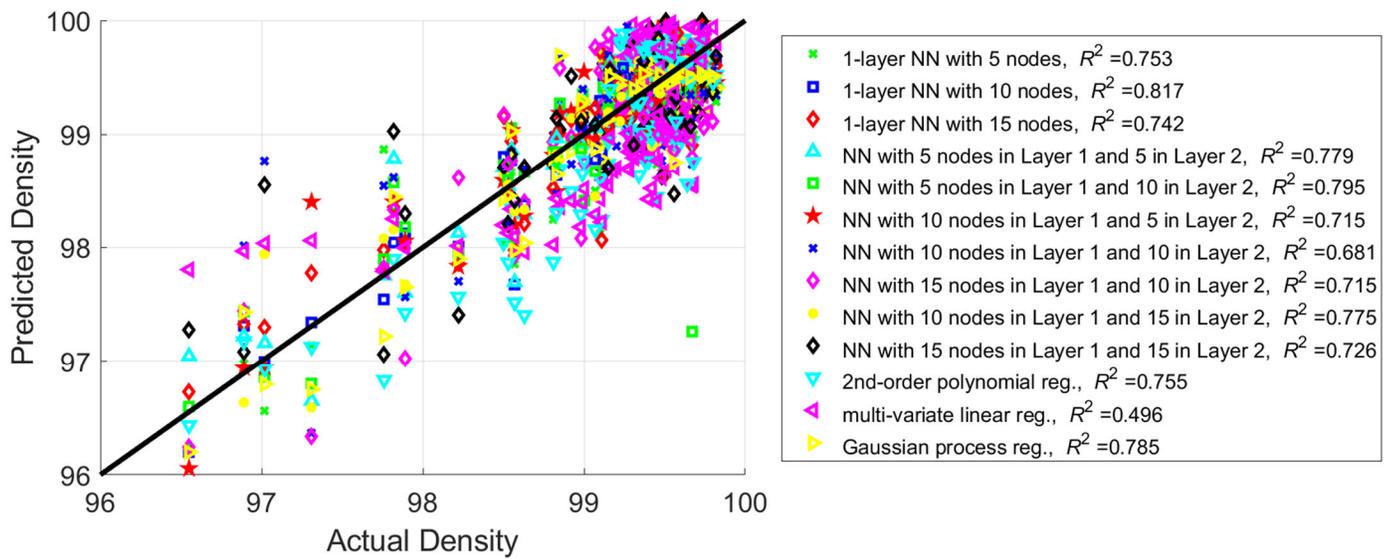

**Figure S11**: Prediction accuracy of the neural network solutions compared to that of the multi-variate linear regression, 2nd-order polynomial regression, and Gaussian process regression for the test set (Inconel 718, Build Run 1, ten-fold cross-validation, boundary conditions excluded in the neural network solutions).

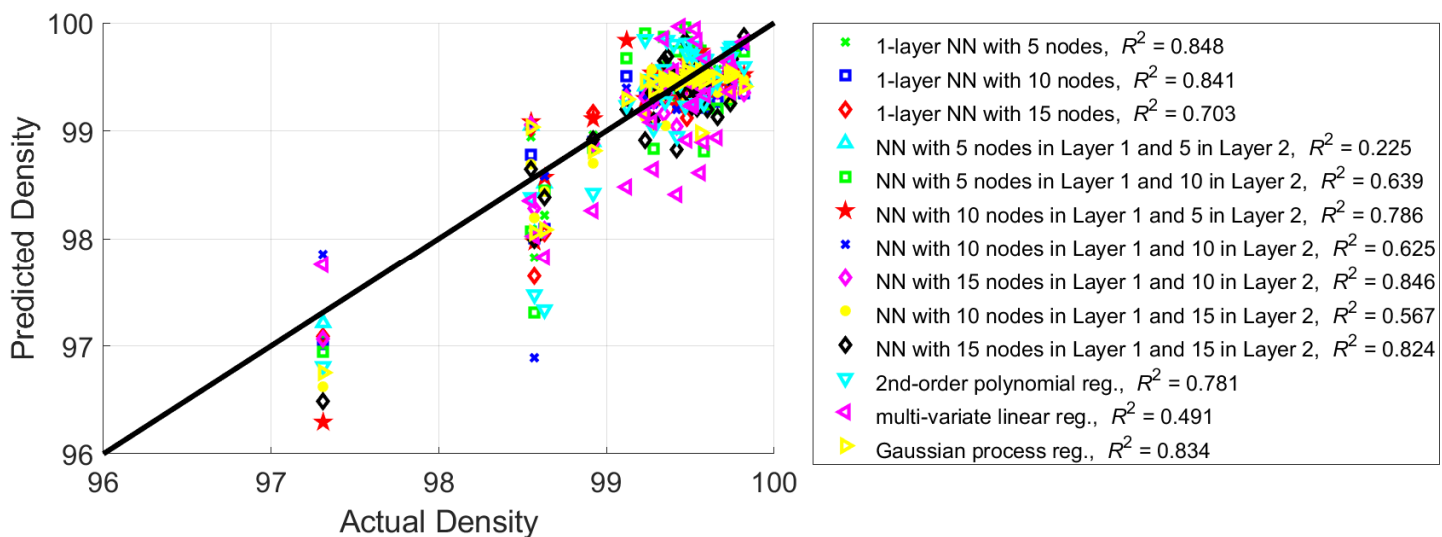

**Figure S12**: Prediction accuracy of the neural network solutions compared to that of the multi-variate linear regression, 2nd-order polynomial regression, and Gaussian process regression for the test set (Inconel 718, Build Run 1, no cross-validation, boundary conditions excluded in the neural-network solutions).

**Supplementary Notes**

1. <u>Further Background and Specifics on Implementation of the Cross-Validation</u>

To facilitate comparison with [4], we looked at three cases:
1. No cross-validation.
2. 5-fold cross-validation.
3. 10-fold cross-validation.

We started out with the case of no cross-validation, with 80% of the input data set from a given build run reserved for training, but with 20% of the input data set assigned to testing, in accordance with the classical 80/20 rule of data science [5, 6]. We then implemented the case of 5-fold cross-validation, where for each split 80% of the input data set from a given build run was reserved for training, but where 20% of the input data set was assigned to testing. We then implemented the case of 10-fold cross-validation, where only 10% of the input data set was reserved for testing for each split.

Consistent with the explanations from the main manuscript, the cross-validation was implemented in Matlab as follows:

**Fold   = 1;**                                              % In the case of no cross-validation      (1)
**Fold   = 5;**                                              % In the case of 5-fold cross-validation  (2)
**Fold   = 10;**                                             % In the case of 10-fold cross-validation (3)
**indices = crossvalind( 'Kfold', length(input_data_set (:,1)), Fold );**                              (4)

We then obtain the overall performance by averaging the results obtained from each of the folds. For illustration, in the case of the $2^{nd}$-order polynomial regression, this can be implemented as follows:

**for i = 1 : Fold,**
  **[ part omitted ]**
  **regPoly2 = MultiPolyRegress( input_data_train, output_data_set_train', 2 );**
  **noisy_output_modeled_2ndOrderPoly_test =**
              **feval( regPoly2.PolynomialExpression, x_data_test, y_data_test );**
  **R_sparse_test_2ndOrderPoly            =**
              **corrcoef( output_data_set_test, noisy_output_modeled_2ndOrderPoly_test );**
  **Rsup2_sparse_test_2ndOrderPoly(i)     =**
              **R_sparse_test_2ndOrderPoly(2,1) * R_sparse_test_2ndOrderPoly(2,1);**
**end**

The quantity reported under the rows titled "$2^{nd}$-order polynomial reg." in Tables S4 –S5 is **mean(Rsup2_sparse_test_2ndOrderPoly)**. A corresponding approach is applied for the other methods. In k-fold-cross-validation, the "correct" scheme seems to compute the metric of interest (say, the accuracy) from the testing split for each fold, and then return the mean of the testing splits, averaged across the folds, as the final metric [7, 8].

2. <u>Further Background and Specifics on Gaussian Process Regression (and Bayesian Regression)</u>

2.1  Starting Point: The Fundamentals of Bayesian Regression
Bayesian regression involves a probabilistic modeling approach, where a posterior probability distribution over the model (parameters) is determined, and where this posterior probability distribution is updated, whenever new data points are observed [9]. Bayesian regression provides one with a relevant regression model, $D$, along with a probability density function, $p(\mathbf{f}_{train}|\mathbf{X}_{train},D)$, for the observations, $\mathbf{y}_{train} = \mathbf{f}_{train} = f(\mathbf{X}_{train})$, given the model $D$ and the input data (training set) $\mathbf{X}_{train}$. Given the regression model, $D$, and some new $\mathbf{X}^*_{test}$ (a testing set), Bayesian regression models can return (random) values drawn from the posterior probability distribution $p(\mathbf{f}^*_{test}|\mathbf{X}^*_{test}, \mathbf{X}_{train}, \mathbf{y}_{train}, D)$. In this sense, the Bayesian regression techniques can return predictions of yet-to-be-observed function values $\mathbf{f}^*_{test} = f(\mathbf{X}^*_{test})$. Once the actual observations from the test set, $\mathbf{y}^*_{test}$, become available, the predictions $\mathbf{f}^*_{test}$ can be compared to these observations. Bayesian regression techniques are probabilistic (non-deterministic). But with that said, although $p(\mathbf{f}^*_{test}|\mathbf{X}_{train}, D)$ is not deterministic, the mean $\mathbb{E}[\mathbf{f}_{train}|\mathbf{X}_{train}, D]$ is indeed deterministic [10-17].

### 2.2 Gaussian Process Regression in Context with Bayesian Regression (Kernel View – Noiseless Case)

#### 2.2.1 *Objective*

Given some training data ($\mathbf{X}_{train}$, $\mathbf{y}_{train}$), the objective of Gaussian process regression (GPR) is to determine a function (a structure or a mapping) that fits these data points. One is trying to determine the function $f(\cdot)$ that yields

$$\mathbf{y}_{train} = \mathbf{f}_{train} = f(\mathbf{X}_{train}). \tag{5}$$

Oftentimes, there are many functions that can fit through the known set of ($\mathbf{X}_{train}$, $\mathbf{y}_{train}$).

#### 2.2.2 *Main Assumption*

The main assumption, captured in Eq. (5), involves the observations exactly matching with the function values evaluated. There is no mismatch. This is referred to as a noiseless case ($\beta^{-1} = \sigma^2 = 0$, in the terminology of [18, 19]). The Gaussian process can also be generalized, such that the function values evaluated pass close to the observations, but not exactly. The latter is referred to as a noisy case (where $\beta^{-1} = \sigma^2 > 0$).

#### 2.2.2 *Key Idea*

GPR involves a non-parametric approach, in that the GPR determines a distribution over the possible function mappings that are consistent with the observed data [9]. As with all Bayesian regression methods, the GPR begins with a prior distribution, which then gets updated as new data points are observed, producing a posterior distribution over functions, as noted in Section 2.1 [9].

#### 2.2.3 *Summary of the Algorithm*

1. We are given a set of training points, $\mathbf{X}_{train}$, together with corresponding observations, $\mathbf{y}_{train}$.
2. We define a Gaussian process as an infinite space of Gaussian variables, where we can take any subset of points in that space, and they all distribute Gaussian. And taken together, they distribute as multi-variate Gaussian distributions [20]:

$$f(\mathbf{x}) \sim GP(\mathbb{E}[f(\mathbf{x})], \text{cov}(f(\mathbf{x}), f(\mathbf{x}'))) = GP(m(\mathbf{x}), k(\mathbf{x}, \mathbf{x}')) \tag{6}$$

3. We calculate the kernel function (~ covariance matrix), $k(\mathbf{x}_{train}, \mathbf{x}'_{train})$, and the mean function, $m(\mathbf{x}_{train})$. Here, we are going to assume exponential kernel functions of the following form [18]:

$$k(\mathbf{x}_n, \mathbf{x}_m) = \theta_0 \exp\left(-\frac{1}{2\,\theta_1^2} \|\mathbf{x}_n - \mathbf{x}_m\|^2\right) + \theta_2 + \theta_3\, \mathbf{x}_n^T \mathbf{x}_m \tag{7}$$

So for each point in the training data set, we calculate the kernel function with respect to each of the other points in the training data set [18]:

$$\mathbf{K}(\mathbf{X}_{train}, \mathbf{X}_{train}) = \begin{pmatrix} k(\mathbf{x}_1, \mathbf{x}_1) & \cdots & k(\mathbf{x}_1, \mathbf{x}_n) \\ \vdots & \ddots & \vdots \\ k(\mathbf{x}_n, \mathbf{x}_1) & \cdots & k(\mathbf{x}_n, \mathbf{x}_n) \end{pmatrix} \tag{8}$$

The kernel function, $k(\mathbf{x}_i, \mathbf{x}_j)$, specifies how much the data points $\mathbf{x}_i$ and $\mathbf{x}_j$ are related to one another. The kernel function, $k(\mathbf{x}_i, \mathbf{x}_j)$, provides smoothness. Data points, that are very close to one another, will have almost the same kernel value. But data points, that are far apart, will not be related any more [18, 19].

4. The joint distribution of the test points $\mathbf{f}^*_{test}$ (at $\mathbf{X}^*$) and the training points $\mathbf{f}_{train}$ (at $\mathbf{X}$), according to our Gaussian process, is given by

$$\begin{bmatrix} \mathbf{f}_{train} \\ \mathbf{f}^*_{test} \end{bmatrix} = N\left( \begin{bmatrix} \mathbf{m}_{train} \\ \mathbf{m}_{test} \end{bmatrix}, \begin{bmatrix} \mathbf{K}(\mathbf{X}_{train}, \mathbf{X}_{train}) & \mathbf{K}(\mathbf{X}_{train}, \mathbf{X}^*_{test}) \\ \mathbf{K}(\mathbf{X}^*_{test}, \mathbf{X}_{train}) & \mathbf{K}(\mathbf{X}^*_{test}, \mathbf{X}^*_{test}) \end{bmatrix} \right) \quad (9)$$

5. We then obtain the posterior distribution

$$p(\mathbf{f}^*_{test} | \mathbf{X}^*_{test}, \mathbf{X}_{train}, \mathbf{y}_{train}) \sim N(\boldsymbol{\mu}^*, \boldsymbol{\Sigma}^*) \quad (10)$$

where

$$\boldsymbol{\mu}^*_{test} = \mathbf{K}(\mathbf{X}^*_{test}, \mathbf{X}_{train})\, \mathbf{K}(\mathbf{X}_{train}, \mathbf{X}_{train})^{-1}\, \mathbf{y}_{train} \quad (11)$$

$$\boldsymbol{\Sigma}^*_{test} = \mathbf{K}(\mathbf{X}^*_{test}, \mathbf{X}^*_{test}) - \mathbf{K}(\mathbf{X}^*_{test}, \mathbf{X}_{train})\, \mathbf{K}(\mathbf{X}_{train}, \mathbf{X}_{train})^{-1}\, \mathbf{K}(\mathbf{X}_{train}, \mathbf{X}^*_{test}) \quad (12)$$

In Eq. (12), the term $\mathbf{K}(\mathbf{X}^*_{test}, \mathbf{X}_{train})\, \mathbf{K}(\mathbf{X}_{train}, \mathbf{X}_{train})^{-1}$ represents a weighting matrix. The term $\mathbf{K}(\mathbf{X}_{train}, \mathbf{X}_{train})^{-1}$ captures interactions between the input data in the training data set, $\mathbf{X}_{train}$, specifically how much the data points in the training set are related to one another, how close the data points are to one another, etc. The term $\mathbf{K}(\mathbf{X}^*_{test}, \mathbf{X}_{train})$ captures how much of the new points we want to sample. How much are these new data points related, or how much are they correlated, to the data points that we do have in our training data set? The term $\mathbf{K}(\mathbf{X}^*_{test}, \mathbf{X}^*_{test})$ captures the relationships between the new points and themselves. The term $\mathbf{K}(\mathbf{X}^*_{test}, \mathbf{X}_{train})\, \mathbf{K}(\mathbf{X}_{train}, \mathbf{X}_{train})^{-1}\, \mathbf{K}(\mathbf{X}_{train}, \mathbf{X}^*_{test})$, which captures the relationship between the new and the old data points, reduces the variance for the data points that are close to the data points that we already know the values of [18, 19].

6. For new data points of interest, $\mathbf{X}^*_{test}$, we can arrive at predictions for these points by pulling out samples from the posterior distribution $p(\mathbf{f}^*_{test} | \mathbf{X}^*_{test}, \mathbf{X}_{train}, \mathbf{y}_{train})$. We can draw samples from the posterior distribution at the testing points $\mathbf{X}^*_{test}$ and return as predictions. Or we can simply return as the predictions the mean for data points from the test set, as well as corresponding confidence intervals, as explained in Step 7.

7. We can, furthermore, quantify the confidence associated with these predictions by calculating

$$\boldsymbol{\sigma}^*_{test} = \text{diag}(\boldsymbol{\Sigma}^*_{test}) \quad (13)$$

For a given point $i$ in the test set, the confidence interval associated with the prediction can, for example, be specified as $\pm 3\, \sigma^*_{test,i}$ around the predictive mean, $\mu^*_i$. The confidence intervals will be zero, at the points from the training set, but will grow as one goes further away from these known points.

### 2.2.4 Important Interpretation

The Gaussian processes provide a very flexible way to derive predictive distributions in a non-parametric way by using a kernel that describes the Gaussian process. The predictive distribution in Eq. (10) can be thought of as a posterior Gaussian process, where one conditions the function evaluation, for unseen data points, on the data points in the training set. By vectorizing the multi-variate Gaussian process as shown in Eq. (9), one can condition the predictions on the existing data set and obtain the conditional distribution in Eq. (10). This conditional distribution still follows a Gaussian distribution. with a particular mean $\boldsymbol{\mu}^*$, listed in Eq. (11), and a particular covariance matrix $\boldsymbol{\Sigma}^*$, listed in Eq. (12), both of which depend on the kernel. The predictive distribution in Eq. (10) describes a non-parametric model, because it does not rely on any explicit parametrization. Instead, the predictive distribution in Eq. (10) is fully characterized by the kernel functions [18].

### 2.3 Gaussian Process Regression in Context with Bayesian Regression (Kernel View – With Noise)

1. As before, a Gaussian process is a collection of random variables, any finite number (any subset) of which is jointly Gaussian distributed.

2. Furthermore, as before, the Gaussian process regression yields a distribution over functions. The Gaussian processes define distributions over functions, that we can use to model the data.
3. In the noisy case, we assume that we have observed a set of input-output pairs (a training set), $\{(\mathbf{x}_i, f_i)\}_{i=1}^{N}$, and we assume the observation model listed below:
$$f_i = f(\mathbf{x}_i) = y(\mathbf{x}_i) + \varepsilon, \qquad \varepsilon \sim N(0, \beta^{-1}) \tag{14}$$
This is a forward model with $\varepsilon$ representing random measurement noise. Here, the Gaussian process is generalized, such that the Gaussian process assumed for the model, $y(\mathbf{x}_i)$, does not have to pass exactly through the observation points, $f(\mathbf{x}_i)$, but is expected to closely approximate the observation points.
4. The training set and the testing set are defined in the same way as before.
5. Since we are assuming a Gaussian process for the model $y(\mathbf{x}_i)$, the vector $\mathbf{y} = [\ y(\mathbf{x}_1), y(\mathbf{x}_2), \ldots, y(\mathbf{x}_N)\ ]^T$ will be given by the jointly Gaussian distribution
$$\mathbf{y} = \begin{bmatrix} y(\mathbf{x}_1) \\ \vdots \\ y(\mathbf{x}_N) \end{bmatrix} \sim N\left(\begin{bmatrix} \mathbb{E}[y(\mathbf{x}_1)] \\ \vdots \\ \mathbb{E}[y(\mathbf{x}_N)] \end{bmatrix}, \begin{bmatrix} k(\mathbf{x}_1,\mathbf{x}_1) & \cdots & k(\mathbf{x}_1,\mathbf{x}_N) \\ \vdots & \ddots & \vdots \\ k(\mathbf{x}_N,\mathbf{x}_1) & \cdots & k(\mathbf{x}_N,\mathbf{x}_N) \end{bmatrix}\right), \tag{15}$$
6. As before, we are going to assume exponential kernel functions of the form specified in Eq. (7) [18].
7. Similarly, because $\mathbf{y}$ is modeled as a Gaussian process, and because the measurement noise $\boldsymbol{\varepsilon}$ is Gaussian, the sum $\mathbf{f} = \mathbf{y} + \boldsymbol{\varepsilon}$, will also produce a Gaussian distributed random vector.
8. So as before,
$$f(\mathbf{x}) \sim GP(\mathbb{E}[f(\mathbf{x})], \text{cov}(f(\mathbf{x}), f(\mathbf{x}'))) \tag{16}$$
where the covariance matrix $\mathbf{K} = \text{cov}(f(\mathbf{x}), f(\mathbf{x}'))$ is also referred to as Gram matrix [21].
9. Ref. [20, 21] contain nice illustration of how a function $f(\cdot)$ can be vectorized into an $N$-dimensional random vector $\mathbf{f} = [f(\mathbf{x}_1), f(\mathbf{x}_2), \ldots, f(\mathbf{x}_N)]$ with the distribution
$$\mathbf{f} \sim N(m(\mathbf{x}), K(\mathbf{X}, \mathbf{X}) + \beta^{-1}\mathbf{I}) \tag{17}$$
In other words, the vector $\mathbf{f} = [\ f(\mathbf{x}_1), f(\mathbf{x}_2), \ldots, f(\mathbf{x}_N)\ ]$ is going to be given by the jointly Gaussian distribution
$$p\left(\begin{bmatrix} f(\mathbf{x}_1) \\ \vdots \\ f(\mathbf{x}_N) \end{bmatrix}\right) = N\left(\begin{bmatrix} m(\mathbf{x}_1) \\ \vdots \\ m(\mathbf{x}_N) \end{bmatrix}, \begin{bmatrix} k(x_1,x_1) + \beta^{-1} & \cdots & k(x_1,x_N) \\ \vdots & \ddots & \vdots \\ k(x_N,x_1) & \cdots & k(x_N,x_N) + \beta^{-1} \end{bmatrix}\right) \tag{18}$$
10. According to the marginalization property of Gaussian distributions, we can split our observation set into two parts, e.g., a training set and a testing set, $[\ f(\mathbf{x}_1), f(\mathbf{x}_2), \ldots, f(\mathbf{x}_N)\ ] = [\mathbf{f}_1, \mathbf{f}_2]$, each of which will be jointly Gaussian distributed:
$$p\left(\begin{bmatrix} \mathbf{f}_1 \\ \mathbf{f}_2^* \end{bmatrix}\right) = N\left(\begin{bmatrix} \mathbf{m}_1 \\ \mathbf{m}_2 \end{bmatrix}, \begin{bmatrix} \mathbf{K}_{11} + \beta^{-1}\mathbf{I} & \mathbf{K}_{12} \\ \mathbf{K}_{21} & \mathbf{K}_{22} + \beta^{-1}\mathbf{I} \end{bmatrix}\right) \Rightarrow p(\mathbf{f}_1) = N(\mathbf{m}_1, \mathbf{K}_{11} + \beta^{-1}\mathbf{I}) \tag{19}$$
This is the main point of a Gaussian process: Each subset is also going to be a Gaussian distributed. Moreover, the covariance matrices in Eq. (18) and Eq. (19) are always going to be positive definite.
11. Further, according to the marginalization property of Gaussian distributions, the joint distribution of test points $\mathbf{f}_{\text{test}}^*$ (to be evaluated at $\mathbf{X}_{\text{test}}^*$) and the training points $\mathbf{f}_{\text{train}}$ (observed at $\mathbf{X}_{\text{train}}$), is given by [18]
$$\begin{bmatrix} \mathbf{f}_{\text{train}} \\ \mathbf{f}_{\text{test}}^* \end{bmatrix} = N\left(\begin{bmatrix} \mathbf{m}_{\text{train}} \\ \mathbf{m}_{\text{test}} \end{bmatrix}, \begin{bmatrix} K(\mathbf{X}_{\text{train}}, \mathbf{X}_{\text{train}}) + \beta^{-1}\mathbf{I} & K(\mathbf{X}_{\text{train}}, \mathbf{X}_{\text{test}}^*) \\ K(\mathbf{X}_{\text{test}}^*, \mathbf{X}_{\text{train}}) & K(\mathbf{X}_{\text{test}}^*, \mathbf{X}_{\text{test}}^*) + \beta^{-1}\mathbf{I} \end{bmatrix}\right) \tag{20}$$
12. Then, using the conditioning property of Gaussian distributions (through conditional vectorization), we can formulate these conditional probabilities for the function values $\mathbf{f}_{\text{test}}^*$ at these unseen data points $\mathbf{X}^*$, given the original (training) data set, as follows:
$$p(\mathbf{f}_{\text{test}}^* | \mathbf{X}_{\text{test}}^*, \mathbf{X}_{\text{train}}, \mathbf{f}_{\text{train}}) \sim N(\boldsymbol{\mu}^*, \boldsymbol{\Sigma}^*) \tag{21}$$

with
$$\boldsymbol{\mu}^* = \mathbf{K}(\mathbf{X}_{\text{test}}^*, \mathbf{X}) \, (\mathbf{K}(\mathbf{X}_{\text{train}}, \mathbf{X}_{\text{train}}) + \beta^{-1} \mathbf{I})^{-1} \, \mathbf{y}_{\text{train}} \quad (22)$$

$$\boldsymbol{\Sigma}^* = \mathbf{K}(\mathbf{X}_{\text{test}}^*, \mathbf{X}_{\text{test}}^*) + \beta^{-1} \mathbf{I} - \mathbf{K}(\mathbf{X}_{\text{test}}^*, \mathbf{X}_{\text{train}}) \, (\mathbf{K}(\mathbf{X}_{\text{train}}, \mathbf{X}_{\text{train}}) + \beta^{-1} \mathbf{I})^{-1} \, \mathbf{K}(\mathbf{X}_{\text{train}}, \mathbf{X}_{\text{test}}^*) \quad (23)$$

Eq. (21) captures a posterior distribution, informed by the training set, which we can use to make new predictions about new data points of interest, $\mathbf{X}_{\text{test}}^*$. The posterior distribution in Eq. (21) specifies what the function values $\mathbf{f}_{\text{test}}^*$ should look like, given the already observed training data points, $\mathbf{X}_{\text{train}}$ and $\mathbf{f}_{\text{train}}$. Through Eq. (21), we have derived a non-parametric model. The model is non-parametric, because the Gaussian processes generate functions with certain characteristics defined by the kernel, but without use of true or explicit parametrization [18].

13. We can, furthermore, quantify the confidence associated with these predictions by calculating

$$\boldsymbol{\sigma}_{\text{test}}^* = \text{diag}(\boldsymbol{\Sigma}_{\text{test}}^*) \quad (24)$$

For a given point $i$ in the test set, the confidence interval associated with the prediction can, for example, be specified as $\pm 3 \, \sigma_{\text{test},i}^*$ around the predictive mean, $\mu_i^*$. The confidence intervals will be zero, at the points from the training set, but will grow as one goes further away from these known points. So the probabilistic modeling allows us to determine regions of large uncertainty. Information about the regions with the large uncertainty can be used to inform the sampling process. So if we want to improve the accuracy of our model, then it makes sense to gather new data points in the regions, where the uncertainty is large [18]. This is referred to as active learning. The active learning consists of first identifying the uncertain regions, and then gathering more data for these particular data points (in regions with large uncertainty) [18].

14. For additional information, refer to [18-23].

## 2.4 Prior View and Min MSE View

The prior view and min. MSE view capture alternative approaches for deriving the posterior probability or likelihood distributions for the GPR (see Eq. (21)), complementing the kernel view outlined above. For specifics, refer to [22, 23].

## 2.5 Optimization of Hyperparameters

In order to construct the posterior distribution in Eq. (10) – (12), for the noiseless case, or Eq. (21) – (23), for the noisy case, using the kernels from Eq. (7), the values for the hyperparameters, $\boldsymbol{\theta} = [\theta_0, \theta_1, \theta_2, \theta_3]$, need to be specified. While a reasonable choice of $\theta_0, \theta_1, \theta_2$ and $\theta_3$ may enable reasonable approximation of many functions, proper optimization of these hyperparameters will yield the best model results.

The simplest approach to the optimization to the optimization of the hyperparameters is based on looking at the training observations, for which we already know the distribution [18]:

$$\mathbf{f} \sim N(\mathbf{0}, \boldsymbol{C}_{\theta}(\mathbf{X}, \mathbf{X})) = \frac{1}{(2\pi)^{N/2} \, |\boldsymbol{C}_{\theta}|^{1/2}} \exp\left(-\frac{1}{2} \mathbf{f}^T \boldsymbol{C}_{\theta}^{-1} \mathbf{f}\right) \quad (25)$$

where $\boldsymbol{C}_{\theta}(\mathbf{X}, \mathbf{X}) = \mathbf{K}_{\theta}(\mathbf{X}, \mathbf{X}) + \beta^{-1} \mathbf{I}$. In other words, we can evaluate the probability of such a vector of observations being generated by the Gaussian process. Using Eq. (25), one can then construct the maximum likelihood estimate [18]

$$\max_{\theta} \ln(p(\mathbf{f}|\mathbf{X}, \boldsymbol{\theta})) = \max_{\theta} -\frac{1}{2} \ln|\boldsymbol{C}_{\theta}| - \frac{1}{2} \mathbf{f}^T \boldsymbol{C}_{\theta}^{-1} \mathbf{f} - \frac{N}{2} \ln(2\pi) \quad (26)$$

Although Eq. (26) may look somewhat complicated, one can arrive at a numerical solution for $\boldsymbol{\theta}$. One, for example, can carry out gradient descent optimization of the log-likelihood with respect to the model parameters, $\boldsymbol{\theta} = [\theta_0, \theta_1, \theta_2, \theta_3]$ [18].

## 3 Towards further assessment of the input data needed for accurate NN modeling (training)

As noted in the main manuscript, the extent of the input data needed for accurate neural network (NN) construction of the multi-dimensional surfaces may in general depend on the correlation of the input predictors (features) to the quantity being predicted (on the problem at hand), on the number of dimensions involved (on the structure of the NN) and possibly also on the training algorithm used. Of course, in the extreme case of the input predictors comprising of pure noise, i.e., with no correlation to the quantity being predicted, then no NN will generalize well, even when provided with arbitrary large volumes of input data.

As stated in Steingrimsson et al. [24], the complexity of a probabilistic model controls how much data is needed for the model to be suitable. Estimates for the data needed is often presented in the form of bounds, of the form $m \leq O(\cdot)$, for the so-called *sample-complexity* [24]. The *sample-complexity* represents the number of samples, which need to be drawn from a given probability density function (PDF), in order to construct an estimated PDF that approximates the true PDF within a given $\delta$ of accuracy [25]. The *sample-complexity* differs from the *algorithm complexity*, which specifies how the number of floating-point operations, that are needed to compute the output of an algorithm, scales with the size of the vector input provided to the algorithm [24].

Determination of the input data needed for accurate NN modeling is a profound, open question, with no simple answer available (at least not yet). As one works towards a solution, it is important to understand what type of probability distribution the machine learning (ML), or Bayesian inference, algorithm has been designed to approximate. Steingrimsson et al. [24] offer a probabilistic interpretation to the NN learning (training) process. Oftentimes, the ML or Bayesian inference algorithms are looking to approximate a Gaussian conditional PDF. In this case, one can employ well known results for the distribution and confidence interval for the mean and standard deviation of a Gaussian PDF, for the purpose of estimating the *sample-complexity* of Bayesian inferencing algorithms, since the Gaussian distribution is fully defined by a mean and a variance [24]. To estimate such a Gaussian PDF, one needs to draw large enough number of samples estimate the mean and variance within a given degree of accuracy. Eq. (27) prescribes a 100 (1 - $\alpha$)% confidence interval for the unknown mean, $\mu$, of a Gaussian distribution, given $n$ observations drawn and an unknown standard deviation, $s$ [26]:

$$\bar{x} - t_{\frac{\alpha}{2}, n-1} \frac{s}{\sqrt{n}} \leq \mu \leq \bar{x} + t_{\frac{\alpha}{2}, n-1} \frac{s}{\sqrt{n}}, \qquad (27)$$

Here, $\bar{x}$ represents the sample mean from the $n$ observations and $t_{\frac{\alpha}{2}, n-1}$ the upper (1 - $\alpha$)/2 critical value for a *t*-distribution with $n$ - 1 degrees of freedom. In other words [26],

$$P\left(-t_{\frac{\alpha}{2}, n-1} \leq \frac{\bar{x} - \mu}{\frac{s}{\sqrt{n}}} \leq +t_{\frac{\alpha}{2}, n-1}\right) = 1 - \alpha \qquad (28)$$

As larger number of samples are drawn from the distribution, $n$ increases and the bound in Eq. (27) becomes tighter. Similarly, Eq. (29) prescribes a 100 (1 - $\alpha$)% confidence interval for the variance, $\sigma^2$, of a Gaussian PDF [26]:

$$\frac{(n-1)S^2}{\chi^2_{\frac{\alpha}{2}, n-1}} \leq \sigma^2 \leq \frac{(n-1)S^2}{\chi^2_{1-\frac{\alpha}{2}, n-1}}, \qquad (29)$$

Here, $\chi^2_{\frac{\alpha}{2}, n-1}$ represents an upper (1 - $\alpha$)/2 critical value for a chi-squared distribution with $n - 1$ degrees of freedom. In other words [26],

$$P\left(\frac{(n-1)S^2}{\chi^2_{\frac{\alpha}{2}, n-1}} \leq \sigma^2 \leq \frac{(n-1)S^2}{\chi^2_{1-\frac{\alpha}{2}, n-1}}\right) = 1 - \alpha \qquad (30)$$

The authors refer to [27], for in-depth analysis of *sample-complexity* for the special cases of convolutional neural networks (CNNs) and recurrent neural networks (RNNs). In [27], a study is initiated of rigorously characterizing the *sample-complexity* needed for estimating CNNs and RNNs. Purportedly due to a more compact parametric representation compared to their Fully-Connected Neural Network (FNN) counterparts, CNNs and RNNs tend to require fewer training samples to accurately estimate their parameters [24, 27]. In recent years, researchers have made progress in developing theoretical understanding of various aspects of neural networks [27]. This has included development of provable learning algorithms [27-30]. In addition, theoretical understanding has been developed regarding hardness of estimation [27, 31, 32], the landscape of loss functions [27, 33-44], and the dynamics of gradient descent [45-47]. The work of Du et al. [27] differs from previous research work [28-30] in that Du et al. [27] only look at the *sample-complexity* and at the fundamental information theoretic limits of estimating a CNN, not at the computational complexity. Bounds on the ability of neural networks to generalize are often presented in the form [27]

$$L(\theta) - L_{tr}(\theta) \leq D/\sqrt{n}, \tag{31}$$

where $\theta$ represents parameters of the neural network, $L(\cdot)$ and $L_{tr}(\cdot)$ denote population and empirical errors under some additive loss, and $D$ model capacity [27]. For related work, refer to [48-55]. Du et al. demonstrate that the *sample-complexity* needed to learn CNNs and RNNs scales linearly with their intrinsic dimension and is much smaller than for the FNN counterparts [27]. Du et al. present lower bounds, both for CNNs and RNNs, which indicate that their *sample-complexities* are tight up to logarithmic factors [27].

The authors refer to [56], for analysis of size-independent *sample-complexity* of neural networks, to [57], for analysis of initialization-dependent *sample-complexity* of linear predictors and neural networks, to [58], for analysis of *sample-complexity* of one-hidden-layer neural networks, and to [59], for a theoretical perspective of *sample-complexity* for pruned neural networks. In [56], Golowich et al. study *sample-complexity* of neural networks and provide new bounds on their Rademacher-complexity, assuming norm constraints on the parameter matrix of each layer. These complexity bounds exhibit improved dependence on the network depth, compared to previous work, and under some additional assumptions, are fully independent of the network size (both of the NN depth and width). In [57], Magen et al. offer several new results on the *sample-complexity* of vector-valued linear predictors (parameterized by a matrix), and more generally neural networks. In the case of size-independent bounds, where only the Frobenius norm distance of the parameters from some fixed reference matrix, $W_0$, is controlled, Magen et al. illustrate that the *sample-complexity* behavior can be strikingly different from what one may expect considering the well-studied setting of scalar-valued linear predictors [57]. New *sample-complexity* bounds for feed-forward neural networks are presented, some open issues from the literature tackled, and a new convex linear prediction problem established, which is provably learnable without uniform convergence [57]. In [58], Vardi et al. study norm-based uniform convergence founds for neural networks, with the intent of developing tight understanding of how these NNs are affected by the architecture and type of norm constraint, for the relatively simple class of scalar-valued one-hidden-layer NNs and inputs bounded in Euclidean norm. Vardi et al. extend and improve upon previous results by proving that in general controlling the spectral norm of the weight matrix for the hidden layer is insufficient to obtain uniform convergence guarantees (independent of the network width), while a stronger Frobenius norm control is sufficient.

# 4 References

[1] T.G. Spears, S.A. Gold, In-process sensing in selective laser melting (SLM) additive manufacturing, Integrating Materials and Manufacturing Innovation 5(1) (2016) 16-40.

[2] B.A. Steingrimsson, P.K. Liaw, X. Fan, A.A. Kulkarni, D. Kim, Machine learning to accelerate alloy design, Google Patents, 2024.

[3] V.L. Deringer, A.P. Bartók, N. Bernstein, D.M. Wilkins, M. Ceriotti, G. Csányi, Gaussian Process Regression for Materials and Molecules, Chemical Reviews 121(16) (2021) 10073-10141.

[4] V. Maitra, J. Shi, C. Lu, Robust prediction and validation of as-built density of Ti-6Al-4V parts manufactured via selective laser melting using a machine learning approach, Journal of Manufacturing Processes 78 (2022) 183-201.

[5] I. Guyon, A scaling law for the validation-set training-set size ratio, AT&T Bell Laboratories 1(11) (1997).

[6] D.J. MacKay, Bayesian methods for neural networks: Theory and applications, Course notes for Neural Networks Summer School (1995).

[7] scikit-learn, 3.1. Cross-validation: evaluating estimator performance. https://scikit-learn.org/stable/modules/cross_validation.html, 2024 (accessed October 16.2024).

[8] D. Science, In k-fold-cross-validation, why do we compute the mean of the metric of each fold. 2024 (accessed October 16.2024).

[9] K. Bailey, Gaussian Processes for Dummies. https://katbailey.github.io/post/gaussian-processes-for-dummies/, 2019 (accessed October 23.2024).

[10] Wikipedia, Bayesian linear regression. https://en.wikipedia.org/wiki/Bayesian_linear_regression, 2024 (accessed October 23.2024).

[11] M. Monk, (ML 10.1) Bayesian Linear Regression. https://www.youtube.com/watch?v=dtkGq9tdYcI, 2024 (accessed October 23.2024).

[12] M. Monk, (ML 10.2) Posterior for linear regression (part 1). https://www.youtube.com/watch?v=nrd4AnDLR3U, 2024 (accessed October 23.2024).

[13] M. Monk, (ML 10.3) Posterior for linear regression (part 2). https://www.youtube.com/watch?v=qz2U8coNwV4, 2024 (accessed October 23.2024).

[14] M. Monk, (ML 10.4) Predictive distribution for linear regression (part 1). https://www.youtube.com/watch?v=xyuSiKXttxw, 2024 (accessed October 23.2024).

[15] M. Monk, (ML 10.5) Predictive distribution for linear regression (part 2). https://www.youtube.com/watch?v=vTcsacTqlfQ, 2024 (accessed October 23.2024).

[16] M. Monk, (ML 10.6) Predictive distribution for linear regression (part 3). https://www.youtube.com/watch?v=LCISTY9S6SQ, 2024 (accessed October 23.2024).

[17] M. Monk, (ML 10.7) Predictive distribution for linear regression (part 4). https://www.youtube.com/watch?v=g5BjTQghNf8, 2024 (accessed October 23.2024).

[18] E. Bekkers, 12.5 Gaussian Processes: Regression (UvA - Machine Learning 1). https://www.youtube.com/watch?v=fBDlurqqbYY&t=600s, 2020 (accessed October 23.2024).

[19] M. Statistics, Gaussian Process - Regression - Part 1 - Kernel First. https://www.youtube.com/watch?v=lWNy71IC8CU&t=1s, 2024 (accessed October 23.2024).

[20] E. Bekkers, 12.3 Gaussian Processes (UvA - Machine Learning 1). 2020 (accessed October 23.2024).

[21] E. Bekkers, 12.4 Gaussian Processes With An Exponential Kernel (UvA - Machine Learning 1). https://www.youtube.com/watch?v=Ksb8SasoO10&t=19s, 2020 (accessed October 23.2024).

[22] M. Statistics, Gaussian Process - Regression - Part 2 - Prior View. https://www.youtube.com/watch?v=ky_8vPyWmlo, 2024 (accessed October 23.2024).

[23] M. Statistics, Gaussian Process - Regression - Part 3 - Linear Estimator. https://www.youtube.com/watch?v=1y3tcJi6p1g, 2024 (accessed October 23.2024).

[24] B. Steingrimsson, X. Fan, A. Kulkarni, M.C. Gao, P.K. Liaw, Machine Learning and Data Analytics for Design and Manufacturing of High-Entropy Materials Exhibiting Mechanical or Fatigue Properties of Interest, in: J. Brechtl, P.K. Liaw (Eds.), High-Entropy Materials: Theory, Experiments, and Applications, Springer International Publishing, Cham, 2021, pp. 115-238.

[25] V. Vapnik, The nature of statistical learning theory, Springer science & business media, 2013.

[26] H. Pishro-Nik, Introduction to probability, statistics, and random processes, Kappa Research LLC, 2014.
[27] Y.W. S.S. Du, X. Zhai, S. Balakrishnan, R. Salakhutdinov, A. Singh, How Many Samples are Needed to Estimate a Convolutional or Recurrent Neural Network? https://arxiv.org/abs/1805.07883, 2018).
[28] S. Goel, A. Klivans, Eigenvalue decay implies polynomial-time learnability for neural networks, Advances in Neural Information Processing Systems 30 (2017).
[29] S. Goel, A. Klivans, Learning depth-three neural networks in polynomial time, arXiv preprint arXiv:1709.06010 (2017).
[30] Y. Zhang, J.D. Lee, M.J. Wainwright, M.I. Jordan, Learning halfspaces and neural networks with random initialization, arXiv preprint arXiv:1511.07948 (2015).
[31] S. Goel, V. Kanade, A. Klivans, J. Thaler, Reliably Learning the ReLU in Polynomial Time, in: K. Satyen, S. Ohad (Eds.) Proceedings of the 2017 Conference on Learning Theory, PMLR, Proceedings of Machine Learning Research, 2017, pp. 1004--1042.
[32] A. Brutzkus, A. Globerson, Globally Optimal Gradient Descent for a ConvNet with Gaussian Inputs, in: P. Doina, T. Yee Whye (Eds.) Proceedings of the 34th International Conference on Machine Learning, PMLR, Proceedings of Machine Learning Research, 2017, pp. 605--614.
[33] K. Kawaguchi, Deep learning without poor local minima, Advances in neural information processing systems 29 (2016).
[34] A. Choromanska, M. Henaff, M. Mathieu, G.B. Arous, Y. LeCun, The Loss Surfaces of Multilayer Networks, in: L. Guy, S.V.N. Vishwanathan (Eds.) Proceedings of the Eighteenth International Conference on Artificial Intelligence and Statistics, PMLR, Proceedings of Machine Learning Research, 2015, pp. 192--204.
[35] M. Hardt, T. Ma, Identity matters in deep learning, arXiv preprint arXiv:1611.04231 (2016).
[36] B.D. Haeffele, R. Vidal, Global optimality in tensor factorization, deep learning, and beyond, arXiv preprint arXiv:1506.07540 (2015).
[37] C.D. Freeman, J. Bruna, Topology and geometry of half-rectified network optimization, arXiv preprint arXiv:1611.01540 (2016).
[38] I. Safran, O. Shamir, On the Quality of the Initial Basin in Overspecified Neural Networks, in: B. Maria Florina, Q.W. Kilian (Eds.) Proceedings of The 33rd International Conference on Machine Learning, PMLR, Proceedings of Machine Learning Research, 2016, pp. 774--782.
[39] P. Zhou, J. Feng, The landscape of deep learning algorithms, arXiv preprint arXiv:1705.07038 (2017).
[40] Q. Nguyen, M. Hein, The Loss Surface of Deep and Wide Neural Networks, in: P. Doina, T. Yee Whye (Eds.) Proceedings of the 34th International Conference on Machine Learning, PMLR, Proceedings of Machine Learning Research, 2017, pp. 2603--2612.
[41] Q. Nguyen, M. Hein, The loss surface and expressivity of deep convolutional neural networks, (2018).
[42] R. Ge, J.D. Lee, T. Ma, Learning one-hidden-layer neural networks with landscape design, arXiv preprint arXiv:1711.00501 (2017).
[43] I. Safran, O. Shamir, Spurious Local Minima are Common in Two-Layer ReLU Neural Networks, in: D. Jennifer, K. Andreas (Eds.) Proceedings of the 35th International Conference on Machine Learning, PMLR, Proceedings of Machine Learning Research, 2018, pp. 4433--4441.
[44] S. Du, J. Lee, On the Power of Over-parametrization in Neural Networks with Quadratic Activation, in: D. Jennifer, K. Andreas (Eds.) Proceedings of the 35th International Conference on Machine Learning, PMLR, Proceedings of Machine Learning Research, 2018, pp. 1329--1338.
[45] Y. Tian, An Analytical Formula of Population Gradient for two-layered ReLU network and its Applications in Convergence and Critical Point Analysis, in: P. Doina, T. Yee Whye (Eds.) Proceedings of the 34th International Conference on Machine Learning, PMLR, Proceedings of Machine Learning Research, 2017, pp. 3404--3413.
[46] K. Zhong, Z. Song, P. Jain, P.L. Bartlett, I.S. Dhillon, Recovery Guarantees for One-hidden-layer Neural Networks, in: P. Doina, T. Yee Whye (Eds.) Proceedings of the 34th International

Conference on Machine Learning, PMLR, Proceedings of Machine Learning Research, 2017, pp. 4140--4149.
[47] Y. Li, Y. Yuan, Convergence analysis of two-layer neural networks with relu activation, Advances in neural information processing systems 30 (2017).
[48] S. Arora, R. Ge, B. Neyshabur, Y. Zhang, Stronger generalization bounds for deep nets via a compression approach, International Conference on Machine Learning, PMLR, 2018, pp. 254-263.
[49] M. Anthony, P.L. Bartlett, P.L. Bartlett, Neural network learning: Theoretical foundations, Cambridge University Press, Cambridge, 1999.
[50] P.L. Bartlett, D.J. Foster, M.J. Telgarsky, Spectrally-normalized margin bounds for neural networks, Advances in neural information processing systems 30 (2017).
[51] P.L. Bartlett, N. Harvey, C. Liaw, A. Mehrabian, Nearly-tight VC-dimension and pseudodimension bounds for piecewise linear neural networks, Journal of Machine Learning Research 20(63) (2019) 1-17.
[52] B. Neyshabur, S. Bhojanapalli, N. Srebro, A pac-bayesian approach to spectrally-normalized margin bounds for neural networks, arXiv preprint arXiv:1707.09564 (2017).
[53] K. Pitas, M. Davies, P. Vandergheynst, Pac-bayesian margin bounds for convolutional neural networks, arXiv preprint arXiv:1801.00171 (2017).
[54] X. Li, J. Lu, Z. Wang, J. Haupt, T. Zhao, On tighter generalization bound for deep neural networks: Cnns, resnets, and beyond, arXiv preprint arXiv:1806.05159 (2018).
[55] Y. Li, Y. Liang, Learning overparameterized neural networks via stochastic gradient descent on structured data, Advances in neural information processing systems 31 (2018).
[56] N. Golowich, A. Rakhlin, O. Shamir, Size-Independent Sample Complexity of Neural Networks, in: B. Sébastien, P. Vianney, R. Philippe (Eds.) Proceedings of the 31st Conference On Learning Theory, PMLR, Proceedings of Machine Learning Research, 2018, pp. 297--299.
[57] R. Magen, O. Shamir, Initialization-Dependent Sample Complexity of Linear Predictors and Neural Networks, Advances in Neural Information Processing Systems 36 (2024).
[58] G. Vardi, O. Shamir, N. Srebro, The sample complexity of one-hidden-layer neural networks, Advances in Neural Information Processing Systems 35 (2022) 9139-9150.
[59] S. Zhang, M. Wang, S. Liu, P.-Y. Chen, J. Xiong, Why lottery ticket wins? a theoretical perspective of sample complexity on sparse neural networks, Advances in Neural Information Processing Systems 34 (2021) 2707-2720.